\documentclass[aps,prd,twocolumn,reprint,amsmath,amssymb,preprintnumbers,superscriptaddress,nobibnotes,nofootinbib,floatfix]{revtex4-1}

\usepackage{graphicx}
\usepackage{dcolumn}
\usepackage{bm}
\usepackage[colorlinks=true,allcolors=Purple,pdfusetitle]{hyperref}
\usepackage{rotating}
\usepackage{multirow}
\usepackage{graphicx}
\usepackage[dvipsnames]{xcolor}
\usepackage{esdiff}

\newcommand{\nocontentsline}[3]{}
\newcommand{\tocless}[2]{\bgroup\let\addcontentsline=\nocontentsline#1{#2}\egroup}

\newcommand{\orcid}[1]{\href{https://orcid.org/#1}{\includegraphics[width=10pt]{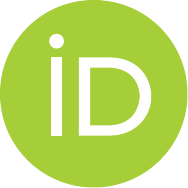}}}

\begin{document}

\title{Distinctive nuclear signatures of low-energy atmospheric neutrinos}

\author{Anna M. Suliga \orcid{0000-0002-8354-012X}\,}
\email{asuliga@berkeley.edu}
\affiliation{Department of Physics, University of California Berkeley, Berkeley, California 94720, USA}
\affiliation{Department of Physics, University of Wisconsin--Madison, Madison, Wisconsin 53706, USA}

\author{John F. Beacom \orcid{0000-0002-0005-2631}\,}
\email{beacom.7@osu.edu}
\affiliation{Center for Cosmology and AstroParticle Physics (CCAPP), Ohio State University, Columbus, OH 43210, USA}
\affiliation{Department of Physics, Ohio State University, Columbus, OH 43210, USA}
\affiliation{Department of Astronomy, Ohio State University, Columbus, OH 43210, USA}

\date{August 31, 2023}

\preprint{N3AS-23-002}
\preprint{INT-PUB-23-020}

\begin{abstract}
New probes of neutrino mixing are needed to advance precision studies.  One promising direction is via the detection of low-energy atmospheric neutrinos (below a few hundred MeV), to which a variety of near-term experiments will have much-improved sensitivity.  Here we focus on probing these neutrinos through distinctive nuclear signatures of exclusive neutrino-carbon interactions --- those that lead to detectable nuclear-decay signals with low backgrounds --- in both neutral-current and charged-current channels.  The neutral-current signature is a line at 15.11 MeV and the charged-current signatures are two- or three-fold coincidences with delayed decays.  We calculate the prospects for identifying such events in the Jiangmen Underground Neutrino Observatory (JUNO), a large-scale liquid-scintillator detector.  A five-year exposure would yield about 16 neutral-current events (all flavors) and about 16 charged-current events (mostly from $\nu_e + \bar{\nu}_e$, with some from $\nu_\mu + \bar{\nu}_\mu$), and thus roughly 25\% uncertainties on each of their rates.  Our results show the potential of JUNO to make the first identified measurement of sub-100~MeV atmospheric neutrinos.  They also are a step towards multi-detector studies of low-energy atmospheric neutrinos, with the goal of identifying additional distinctive nuclear signatures for carbon and other targets.
\end{abstract}

\maketitle

%%%%%%%%%%%%%%%%%%%%%%%%%%%%%%%%%%%%%%%%%%%%%%%%%%%%%%%%%%
%%%%%%%%%%%%%%%%%%%%%%%%%%%%%%%%%%%%%%%%%%%%%%%%%%%%%%%%%%

\section{Introduction}
\label{sec:Introduction}

Understanding neutrino flavor mixing is critical to our framework for the weak sector.  Past measurements~\cite{Haxton:2012wfz, Kajita:2014koa, Vitagliano:2019yzm, ParticleDataGroup:2022pth, deGouvea:2022gut} of mixing among neutrinos established several key points, including that at least two of their masses are nonzero, that their mass-squared splittings are small but their mixing angles are large, and that their mixing is affected by the presence of matter.  Future precise mixing measurements~\cite{IceCube-Gen2:2020qha, KM3NeT:2021ozk, DUNE:2022aul, JUNO:2022mxj, Hyper-Kamiokande:2022smq, Denton:2022een, Huber:2022lpm} will be used to pursue even more ambitious goals, among them determining the mass ordering, breaking an octant degeneracy for one angle, and probing the size of CP-violating effects.

These new neutrino-mixing measurements will be conducted with a variety of approaches, which will be valuable for improved probes of mixing and for testing new physics~\cite{Abdullahi:2022jlv, Arguelles:2022tki, Berryman:2022hds, Gerbino:2022nvz, Acero:2022wqg, DUNE:2022aul}.  The sources used will include MeV-range reactor antineutrinos, MeV-range solar neutrinos, GeV-range accelerator neutrinos and antineutrinos (separately), and GeV-range atmospheric neutrinos and antineutrinos (together).  For these measurements, depending on the setup, matter effects on mixing due to the electron density will be vanishing, moderately important, or dominant. If MeV-range supernova neutrinos are detected, there will also be matter effects on mixing due to the neutrino density~\cite{Duan:2010bg, Chakraborty:2016yeg, Tamborra:2020cul, Richers:2022zug, Patwardhan:2022mxg, Volpe:2023met}.

In this paper, building on earlier work~\cite{Gaisser:1986bv, Nussinov:2000qc, Kolbe:2002gk, Kelly:2019itm, Newstead:2020fie, Denton:2021mso, Zhuang:2021rsg, Kelly:2023ugn}, we make a general point that the collection of future detectors should enable measuring neutrino mixing in yet another way, using low-energy (below a few hundred MeV) atmospheric neutrinos.  In this energy range, which has barely been explored in experimental analyses, the key physics advantage is potential sensitivity to CP-violating effects~\cite{Peres:2003wd, Mena:2008rh, Akhmedov:2012ah, Kelly:2019itm, Ioannisian:2020isl}.  While this measurement will be highly challenging, measuring CP violation is extremely important and no approach is easy, so it will be valuable to have multiple methods.  Here the obstacles include large uncertainties on the unoscillated fluxes, the neutrino-nucleus interactions, and even the detector capabilities.  However, we are optimistic that a broad collection of measurements can reduce these individual uncertainties --- which would be valuable in its own right --- and thus enable new probes of neutrino mixing.  This will also improve the sensitivity of searches for which atmospheric-neutrino events are backgrounds.

As a specific step towards this goal, in this paper we focus on low-energy atmospheric-neutrino interactions in the Jiangmen Underground Neutrino Observatory (JUNO)~\cite{JUNO:2015zny, JUNO:2015sjr, JUNO:2021vlw}, a liquid-scintillator detector with a fiducial mass of 20 kton that will start taking data in 2024. Building on earlier work that considered {\it inclusive} {(and certain exclusive)} atmospheric-neutrino interactions in JUNO~\cite{Cheng:2020aaw, Cheng:2020oko, JUNO:2021tll}, here we provide new predictions for the {\it exclusive} channels for which the neutrino-nucleus interactions and detector response are best understood, though not completely.  We focus on the channels corresponding to superallowed transitions within the $A = 12$ triad [carbon (C), nitrogen (N), and boron (B)]~\cite{Fukugita:1988hg, Engel:1996zt, Kubodera:1993rk, Kolbe:1996km}. Our work is complementary to that of Ref.~\cite{Cheng:2020aaw}, which did not call attention to these channels. For these channels, the detector backgrounds are low and there is some ability to separate flavor as well as neutrinos and antineutrinos.

%%%%%%%%%%%%%%%%%%%%%%%%%%%%%%%%%%%%%%%%%%%%%
\begin{figure*}[t]
\centering
\includegraphics[width=0.99\columnwidth]{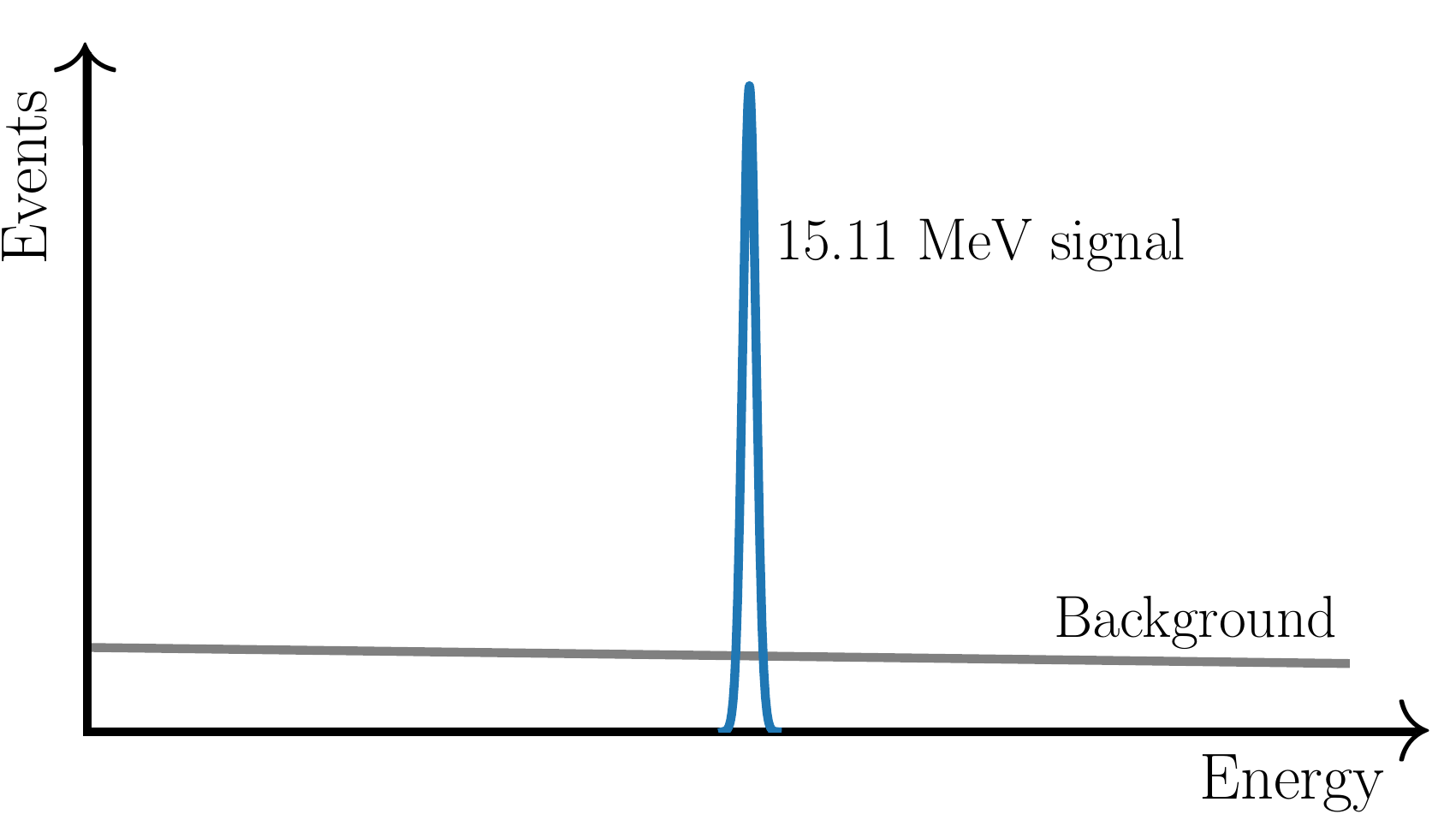}
\includegraphics[width=0.99\columnwidth]{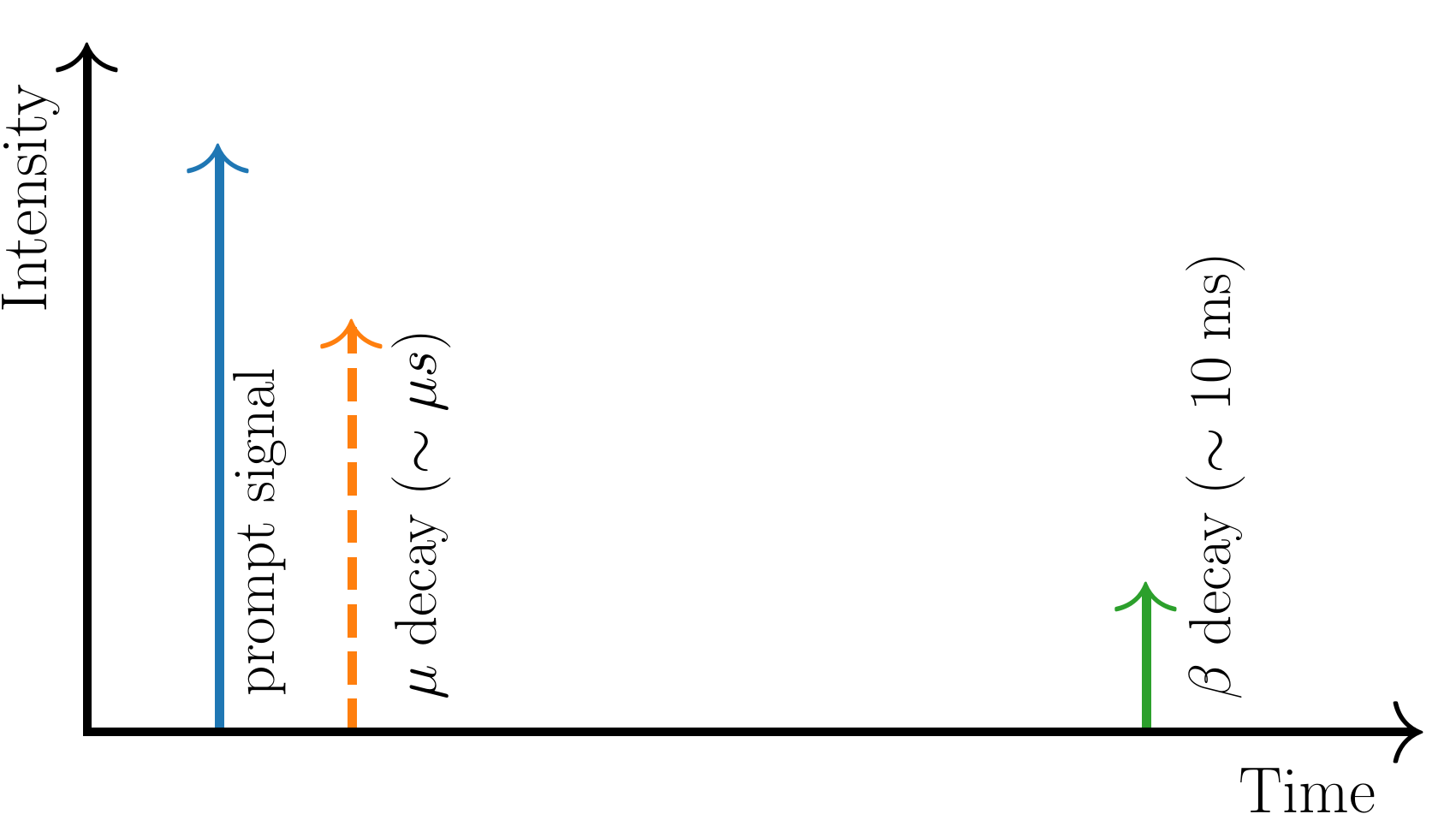}
\caption{
Schematic illustration of background rejection for the considered exclusive neutrino-carbon interactions.  {\it Left panel:} For the NC channels, the primary feature is the line-spectrum nature of the signal compared to the continuum backgrounds.  {\it Right panel:} For the CC channels, the primary feature is the two-part (for $\nu_e$ and $\bar{\nu}_e$) or three-part (for $\nu_\mu$ and $\bar{\nu}_\mu$) nature of the signal, for which there are essentially no backgrounds.
}
\label{fig:schematic}
\end{figure*}
%%%%%%%%%%%%%%%%%%%%%%%%%%%%%%%%%%%%%%%%%%%%%

Figure~\ref{fig:schematic} schematically illustrates the special detection features of the exclusive channels we consider.  First, there are two neutral-current (NC) channels, sensitive to all flavors of neutrinos and antineutrinos:
\begin{subequations}
\label{eq:carbon-deex} 
\begin{align}
\nu + {}^{12}\mathrm{C} \rightarrow \nu + {}^{12}\mathrm{C}^{*} & \label{eq:carbon-deex:nu} \\
{}^{12}\mathrm{C}^{*} \rightarrow {}^{12}\mathrm{C} + \gamma & \,, \nonumber \\
\nonumber \\
\bar\nu + {}^{12}\mathrm{C} \rightarrow \bar\nu + {}^{12}\mathrm{C}^{*} & \label{eq:carbon-deex:nubar} \\
{}^{12}\mathrm{C}^{*} \rightarrow {}^{12}\mathrm{C} + \gamma & \,. \nonumber
\end{align}
\end{subequations}
For both neutrinos and antineutrinos, the threshold energy for the transition to reach the dominant $1^+$ excited state is 15.11 MeV; this excitation decays via the emission of a gamma ray of this energy.  The high energy of this line and the excellent energy resolution of JUNO will allow strong background suppression, though this signal gives no information on the neutrino flavor.

Second, there are $\nu_e + \bar{\nu}_e$ charged-current (CC) interactions:
\begin{subequations}
\label{eq:CC-nu} 
\begin{align}
\nu_e + {}^{12}\mathrm{C} \rightarrow e^- + {}^{12}\mathrm{N}_\mathrm{g.s.} &  \label{eq:CC-nue} \\ 
{}^{12}\mathrm{N}_\mathrm{g.s.} \rightarrow {}^{12}\mathrm{C} + e^+ + \nu_e & \,, \nonumber \\ 
\nonumber \\
\bar\nu_e + {}^{12}\mathrm{C} \rightarrow e^+ + {}^{12}\mathrm{B}_\mathrm{g.s.} &  \label{eq:CC-nuebar} \\ 
{}^{12}\mathrm{B}_\mathrm{g.s.} \rightarrow {}^{12}\mathrm{C} + e^- + \bar \nu_e & \,. \nonumber
\end{align}
\end{subequations}
For neutrinos, the threshold energy to reach the ground state is 17.34~MeV; the $^{12}\mathrm{N}_\mathrm{g.s.}$ then beta-decays with a $Q$-value of 17.34~MeV and a halflife of 11~ms~\cite{Kelley:2017qgh}.  For antineutrinos, the threshold energy is 14.39 MeV; the $^{12}\mathrm{B}_\mathrm{g.s.}$ then beta-decays with a $Q$-value of 13.37 MeV and a halflife of 20~ms~\cite{Kelley:2017qgh}.  The two-part coincidence nature of these events will powerfully reduce backgrounds.  There should be some ability to separate neutrino- and antineutrino induced events by taking into account the slightly different nuclear decay properties and through positron tagging (as in Borexino~\cite{Borexino:2011ufb, Borexino:2013zhu}).  Both the $^{12}\mathrm{B}_\mathrm{g.s.}$ and $^{12}\mathrm{N}_\mathrm{g.s.}$ decay primarily ($> 98\%$ of the time) to the ground state of the carbon nucleus~\cite{Kelley:2017qgh}).

Third, there are $\nu_\mu + \bar{\nu}_\mu$ CC interactions:
\begin{subequations}
\label{eq:CC-nubar} 
\begin{align}
& \nu_\mu + {}^{12}\mathrm{C} \rightarrow  \mu^- + {}^{12}\mathrm{N}_\mathrm{g.s.} \label{eq:CC-numu} \\
& \mu^- \rightarrow e^- + \bar\nu_e + \nu_\mu  \ \nonumber \\
& {}^{12}\mathrm{N}_\mathrm{g.s.} \rightarrow {}^{12}\mathrm{C} + e^+ + \nu_e  \,, \nonumber \\
\nonumber \\
& \bar\nu_\mu + {}^{12}\mathrm{C} \rightarrow \mu^+ + {}^{12}\mathrm{B}_\mathrm{g.s.} \label{eq:CC-numubar} \\
& \mu^+ \rightarrow e^+ + \nu_e + \bar\nu_\mu\nonumber \\
& {}^{12}\mathrm{B}_\mathrm{g.s.} \rightarrow {}^{12}\mathrm{C} + e^- + \bar \nu_e \,. \nonumber
\end{align}
\end{subequations}
For neutrinos, the threshold energy is 122.49~MeV; for antineutrinos, it is 119.54~MeV.  The muons will come to rest and produce decay electrons or positrons (with halflife 1.4$\:\mu$s, taking into account $\mu^{-}$ capture in carbon-based liquid scintillator~\cite{Measday:2001yr, LSND:2002oco, DoubleChooz:2015jlf}) with a maximum kinetic energy of 52.3~MeV. The details of the nuclear decays are the same as above.  The three-part coincidence nature of these events will even more powerfully reduce backgrounds.  There should be some ability to separate neutrino- and antineutrino-induced events via the points noted above, plus the fact that about 8\% of $\mu^-$ will undergo nuclear capture~\cite{Measday:2001yr, LSND:2002oco, DoubleChooz:2015jlf}.

The exclusive channels considered here correspond to a small fraction of the total cross section, which includes many other final states, some of which may also have distinctive nuclear signatures (see Sec.~\ref{sec:cross-sections})

This remainder of this paper is organized as follows.  In Sec.~\ref{sec:atm-flux}, we review the atmospheric-neutrino fluxes, mixing effects, and uncertainties.  In Sec.~\ref{sec:cross-sections}, we detail the calculations and measurements of the exclusive neutrino-carbon interactions.  In Sec.~\ref{sec:Detector}, we define our detection strategy and present our main calculations of signals and backgrounds for JUNO.  We conclude by summarizing our results and placing them in a larger context.  In our calculations, we aim for a precision on the event yields of a few tens of percent, which is appropriate given the existing uncertainties and the expected statistics; ultimately, this should be improved.

%%%%%%%%%%%%%%%%%%%%%%%%%%%%%%%%%%%%%%%%%%%%%%%%%%%%%%%%%%
%%%%%%%%%%%%%%%%%%%%%%%%%%%%%%%%%%%%%%%%%%%%%%%%%%%%%%%%%%

\section{Atmospheric-neutrino fluxes}
\label{sec:atm-flux}

In this section, we review predictions of the sub-GeV atmospheric-neutrino fluxes and their uncertainties, plus the effects of neutrino mixing.

Atmospheric neutrinos are created primarily by the decays of mesons produced via hadronic cosmic-ray interactions with Earth's atmosphere~\cite{Greisen1960, Gaisserbook, Gaisser:2002jj, Kajita:2014koa, Vitagliano:2019yzm}.  At the low energies we consider, the dominant production processes are pion and muon decays,
\begin{equation}
\label{eq:atm1}
\begin{split}
& \pi^- \rightarrow \bar\nu_\mu \; + \;  \mu^- \\
& \mu^- \rightarrow e^- + \bar\nu_e + \nu_\mu \,,
\end{split}
\end{equation}
for negatively charged pions and analogously for positively charged pions.  Because pions of various types are produced with comparable yields (taking into account the neutrons within nuclei in both the beam and target) and because the muons mostly decay before reaching Earth, the initial flavor ratios for the sums of neutrinos and antineutrinos are close to $f_{\nu_e}:f_{\nu_\mu}:f_{\nu_\tau}=1:2:0$~\cite{Gaisserbook}; each flux ratio representing its part of the total flux.

Despite the simplicity of those points, it is challenging to accurately predict the low-energy atmospheric-neutrino fluxes, due to hadronic-interaction uncertainties and the need to take non-collinear propagation into account.  In addition, at low energies, two effects suppress the fluxes.  One effect is solar modulation, the result of cosmic rays interacting with the solar wind~\cite{Potgieter:2013pdj, Moraal:2013jxa, Li:2022zio}, the strength of which varies over the 11-year solar cycle, giving larger fluxes near solar minimum.  The scale of this variation in the atmospheric-neutrino flux for JUNO is $\sim\pm$5--30\%~\cite{Cheng:2020aaw, Zhuang:2021rsg}.  Another effect is due to Earth's geomagnetic cutoff, which prevents low-energy cosmic rays from reaching the atmosphere, and which leads to location-dependent differences in the atmospheric-neutrino flux up to $\sim\pm25$\%~\cite{Honda:2001xy, Barr:2004br, Zhuang:2021rsg, Kelly:2023ugn}.

Full 3-d modeling of the low-energy atmospheric-neutrino flux, including the listed effects, has been performed by several groups~\cite{Honda:2001xy, Liu:2002sq, Battistoni:2005pd, Wentz:2003bp, Zhuang:2021rsg}.  We use the results of Ref.~\cite{Zhuang:2021rsg}, one of only two (see also Ref.~\cite{Battistoni:2005pd}) to go below 100 MeV.  These results are in reasonable accord with each other below 100 MeV and with others above 100 MeV.  The fluxes we use are calculated for the location of the China Jinping Underground Laboratory, which is approximately 400~km from the JUNO site; these fluxes are close to those adopted for JUNO in Ref.~\cite{JUNO:2021tll}. We have checked that the differences are negligible with fluxes generated for the JUNO location from Ref.~\cite{Yi-Zhuang:2023}. While Ref.~\cite{Zhuang:2021rsg} provides only the total flux, we assume initial flavor ratios near $1:2:0$, following the references above.

Because the statistics will be low, we use the time-averaged flux.  And because it is difficult to measure the direction of events in scintillator detectors, due to the isotropic nature of the light emission (possible exceptions are noted below), we use the direction-integrated flux.  

\begin{figure}[t]
\centering
\includegraphics[width=0.99\columnwidth]{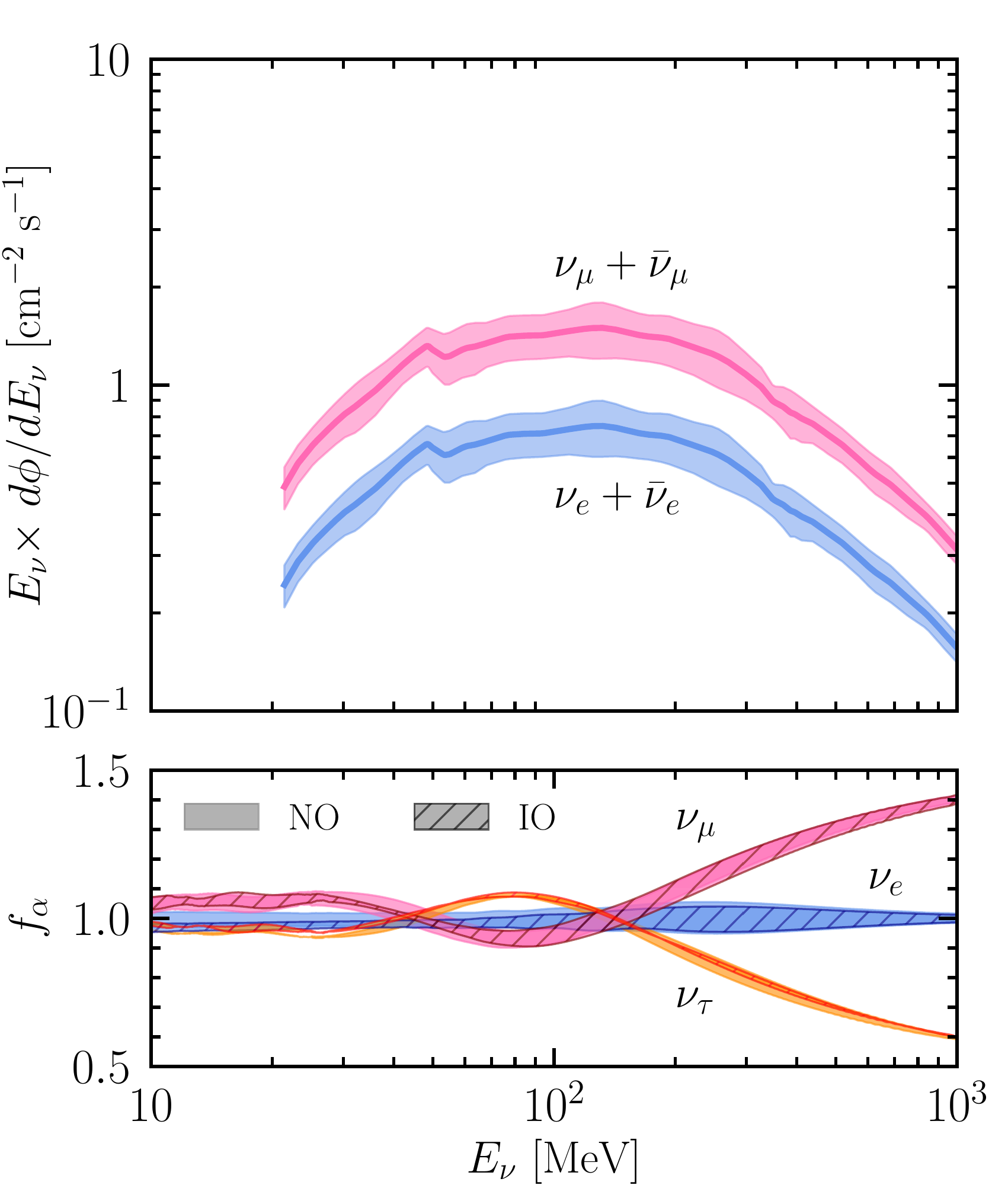}
\caption{
{\it Upper panel:} The time-averaged and direction-integrated atmospheric-neutrino spectra (before mixing) for JUNO~\cite{Zhuang:2021rsg}. 
{\it Lower panel:} The flux ratios under different mixing scenarios, starting from $1:2:0$; the main effect of mixing is to decrease the $\nu_\mu$ flux and increase the $\nu_\tau$ flux.
}
\label{fig:flux}
\end{figure}

Figure~\ref{fig:flux} (upper panel) shows the unmixed spectra; we use the prediction bands from the two upper panels of Fig.~10 in Ref.~\cite{Zhuang:2021rsg}.  (We note that Ref.~\cite{Zhuang:2021rsg} uses somewhat smaller flux uncertainties than Ref.~\cite{Cheng:2020aaw}.)  To match the logarithmic energy scale on the $x$-axis, we plot $E_\nu d\phi/dE_\nu = (2.3)^{-1} d\phi/d\log_{10} E_\nu$, so that the height of a point on the curve is proportional to its contribution to the integrated flux.  The overall shape of the spectrum follows primarily from basic considerations.  At higher energies (above a few hundred MeV), the neutrino spectra follow the power-law shapes of their parent pion and cosmic-ray spectra, with an average relation of $E_\nu \simeq 0.05 E_{\rm p}$, a consequence of pions carrying $\sim$1/5 of the proton energy and neutrinos carrying $\sim$1/4 of the pion energy~\cite{Gaisserbook}.  At lower energies, the neutrino spectra display a ``pion bump" feature analogous to that familiar from astrophysical gamma-ray production via $p + p \rightarrow p + p + \pi^0$ followed by $\pi^0 \rightarrow \gamma + \gamma$~\cite{Steckerbook, Kelner:2006tc}.  This bump arises due to the threshold effects and kinematics of pion production at low energies.

%%%%%%%%%%%%%%%%%%%%%%%%%%%%%%%%%%%%%%%%%%%%%
%%%%%%%%%%%%%%%%%%%%%%%%%%%%%%%%%%%%%%%%%%%%% 

Figure~\ref{fig:flux} (lower panel) shows the effects of neutrino mixing, which we take into account using the {\tt nuCRAFT} code~\cite{Wallraff:2014qka}, which calculates vacuum mixing as well as small corrections due to matter effects~\cite{Wolfenstein:1977ue, Mikheyev:1985zog}, including parametric resonances~\cite{Akhmedov:1988kd, Krastev:1989ix}.  We average over the production height, assuming a flat distribution between 10--40~km at production in the atmosphere~\cite{Wallraff:2014qka}, which is adequate for our purposes.  The bands for the two considered oscillation scenarios --- normal ordering (NO) and inverted ordering (IO) --- were estimated simply by plugging in the $\pm 3\sigma$ mixing-parameter values from Ref.~\cite{ParticleDataGroup:2022pth}; they are hardly different between the two orderings.

The behavior of the flavor ratios with energy can be mostly understood from successive applications of two-flavor vacuum mixing.  At high energies, e.g., $\sim50~\mathrm{GeV}$, much beyond the range of this figure, neutrino mixing effects are modest because $\Delta m^2 L/E$ is small for all neutrino directions.  (At higher energies, the initial flavor ratios are different from $1:2:0$ because muon decays in flight are suppressed~\cite{Gaisser:2002jj}.)  In the 1000-MeV range and lower, ``atmospheric" mixing (with $\Delta m^2 = 2.5 \times 10^{-3} \; \mathrm{eV}^2$ and a near-maximal mixing angle) become increasingly important for most directions, changing the flavor ratios to about $1:1:1$.  In the 100-MeV range and lower, there are some additional effects caused by ``solar" mixing with ($\Delta m^2 = 7.4 \times 10^{-5}\;\mathrm{eV}^2$ and two relatively large mixing angles), but these are modest because the flavor ratios were already nearly equilibrated.  In Appendix~\ref{app:atm-osc}, we give further details on the mixing results.

%%%%%%%%%%%%%%%%%%%%%%%%%%%%%%%%%%%%%%%%%%%%%%%%%%%%%%%%%%%%%%
%%%%%%%%%%%%%%%%%%%%%%%%%%%%%%%%%%%%%%%%%%%%%%%%%%%%%%%%%%%%%%

\section{Exclusive neutrino-nuclear cross sections}
\label{sec:cross-sections}

%%%%%%%%%%%%%%%%%%%%%%%%%%%%%%%%%%%%%%%%%%%%%
\begin{figure*}[t]
\centering
\includegraphics[width=0.66\columnwidth]{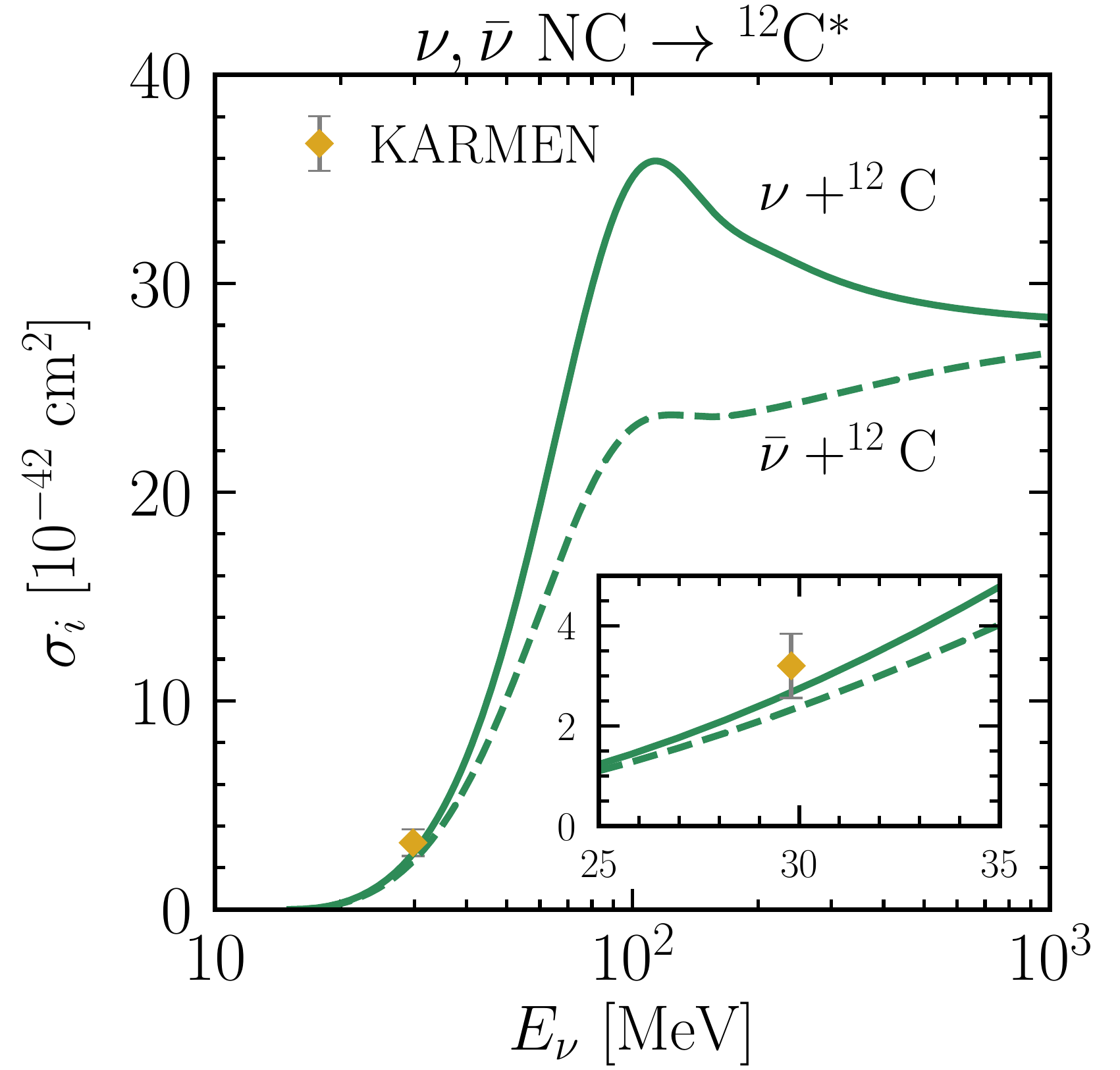}
\includegraphics[width=0.66\columnwidth]{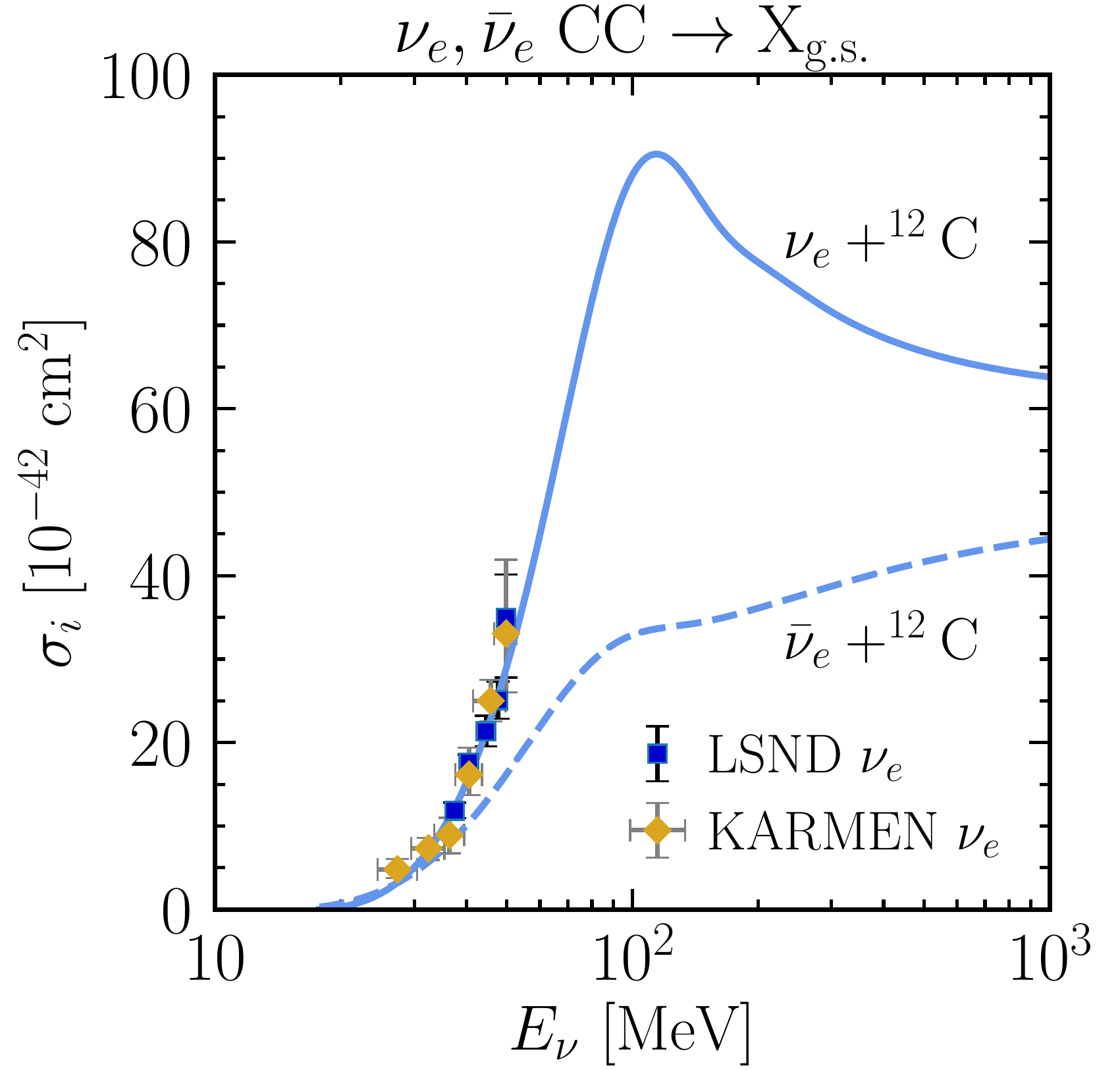}
\includegraphics[width=0.66\columnwidth]{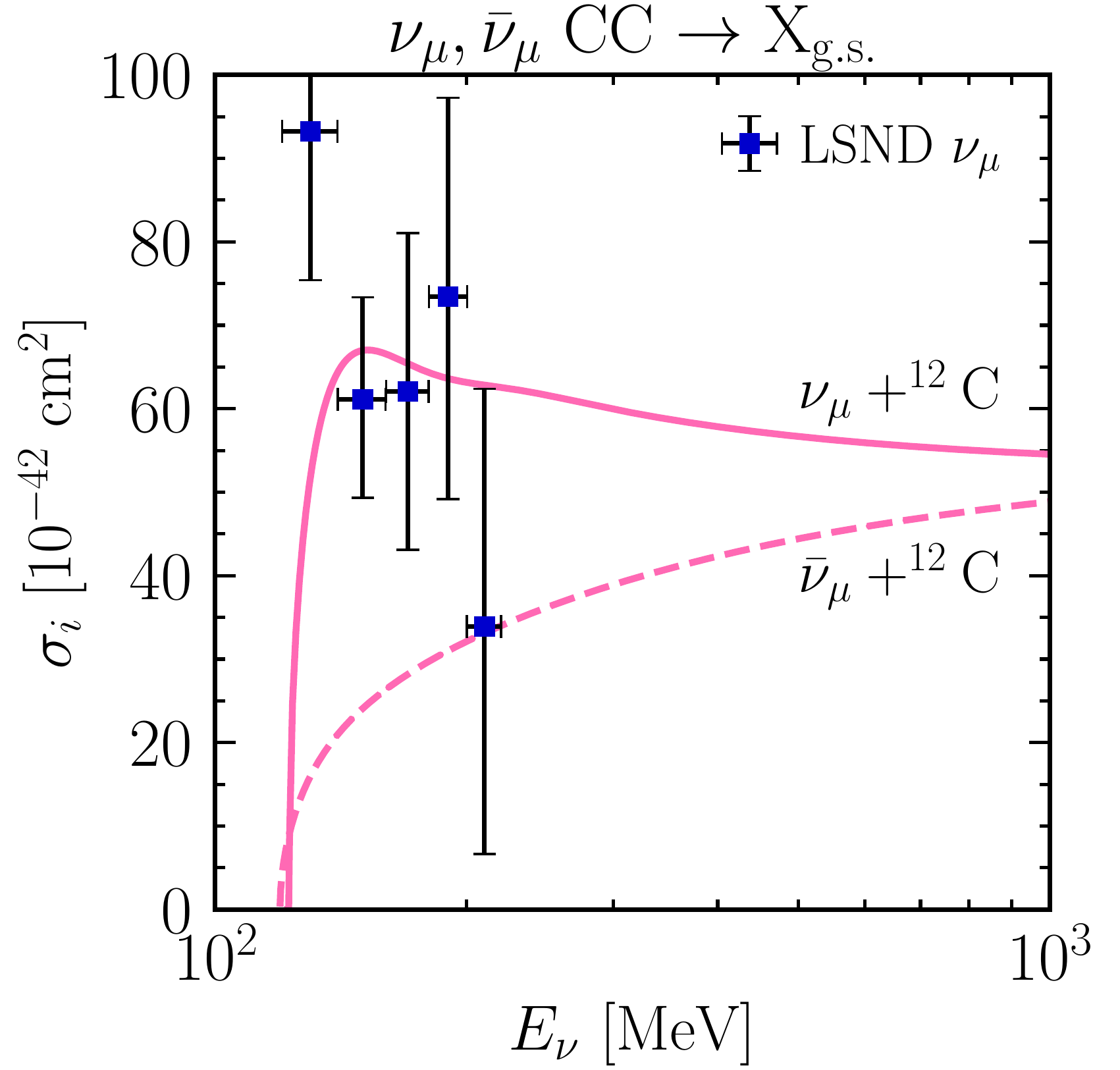}
\caption{
Calculated neutrino-nucleus cross sections compared to data.  {\it Left panel:} NC (all-flavor) transitions to the 15.11-MeV state in $^{12}$C, with data from KARMEN ($\nu_\mu$)~\cite{ KARMEN:1998xmo}.  {\it Middle panel:} CC $\nu_e$ and $\bar{\nu}_e$ transitions to the ground states of $^{12}$N and $^{12}$B, with data from KARMEN~\cite{Zeitnitz:1994kz} and LSND (both $\nu_e$ only)~\cite{LSND:1997lta, LSND:2001fbw}.  {\it Right panel:} CC $\nu_\mu$ and $\bar{\nu}_\mu$ transitions to the ground states of $^{12}$N and $^{12}$B, with data from LSND ($\nu_\mu$ only)~\cite{LSND:2002oco}. Note differences in the axis ranges.
}
\label{fig:cross-sections}
\end{figure*}
%%%%%%%%%%%%%%%%%%%%%%%%%%%%%%%%%%%%%%%%%%%%%

In this section, we detail the considered exclusive neutrino-carbon cross sections.  In Secs.~\ref{sec:cross-section-NC}--\ref{sec:CC-nu-mu}, we review (following especially Refs.~\cite{Fukugita:1988hg, Kubodera:1993rk}) the formulas used to calculate these, and in Sec.~\ref{sec:cros-section-measurements}, we summarize the corresponding experimental measurements.

We focus only on the superallowed transitions from the $0^+$ ground state in ${}^{12}$C to the $1^+$ states in ${}^{12}$C, ${}^{12}$N and ${}^{12}$B because they give distinctive signals~\cite{Fukugita:1988hg, Engel:1996zt, Kubodera:1993rk, Kolbe:1996km}.  For the CC reactions, we neglect transitions to excited states in $^{12}$B or $^{12}$N, as these mostly have low nucleon-separation energies and thus lead to different final states, which they themselves have distinctive nuclear signatures.

In this paper, we estimate the cross sections using the elementary particle treatment (EPT)~\cite{Akihiko:1964, Kim:1965zzc, Kubodera:1973kv, Hwang:1977fi, Nozawa:1983imd}.  While the EPT method has known problems (e.g., poorly taking into account high-momentum modes, neglecting nucleon-nucleon correlations, etc.), it is adequate for our purposes here, as there are other input and statistical (shown below) uncertainties at the level of a few tens of percent.  In addition, this paper is a first step in the dialog with the JUNO collaboration about realistic detection prospects.  In future work, we will seek to refine the cross-section calculations using modern methods, such as chiral perturbation theory currents or large basis shell models~\cite{Leitner:2006ww, Benhar:2013bwa, Ruso:2022qes}.

%%%%%%%%%%%%%%%%%%%%%%%%%%%%%%%%%%%%%%%%%%%%%%%%%%%%%%%%%%%%%%

\subsection{NC cross sections for all flavors}
\label{sec:cross-section-NC}

The NC cross section for the excitation of the 15.11-MeV state, corresponding to Eqs.~\eqref{eq:carbon-deex:nu}--\eqref{eq:carbon-deex:nubar}, following the EPT method employed in Refs.~\cite{Fukugita:1988hg, Pourkaviani:1990et, Kubodera:1993rk} has the following form:
\begin{equation}
\label{eq:cross-section-NC}
\sigma (E_\nu) = \frac{3G_F^2}{2\pi} F_A^2 (E_\nu^\prime)^2 I \,,
\end{equation}
where $G_F$ is the Fermi constant, $E_\nu$ is the incident neutrino energy, $E_\nu^\prime = E_\nu - \Delta M$ is the outgoing neutrino energy, and $F_A$ is the axial form factor.  The $I$ factor is
\begin{equation}
\label{eq:cross-section-integral}
    I = \frac{1}{2} \int_{-1}^{1} dz f(\mathbf{q}^2) \left( A + B + C \right) \,,
\end{equation}
where $z = \cos\theta$ is the angle between the incoming and outgoing neutrino momenta and $q^2 = \Delta M^2 - \mathbf{q}^2$, where $\mathbf{q}^2 = \Delta M^2 + 2 E_\nu E_\nu^\prime \left(1 - \cos\theta \right)$ being the three momentum transfer.  The assumed dependence of the axial-vector form factor on the momentum transfer is
\begin{equation}
\label{eq:form-factor}
f(\mathbf{q}^2) = \left(\frac{F_A(q)}{F_A(0)}\right)^2 = \left( 1 - \frac{1-\rho}{6(b|\mathbf{q}|)^2}\right)^2 \exp\left(-\frac{(b|\mathbf{q}|)^2}{2}\right) \,,
\end{equation}
with $F_A(0) \equiv F_A = 0.711 \pm 0.024$, $b=1.881 \pm 0.053$~fm, and $\rho=0.23 \pm 0.2$~\cite{Chertok:1973, Haxton:1977jr, Fukugita:1988hg}.  The three terms $A, B, C$ are:
\begin{equation}
\label{eq:A}
A = 1-\frac{z}{3} \; \pm \; \frac{4}{3}(E_\nu + E_\nu^\prime) (1-2\sin^2\theta_W) (1-z)\; \frac{F_M}{F_A} \,,
\end{equation}
\begin{equation}
\label{eq:B}
B = \frac{2}{3} \left( E_\nu^\prime E_\nu  (1-z^2) + (1-z) \, \mathbf{q}^2\right)(1-2\sin^2\theta_W)^2 \left(\frac{F_M}{F_A}\right)^2 \,,
\end{equation}
\begin{equation}
\label{eq:C}
C = -\frac{2}{3} \Delta M (1+z)\frac{F_T}{F_A} \; + \;  \frac{1}{3} (1 + z)\, \mathbf{q}^2  \left(\frac{F_M}{F_A}\right)^2 \,,
\end{equation}
where the $\pm$ refers to neutrinos and antineutrinos, respectively.
At zero momentum transfer, the magnetic form factor is $F_M(0) \equiv F_M = (1.516\pm0.016)\times10^{-3}~\mathrm{MeV}^{-1}$ and the tensor form factor is $F_T(0)/F_A(0) \equiv F_T/F_A = (2.01 \pm 0.23)\times 10^{-3}~\mathrm{MeV}^{-1}$~\cite{Fukugita:1988hg}.  We assume that $F_M$ and $F_T$ have the same dependence on the momentum transfer as $F_A$, following Eq.~\eqref{eq:form-factor}, which is adequate at the low energies we consider~\cite{Nozawa:1983imd}.  There is no experimental data on the momentum-transfer dependence of $F_T$, but its contribution to the cross section is negligible.

Figure~\ref{fig:cross-sections} (left panel) shows the calculated exclusive NC cross sections for neutrinos and antineutrinos as a function of the incoming neutrino energy. These grow steeply ($\propto E_\nu^2$) until $\sim$50-MeV, above which the nuclear coherence is lost for an increasing range of the momentum transfer values.  The neutrino cross sections are larger than the antineutrino cross sections, mainly due to favorable vector-axial vector interference via Eq.~\eqref{eq:A}.

%%%%%%%%%%%%%%%%%%%%%%%%%%%%%%%%%%%%%%%%%%%%%%%%%%%%%%%%%%%%%%
\subsection{CC cross sections for electron neutrinos}
\label{sec:e-neutrinos-cross-sections}

The cross sections for the $\nu_e$ and $\bar\nu_e$ exclusive interactions with ${}^{12}\mathrm{C}$, corresponding to Eqs.~\eqref{eq:CC-nue}--\eqref{eq:CC-nuebar}, are calculated using a similar EPT method~\cite{Fukugita:1988hg, Pourkaviani:1990et, Fukugita:1993fr, Kubodera:1993rk}. The cross section is
\begin{equation}
\label{eq:cross-section-CC}
\sigma (E_\nu) = \frac{3 G_F^2}{\pi} \cos\theta^2_C F_A^2 E_e p_e I \mathcal{F}^{\pm}(Z, E_e) \,,
\end{equation}
where $\theta_C$ is the Cabibbo angle, and $E_e$ and $p_e$ are the energy and momentum of the outgoing electron or positron. In $I$ for the CC case instead of NC case, $E_\nu^\prime$ is replaced by $E_e$, $\sin^2\theta_W$ is replaced by 0, and $\mathbf{q}^2$ is changed to $\mathbf{q}^2 = \Delta M^2 + 2 E_\nu E_e\left(1 - \cos\theta \right)- m_e^2$. The Fermi function accounting for the Coulomb correction for electrons ($+$) and positrons ($-$) is given by
\begin{equation}
\label{eq:correction-Fermi}
    \mathcal{F}^{\pm}(Z, E_e)  = 2 (1+S) (2 p_e R)^{S-2} e^{\pm\pi \eta} \frac{|\Gamma(S \pm i\eta)|^2}{\Gamma(2S+1)^2} \,,
\end{equation}
where $Z$ and $R$ are the charge and radius of the created nucleus in natural units, $S = \sqrt{1-Z^2\alpha^2}$, $\eta = Ze^2/\hbar v = 0.0073/v$, and $v = p_e/\sqrt{m_e^2+p_e^2}$~\cite{1934ZPhy...88..161F, doi:10.13182/NSE83-A17574, Venkataramaiah_1985}.  As written, the approximate correction in Eq.~\eqref{eq:correction-Fermi} would need be modified at large momentum transfers so that $\mathcal{F}(Z, \infty) = 1$~\cite{Engel:1996zt}. However, this change would have only a moderate effect ($\lesssim 20\%$) on the calculated cross section at neutrino energies above 100~MeV, so for simplicity we neglect it.

Figure~\ref{fig:cross-sections} (middle panel) shows the calculated CC cross sections for $\nu_e$ and $\bar\nu_e$ as a function of neutrino energy. As in the NC case, these grow with $\propto E_\nu^2$ dependence on the neutrino energy until approximately 50~MeV and are lower for antineutrinos.

%%%%%%%%%%%%%%%%%%%%%%%%%%%%%%%%%%%%%%%%%%%%%%%%%%%%%%%%%%%%%%

\subsection{CC cross sections for muon neutrinos}
\label{sec:CC-nu-mu}

The cross sections for the $\nu_\mu$ and $\bar{\nu}_\mu$ exclusive interactions with ${}^{12}\mathrm{C}$, corresponding to Eqs.~\eqref{eq:CC-numu}--\eqref{eq:CC-numubar}, are calculated in a similar way.  Due to the significantly larger mass of the muon, we use an approximate expression to account for the Coulomb correction given by
\begin{equation}
\label{eq:correction-muon}
    \mathcal{F}^{\pm}(Z, E_\mu) = \left(1 \pm \frac{\langle V \rangle}{E_\mu}\right)\left(E_\mu \pm \langle V \rangle\right) E_\mu^{-1} \,,
\end{equation}
where $E_\mu$ is the muon energy and the average Coulomb potential is $\langle V \rangle = 3Z\alpha / 2R$~\cite{Engel:1996zt}. 

Figure~\ref{fig:cross-sections} (right panel) shows the calculated CC cross sections for $\nu_\mu$ and $\bar{\nu}_\mu$ as a function of incoming neutrino energy.  The threshold for the interaction is close to the muon mass because it is so much larger than the nuclear-excitation scales.

%%%%%%%%%%%%%%%%%%%%%%%%%%%%%%%%%%%%%%%%%%%%%
\begin{figure*}[t]
\centering
\includegraphics[width=0.66\columnwidth]{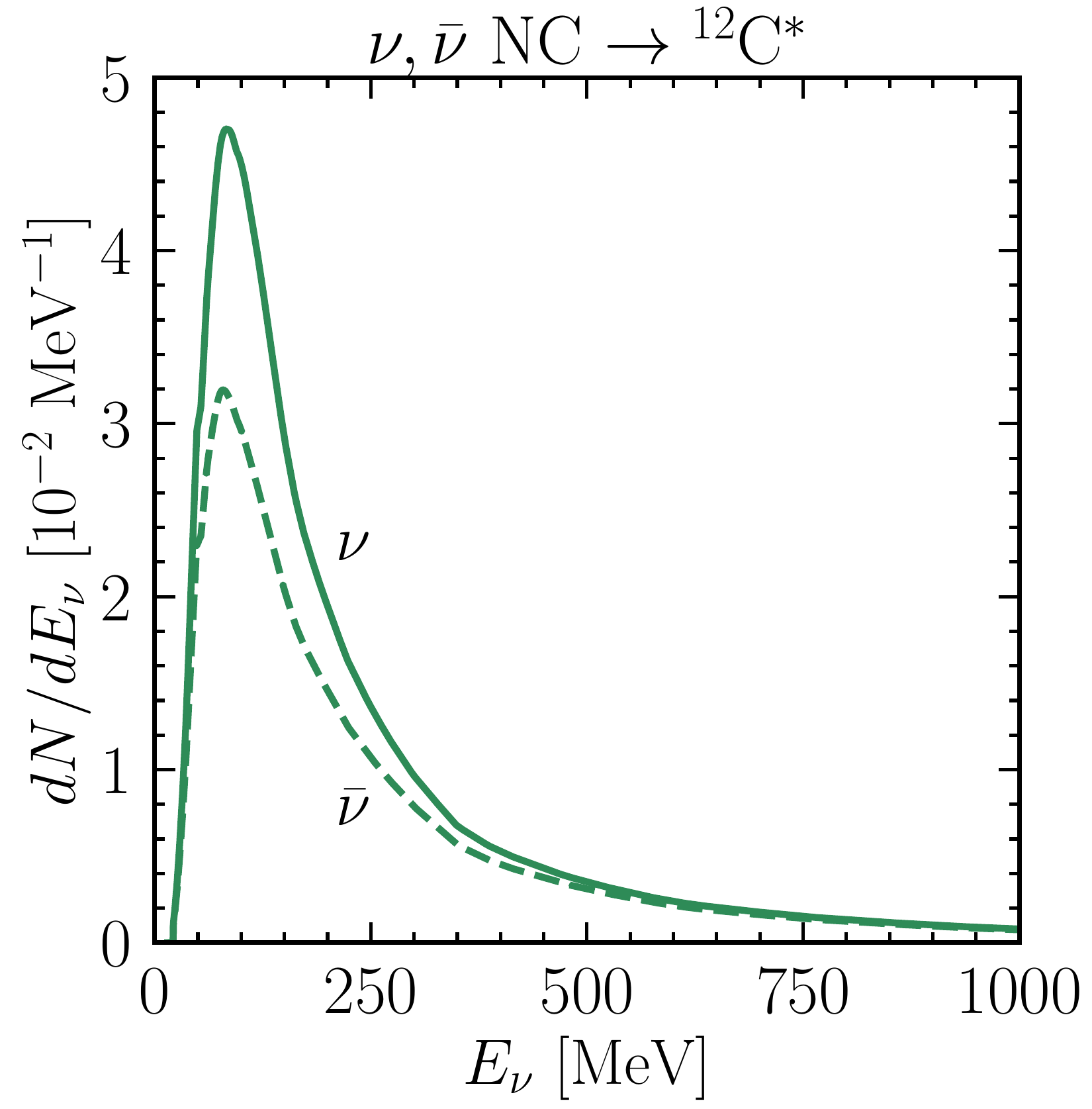}
\includegraphics[width=0.66\columnwidth]{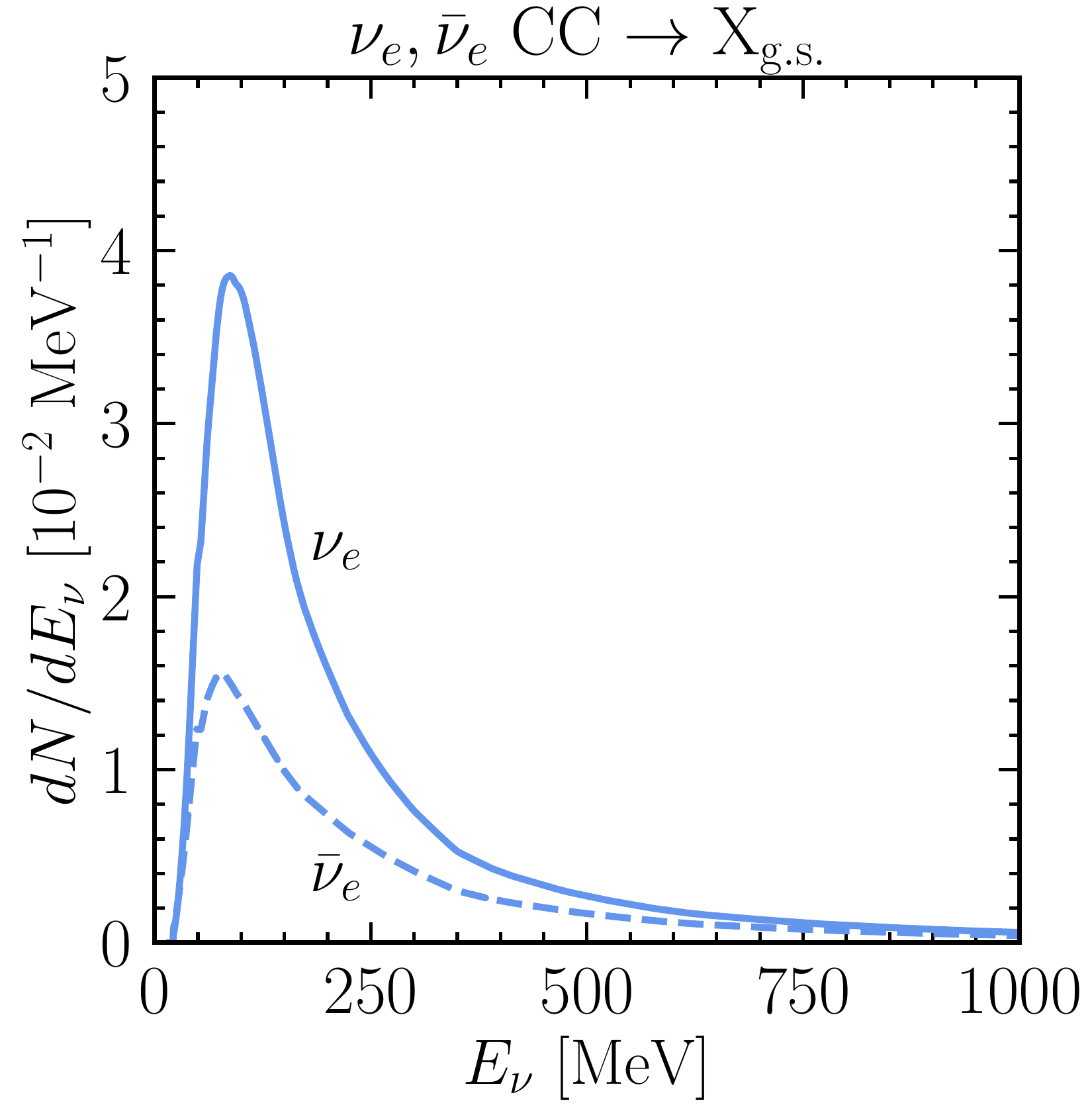}
\includegraphics[width=0.66\columnwidth]{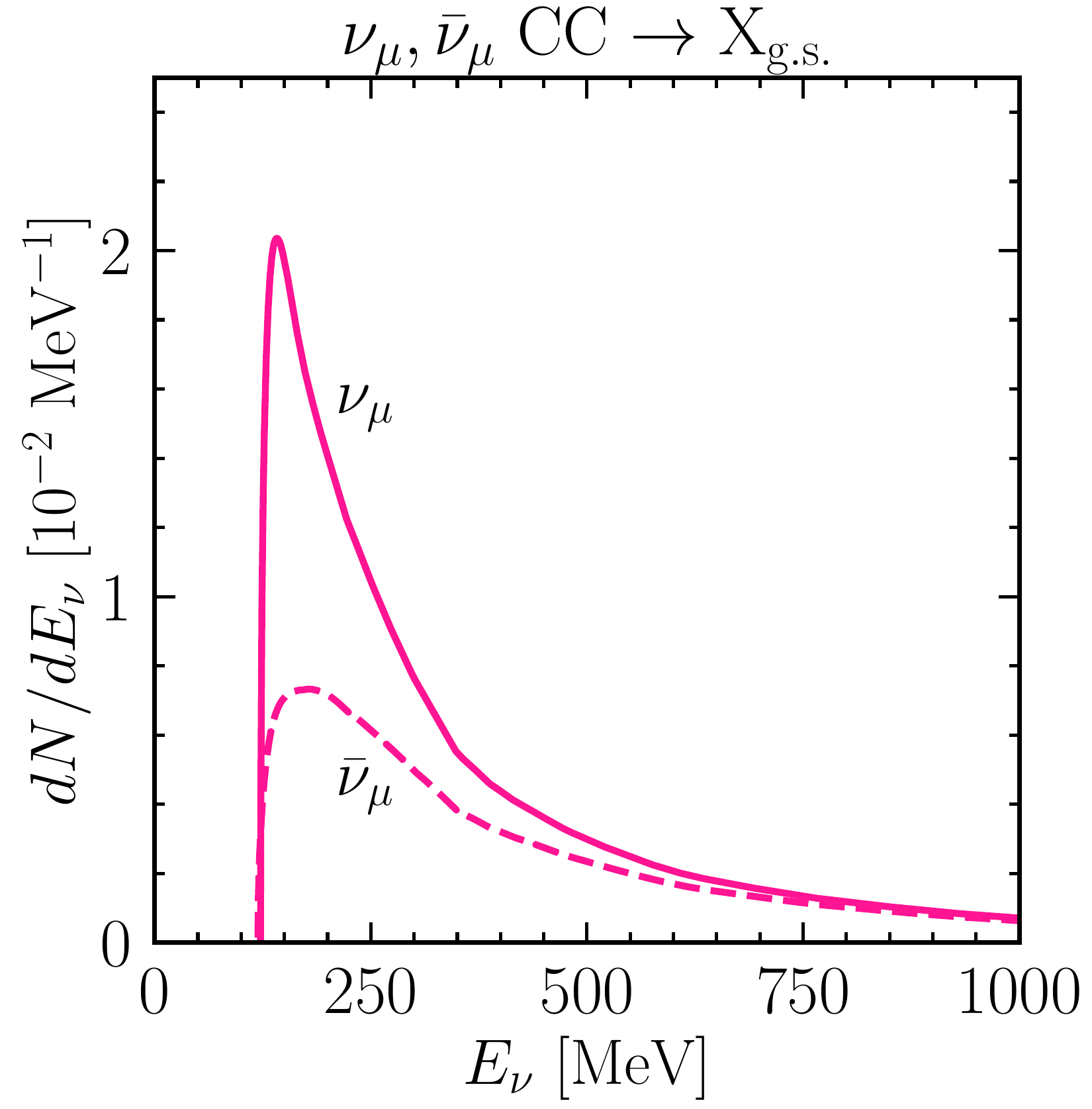}
\caption{
Differential spectra of the parent neutrinos that initiate the various exclusive neutrino-carbon interactions.
{\it Left panel:} NC $\nu$ and $\bar{\nu}$.  {\it Middle panel:} CC $\nu_e$ and $\bar{\nu}_e$.  {\it Right panel:} CC $\nu_\mu$ and $\bar{\nu}_\mu$. Note differences in the $y$-axes.
}
\label{fig:diff-rate}
\end{figure*}
%%%%%%%%%%%%%%%%%%%%%%%%%%%%%%%%%%%%%%%%%%%%%

%%%%%%%%%%%%%%%%%%%%%%%%%%%%%%%%%%%%%%%%%%%%%%%%%%%%%%%%%%%%%%

\subsection{Cross-section measurements}
\label{sec:cros-section-measurements}

Of the exclusive interactions we consider, all but the $\bar\nu_e$ and $\bar\nu_\mu$ CC cross sections have at least some measurements at MeV--GeV energies.  Of course, inclusive neutrino-carbon interactions in this energy range have been measured, e.g., Refs.~\cite{LSND:2001akn, LSND:2001akn, MiniBooNE:2007iti, MiniBooNE:2010bsu, MiniBooNE:2013qnd}, but we focus on exclusive cross sections to specific nuclear final states.

The measurements we discuss were performed with the Liquid Scintillator Neutrino Detector (LSND) at the Los Alamos National Laboratory and the KArlsruhe Rutherford Medium Energy Neutrino experiment (KARMEN) at the Rutherford Appleton Laboratory. These experiments used neutrino fluxes from pion decay in flight (DIF) and pion and muon decay at rest (DAR).  Neither measured the exclusive CC interactions for antineutrino channels, because of the high absorption probability for $\pi^-$ and the smaller cross sections for antineutrinos compared to neutrinos.

For the interactions that were measured, the EPT-predicted cross sections in Eqs.~\eqref{eq:cross-section-NC}--\eqref{eq:cross-section-CC} agree well with data, as shown in Fig.~\ref{fig:cross-sections}. In addition, the EPT-predicted cross sections generally also agree with those obtained with more sophisticated calculations relying on projected random phase and quasiparticle random phase approximations~\cite{Kolbe:1994xb, Engel:1996zt, Kolbe:1996km, Auerbach:1997ay, Vogel:2006sg, Samana:2010up}.  Ultimately, cross-section studies that go beyond the EPT method will be needed.

The exclusive neutrino-carbon interactions we consider here are small fractions of the total cross sections, e.g., approximately a few percent for $E_\nu < 300~\mathrm{MeV}$~\cite{Samana:2010up}.  This suggests that it may be possible to identify other neutrino-carbon final states with distinctive nuclear signatures, which would increase the statistics and better probe the neutrino-nucleus interactions.  We reserve this and studies of other nuclear targets for future work.

\emph{\textbf{Neutral-current interactions:}}
The NC cross section was measured by the KARMEN~\cite{KARMEN:1998xmo, KARMEN:1991vkr, Zeitnitz:1994kz, KARMEN:1994xse}.  Neutrinos were produced through the pion DAR, which leads to monoenergetic $\nu_\mu$ with $E_\nu = 29.8~\mathrm{MeV}$ and well-known spectrum of neutrinos from the following decay of the muon, with $\nu_e$ and $\bar\nu_\mu$ up to energy 52.8~MeV. These two neutrino signals are separable by timing. The cross section measured with only pion-decay neutrinos~\cite{KARMEN:1998xmo} is shown on the left panel of Fig.~\ref{fig:cross-sections}; it agrees well with the EPT predictions.  The energy-averaged cross section measured with only the higher-energy continuum spectrum of muon-decay neutrinos (not shown) is $\langle \sigma \rangle~=~10.4\pm\mathrm{1.0(stat.)\pm 0.9(syst.)}\times10^{-42}$~cm~\cite{KARMEN:1991vkr, KARMEN:1994xse}, which also agrees well with predictions.

\emph{\textbf{Electron-neutrino interactions:}}
KARMEN~\cite{Zeitnitz:1994kz} and LSND~\cite{LSND:1997lta, LSND:2001fbw} made energy-dependent measurements of the exclusive $\nu_e$ cross section using a pion DAR source.  Detection is based on a two-fold coincidence between a prompt electron from the primary interaction (with kinetic energy up to 33.5~MeV) and a delayed positron from the $\beta$-decay of $^{12}$N$_\mathrm{g.s.}$ (with kinetic energy up to 16.3 MeV).   The data points follow closely the EPT predictions, as shown in the middle panel of Fig.~\ref{fig:cross-sections}.

\emph{\textbf{Muon-neutrino interactions:}}
LSND also made energy-dependent measurements of the exclusive $\nu_\mu$ cross section using a pion DIF source~\cite{LSND:2002oco}.  The maximum neutrino energy (approximately 200~MeV) for the measured cross section is dictated by the experimental setup.  Detection is based on a three-fold coincidence between the prompt $\mu^{-}$, its decay electron, and the positron from $\beta$-decay.  Negative muons stop and typically undergo atomic capture, which has only small effects on the muon-decay spectrum~\cite{Czarnecki:2011mx}, but about 8\% of negative muons also undergo nuclear capture, which leads to a very different signal than decay~\cite{LSND:2002oco}.  Although the uncertainties are relatively large, the data agree well with the EPT predictions, as shown in the right panel of Fig.~\ref{fig:cross-sections}.

%%%%%%%%%%%%%%%%%%%%%%%%%%%%%%%%%%%%%%%%%%%%%%%%%%%%%%%%%%%%%%%%%%
%%%%%%%%%%%%%%%%%%%%%%%%%%%%%%%%%%%%%%%%%%%%%%%%%%%%%%%%%%%%%%%%%%

\section{Detection strategies in JUNO}
\label{sec:Detector}

In this section, we describe our detection strategies for distinctive signatures of low-energy atmospheric-neutrino interactions in JUNO.  We review the detector properties in Sec.~\ref{sec:JUNO}, calculate the event rates for the neutrino-carbon reactions in Sec.~\ref{sec:event-rates}, and describe backgrounds and their mitigation in Sec.~\ref{sec:backgrounds}.

%%%%%%%%%%%%%%%%%%%%%%%%%%%%%%%%%%%%%%%%%%%%%%%%%%%%%%%%%%%%%%

\subsection{Basic characteristics of JUNO}
\label{sec:JUNO}

JUNO is a 20-kton liquid-scintillator neutrino detector surrounded by a water Cherenkov detector that vetos atmospheric muons~\cite{JUNO:2015zny, JUNO:2015sjr, JUNO:2021vlw}.  The inner and outer volumes, which are optically isolated, are observed by photomultiplier tubes.  The experiment is located at a depth of approximately 650~m underground in China, with plans to start in 2024~\cite{jie_zhao_2022_6683749, Lin:2022htc, JUNO:2021vlw, JUNO-NOW}.  The primary goal of JUNO is to precisely measure neutrino-oscillation parameters and to contribute to the measurement of the neutrino mass ordering~\cite{JUNO:2015zny, JUNO:2021vlw}.  Key to this are JUNO's excellent energy resolution ($\sim$$3\%/\sqrt{E/\mathrm{MeV}}$) and  position resolution ($\lesssim10~\mathrm{cm}$)~\cite{Qian:2021vnh}, and low backgrounds.

JUNO will detect, with high efficiency, the energies of charged particles above $\sim$0.5~MeV as well as the presence of neutrons via their 2.2-MeV radiative captures on protons~\cite{Fang:2019lej}.  Different charged particles can be identified via pulse shape discrimination (PSD) of the signals received in the photomultiplier tubes~\cite{JUNO:2015zny, JUNO:2021vlw, JUNO:2021tll}.  This is especially effective for separating, say, MeV-range protons from electrons, and follows from the fact that the proton energy-loss rate is enhanced by a factor $\sim 1/\beta^2$, where $\beta$ is the particle speed relative to light.  In addition, muons can be recognized via their delayed decays.

JUNO already plans to measure atmospheric neutrinos via inclusive CC and NC interactions on carbon~\cite{Cheng:2020aaw, Cheng:2020oko, JUNO:2021tll}.  The analysis presented in Ref.~\cite{JUNO:2021tll}, based on PSD, is limited to events with energies above 400~MeV for $\nu_\mu$ because, at lower energies, it is more difficult to distinguish CC and NC events.  It may be possible to do better by searching for delayed muon decays.  We propose that by using the exclusive interactions, with their special detection features, as in Eqs.~\eqref{eq:carbon-deex}--\eqref{eq:CC-nubar}, JUNO will be able to effectively study atmospheric neutrinos at lower energies.

%%%%%%%%%%%%%%%%%%%%%%%%%%%%%%%%%%%%%%%%%%%%%
\begin{figure}[t]
\centering
\includegraphics[width=0.99\columnwidth]{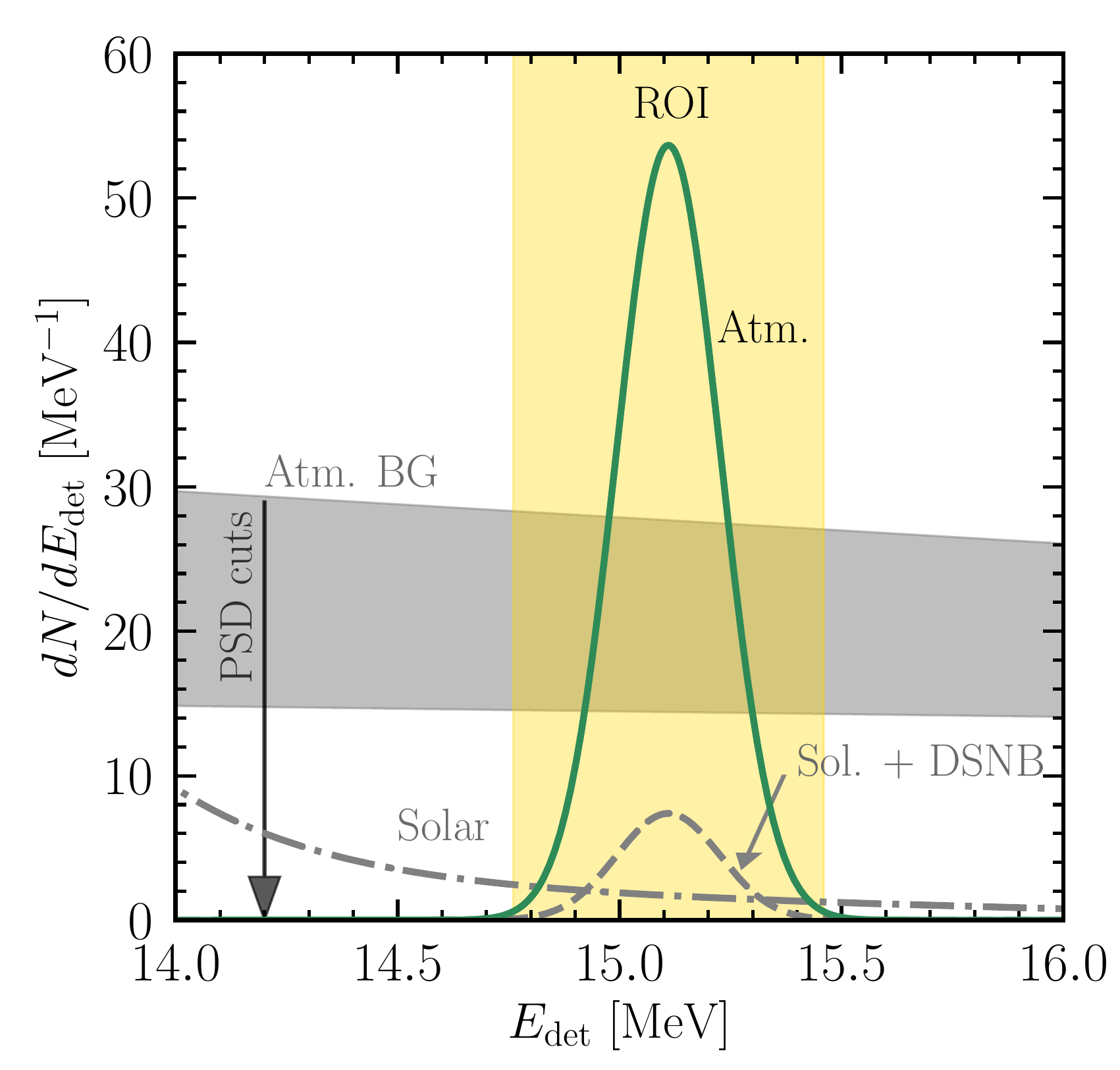}
\caption{
Spectra of the exclusive NC neutrino-carbon signal (green solid line) and the most important backgrounds in JUNO as a function of detected energy, assuming an exposure of 85 kton$\,$yr. The backgrounds are NC atmospheric-neutrino interactions on free and bound protons (gray band), solar neutrinos scattering on electrons (gray dash-dotted lines) and the solar and DSNB excitation of the 15.11-MeV line (gray dashed line). The first source of background can be removed using the PSD techniques.
}
\label{fig:diff-rate-NC}
\end{figure}
%%%%%%%%%%%%%%%%%%%%%%%%%%%%%%%%%%%%%%%%%%%%%

%%%%%%%%%%%%%%%%%%%%%%%%%%%%%%%%%%%%%%%%%%%%%%%%%%%%%%%%%%%%%%

\subsection{Event rates and spectra}
\label{sec:event-rates}

In our calculations, we assume an exposure of 85 kton$\,$yr, corresponding to a 17-kton fiducial volume and five years of runtime at unit efficiency.  This fiducial volume is what JUNO will use to suppress backgrounds for their main analyses; for the distinctive signals we focus on, it may be possible to use a larger volume.  Though the true efficiency will be somewhat less, that can be compensated for by increasing the volume or the runtime.  For each of the considered exclusive interaction channels, Eqs.~\eqref{eq:carbon-deex}--\eqref{eq:CC-nubar}, we calculate the time-integrated differential neutrino event yield via
\begin{equation}
\label{eq:event-rate-Ev}
\frac{dN}{dE_\nu} = T \; N_{t} \; \sigma_{i}(E_\nu) \; \frac{d\phi}{d E_{\nu}} \,,
\end{equation}
where $T$ is the runtime of the detector, $N_t \simeq 7.5 \times 10^{32}$ is the number of targets in 17~kton, and $\sigma_i(E_{\nu}) $ is the cross section for a particular interaction.  For the atmospheric-neutrino fluxes for the CC reactions, we average over the two possible mass orderings shown in Fig.~\ref{fig:flux} because the differences in the solid angle averaged fluxes for the two orderings are small.  Our calculations take into account energy resolution.

Figure~\ref{fig:diff-rate} shows the differential spectra of neutrinos contributing to the total yields.  These spectra peak at energies comparable to the broad peak in the atmospheric-neutrino spectra shown in Fig.~\ref{fig:flux}, but are sharpened by the energy dependence of the cross sections shown in Fig.~\ref{fig:cross-sections}.  Nearly all of the events are caused by neutrinos of energies below a few hundred MeV.  These results should be considered as rough estimates to guide experimental studies.

Figure~\ref{fig:diff-rate-NC} shows the exclusive NC atmospheric-neutrino signal.  We define a generous region of interest (ROI) by $\pm3\sigma$ around the 15.11-MeV peak position, taking the energy resolution into account.  The key backgrounds remaining after basic cuts are detailed in the next subsection.  Nominally, the largest is caused by various inclusive atmospheric-neutrino interaction channels that lead to proton recoils~\cite{Beacom:2003zu, Chauhan:2021fzu}, but we anticipate that these events can be greatly reduced using PSD techniques~\cite{JUNO:2015zny, JUNO:2021vlw, JUNO:2021tll}, though JUNO-led studies will be needed to assess the efficiencies.  The main irreducible background is caused by solar and diffuse supernova neutrino background (DSNB) neutrinos undergoing the same exclusive NC interaction that we consider, plus a contribution from solar neutrino-electron scattering that can be measured separately.  For an exposure of 85 kton$\,$yr, the signal yield is about 16 events and the background yield in the ROI is about 4 events.  It should thus be possible to measure the NC atmospheric-neutrino signal rate to $\sim$25\% precision.

\begin{figure}[t]
\centering
\includegraphics[width=0.99\columnwidth]{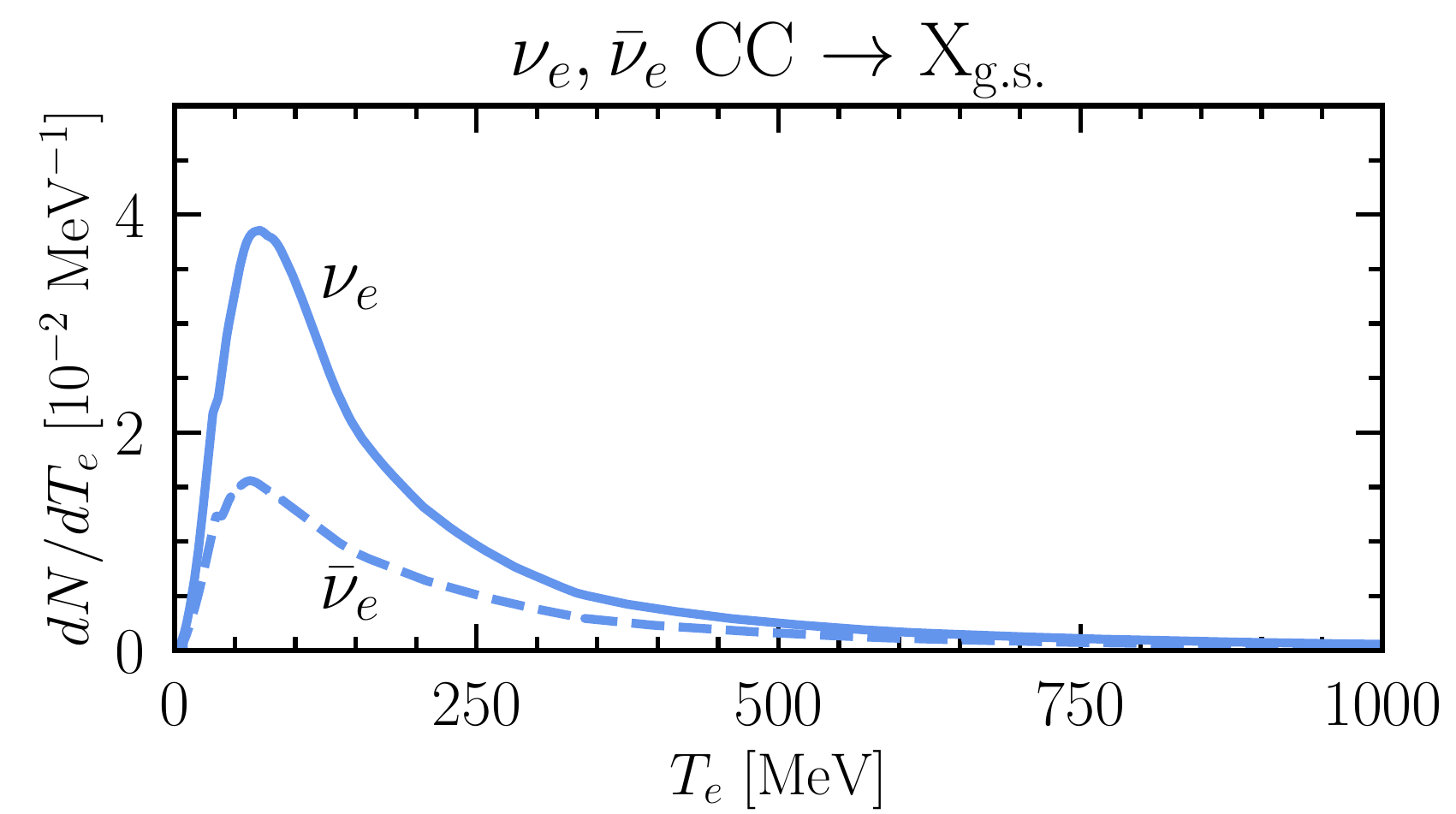}
\includegraphics[width=0.99\columnwidth]{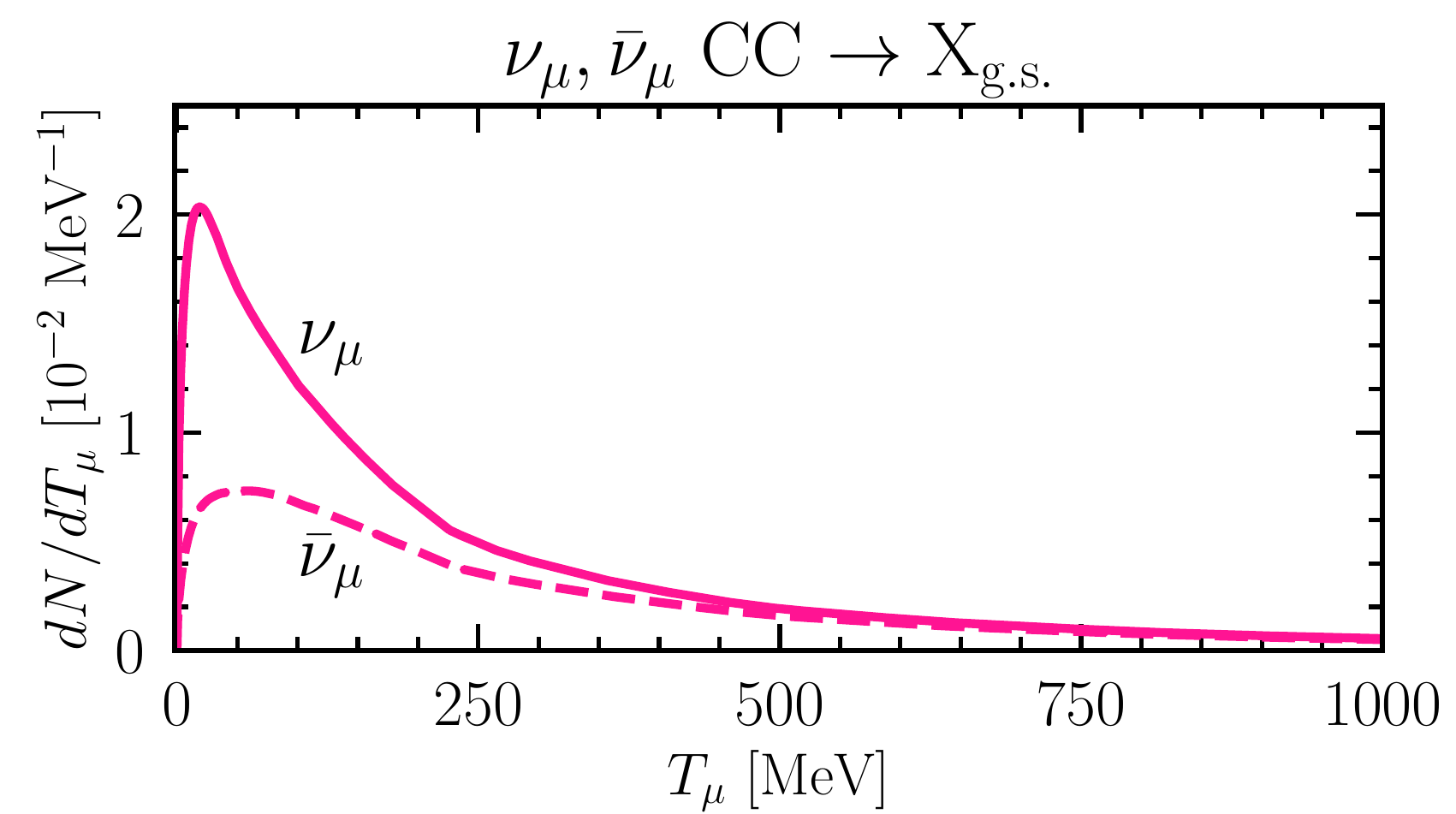}
\caption{
Approximate spectra of the exclusive CC neutrino-carbon signals in JUNO as a function of the prompt lepton energies, assuming an exposure of 85 kton$\,$yr.  For these two-fold and three-fold coincidence events, backgrounds are negligible.  Note differences in the y-axis ranges.
}
\label{fig:diff-rate-CC}
\end{figure}

Figure~\ref{fig:diff-rate-CC} shows the exclusive CC atmospheric-neutrino signals.  The kinetic energies is estimated simply as the total neutrino energy minus the excitation energy minus the mass of the lepton.  This neglects detailed nuclear effects as well as recoil-order kinematics, but it provides adequate guidance to design searches in JUNO that take more detailed calculations of the double-differential cross sections~\cite{Gaisser:1986bv,  Kolbe:1996km, Mintz:1996qh} into account.  The total number of the CC $\nu_e + \bar\nu_e$ events is 7.2 + 3.4 = 10.6 and the total number of CC $\nu_\mu + \bar\nu_\mu$ events is 3.5 + 2.1 = 5.6. As detailed in the next subsection, there are essentially no backgrounds, so it should be possible to measure the total CC atmospheric-neutrino rate (about 16 events) to $\sim$25\% precision as well.  As detailed below, these events have negligible backgrounds due to their two- or three-fold coincidences, which will also allow separation of $\nu_e + \bar\nu_e$ events from $\nu_\mu + \bar\nu_\mu$ events.

While the cross sections considered here are relatively well known theoretically, it would of course be preferable to have precise direct measurements.  For the interaction channels already measured by LSND and KARMEN, here the uncertainties are more or less comparable, depending on the channel.  Even more interesting, here there is reasonable sensitivity to channels not available to those laboratory experiments, due to their lack of appreciable fluxes of $\bar{\nu}_e$, $\bar{\nu}_\mu$, $\nu_\tau$, and $\bar{\nu}_\tau$.  

%%%%%%%%%%%%%%%%%%%%%%%%%%%%%%%%%%%%%%%%%%%%%
%%%%%%%%%%%%%%%%%%%%%%%%%%%%%%%%%%%%%%%%%%%%%

%%%%%%%%%%%%%%%%%%%%%%%%%%%%%%%%%%%%%%%%%%%%%%%%%%%%%%%%%%%%%%

\subsection{Backgrounds in the search energy window}
\label{sec:backgrounds}

For background reduction, JUNO will perform a variety of basic cuts~\cite{JUNO:2015zny, JUNO:2020hqc, Chauhan:2021fzu}.  One class of cuts removes cosmic-ray muons, which are typically throughgoing and produce large light signals.  A second is cuts on the delayed beta decays of unstable nuclei produced through spallation processes initiated by cosmic-ray muons and their showers (standard cuts are expected to induce a detector deadtime of $\simeq$20\%~\cite{JUNO:2015zny, Chauhan:2021fzu}, which could be reduced by using new techniques introduced in Refs.~\cite{Li:2014sea, Li:2015kpa, Li:2015lxa, Super-Kamiokande:2015xra, Super-Kamiokande:2021snn, Nairat:2023xxx}).  A third is cuts on radioactivity (mostly below $\sim$5 MeV) and fast-neutron backgrounds near the detector walls; restricting events to be in the fiducial volume greatly reduces these backgrounds~\cite{JUNO:2020hqc}.  In the following, we discuss the more challenging backgrounds that will remain.  

{\bf{NC-interaction background:}} 
For the NC signal, JUNO will have to contend with all sources of single-fold events near 15.11 MeV.  Nominally, there is a moderately large and uncertain background (see Fig.~\ref{fig:diff-rate-NC}) due to inclusive NC atmospheric-neutrino interactions on free and bound protons that produce one or more proton recoils (that will appear as one)~\cite{Beacom:2003zu, Chauhan:2021fzu}, though PSD cuts can remove these events~\cite{JUNO:2015zny, JUNO:2021vlw, JUNO:2021tll}.  For the proton spectrum, we use the predictions of Ref.~\cite{Chauhan:2021fzu}, which takes into account the quenching of proton light output due to their high ionization rate.  KamLAND has observed events from a closely related channel where a neutrino-induced neutron recoil produces a prompt energy deposition from neutron-proton scattering and then a delayed signal from the neutron capture. The KamLAND data are in reasonable agreement ($\sim2\sigma$) with predictions~\cite{KamLAND:2011bnd, KamLAND:2021gvi}.

Another potential source of background is gamma-ray lines from the de-excitation of the nuclei created by inclusive CC and NC interactions of atmospheric neutrinos and subsequent neutron-nucleus interactions~\cite{Cheng:2020aaw}.  Some of these lines are near 15.11 MeV, though their yields are smaller than that of the signal we consider.  More important, these lines are produced in association with large signals from produced leptons and/or hadrons, whereas the 15.11-MeV signal we focus on has only the gamma-ray line and no other detectable particles.  Similarly, the 15.11-MeV line could be induced by cosmic-ray muons and their secondaries, though not without producing other signals.

Some backgrounds will remain.  Solar-neutrino $^8$B and $hep$ fluxes will induce a continuum spectrum of neutrino-electron-scattering events that is small and which can be checked at energies outside the ROI.  To predict this, we use the fluxes from Ref.~\cite{Vinyoles:2016djt}, cross section from Ref.~\cite{Giunti:2007ry}, and take into account matter-induced mixing effects~\cite{Wolfenstein:1977ue, Mikheyev:1985zog}.  There are also irreducible backgrounds caused by solar and DSNB (calculated as the fiducial models in Refs.~\cite{Moller:2018kpn, Suliga:2021hek, Ziegler:2022ivq}) excitation of the 15.11-MeV line, with the former being about 6 times larger than the latter, and with this overall background being small, so that uncertainties in the DSNB modeling have negligible effects.

{\bf{CC-interactions background:}}
In the CC interaction channels, a two- or three-fold time and space coincidence allows essentially complete removal of backgrounds.  For the two-fold case, an order-of-magnitude estimate shows that their rate is negligible.  The rate in the three-fold case would be even smaller.

For the $\nu_e$ and $\bar{\nu}_e$ signals, we expect a $\sim$100-MeV electron or positron from the neutrino-nucleus interaction followed by another electron or positron at $\sim$10 MeV from the beta decay.  The expected time delay is below 100 ms and the expected distance between these events is below 1~m.  The number of accidental coincidences can be estimated as~\cite{1944Natur.153..592S}
\begin{equation}
\label{eq:accidental-coincidences}
N_\mathrm{ac} \sim N_1 N_2 \;dt/T \; dV/V \\,
\end{equation}
where the $N_1$ is the number of high-energy events and $N_2$~is the number of low-energy events in the $TV =$~85~kton~yr exposure of JUNO.  We estimate $N_1$ from the total number of atmospheric-neutrino events from all channels, which gives approximately 5500 events between energies 100--10000 MeV per 85~kton~yr~\cite{JUNO:2021tll}.  We estimate $N_2$ from the total number of events in 3.5--18 MeV, which is approximately $1.5\times 10^5$ in the same exposure.  These events are dominated by solar neutrinos and spallation backgrounds~\cite{JUNO:2015zny, JUNO:2020hqc, Chauhan:2021fzu}.  We estimate $dt/T \sim 6 \times 10^{-10}$ by comparing 100~ms to 5~yr and $dV/V \sim 2 \times 10^{-4} $ by comparing a sphere of radius 1~m to the fiducial volume.  The expected number of accidental-coincidence events is thus $N_\mathrm{ac} \approx 1 \times 10^{-4}$.  It is thus safe to neglect these as a background for the two-fold signals and even more so for the three-fold signals.

%%%%%%%%%%%%%%%%%%%%%%%%%%%%%%%%%%%%%%%%%%%%%%%%%%%%%%%%%%%%%%
%%%%%%%%%%%%%%%%%%%%%%%%%%%%%%%%%%%%%%%%%%%%%%%%%%%%%%%%%%%%%%

\section{Discussion and Conclusions}
\label{sec:Discussion-Conclusions}

In the coming high-statistics era of neutrino physics, it will be critical to explore all available means of measuring neutrino-mixing parameters.  With many measurements, we will improve the overall precision, which will also allow better sensitivity to new physics.

In this paper, building on earlier work~\cite{Gaisser:1986bv, Nussinov:2000qc, Kolbe:2002gk, Kelly:2019itm, Newstead:2020fie, Denton:2021mso, Zhuang:2021rsg, Kelly:2023ugn, Peres:2003wd, Mena:2008rh, Akhmedov:2012ah, Ioannisian:2020isl, Kelly:2021jfs, Denton:2021rgt, Cheng:2020aaw, Cheng:2020oko, JUNO:2021tll}, we advance the general point that low-energy (below a few hundred MeV) atmospheric neutrinos are an especially promising direction for new measurements of neutrino mixing, due to dramatically improving detector capabilities in the MeV range.  The leading detectors for this purpose (in order of their start dates) are Super-Kamiokande with added gadolinium~\cite{Beacom:2003nk, Super-Kamiokande:2021the, Super-Kamiokande:2023xup}, JUNO~\cite{JUNO:2015zny}, Hyper-Kamiokande~\cite{Abe:2018uyc}, and the Deep Underground Neutrino Experiment (DUNE)~\cite{DUNE:2020ypp}.  Proposed dark-matter detectors such as the DARk matter WImp search with liquid xenoN (DARWIN)~\cite{Aalbers:2016jon} will also be important.  While there are significant uncertainties on the fluxes, cross sections, and even detector responses, we are optimistic that these can be reduced through complementary measurements and targeted theoretical work.

Here we focus on the prospects for distinctive nuclear signatures in the JUNO detector, showing that certain exclusive neutrino-carbon interactions have reasonable yields and low backgrounds.  We calculate results for three channels, the first two of which are familiar from supernova-neutrino detection~\cite{Cadonati:2000kq, Laha:2014yua, Lu:2016ipr}.  First, all-flavor NC interactions that excite the 15.11-MeV state in $^{12}$C, producing a gamma-ray line that stands out from continuum backgrounds.  Second, CC $\nu_e$ and $\bar{\nu}_e$ interactions that produce a prompt $\sim$100-MeV electron or positron from the neutrino-nuclear interaction, followed by a $\sim$10-MeV positron or electron from a $^{12}$N or $^{12}$B beta decay.  The two-fold coincidence in time and space essentially eliminates backgrounds for these atmospheric neutrino interactions.  Third, similar CC interactions, but with $\nu_\mu$ and $\bar{\nu}_\mu$, that produce a three-fold coincidence due to the additional muon decay.

We show that the NC interactions and the combined CC interactions each lead to $\sim$16 events in a five-year (85 kton$\,$yr) exposure of JUNO, and hence nominally to $\sim$25\% uncertainties on the rates of each.  These events would include the first identified measurements of sub-100~MeV atmospheric neutrinos.  While these measurements will not be sufficient on their own to measure neutrino mixing in the face of present uncertainties, these data will be an important step towards that goal, because the cross sections are relatively well known theoretically and because the detector signals are so clean.  If only statistical uncertainties were important, then the NC channel would have $\sim$1$\sigma$ sensitivity to $\nu_\tau$ appearance, which could be a helpful addition to other measurements.

As a point of context, while Super-Kamiokande has detected atmospheric-neutrino-induced events below a visible energy of 100 MeV, e.g., Ref.~\cite{Super-Kamiokande:2023xup}, that does not mean that the parent neutrino energies are below 100 MeV. Indeed, for the dominant component at low energies, due to invisible (sub-Cherenkov) muons that decay at rest, the parent neutrino energies must be a bit above 100~MeV. For the subdominant components --- due to CC $(\nu_e + \bar{\nu}_e)$ interactions and NC all-flavor interactions --- the fraction of parent neutrinos below 100~MeV has not yet been identified, especially in light of large uncertainties on the fluxes, cross sections, and detector response.

Better understanding these interactions will also have other benefits.  One would be probing the physics of atmospheric-neutrino production in its most challenging regime, which is at low energies~\cite{Battistoni:2002ew, Zhuang:2021rsg}.  Another would be better understanding how low-energy atmospheric neutrinos cause important backgrounds for other searches, including the diffuse supernova neutrino background~\cite{Horiuchi:2008jz, Beacom:2010kk, Lunardini:2010ab, Kresse:2020nto, Horiuchi:2020jnc, Suliga:2022ica}, as well as direct-detection dark matter studies as they reach exposures corresponding to the neutrino floor~\cite{Vergados:2008jp, Strigari:2009bq, Billard:2013qya, Baudis:2013qla, Ruppin:2014bra, OHare:2016pjy, Boehm:2018sux, OHare:2020lva, Suliga:2021hek}.

Going forward, it will be important for the JUNO collaboration to perform in-depth studies of the sensitivity to these and potentially other exclusive neutrino-carbon interactions.  As noted above, it is easy to separate $\nu_e + \bar{\nu}_e$ events from $\nu_\mu + \bar{\nu}_\mu$ events, and there may be some sensitivity to separate neutrino from antineutrino events (needed to test CP violation).  It may also be possible to gain some crude directionality (to cancel the flux uncertainty) for the CC channels from exploiting the early Cherenkov light~\cite{BOREXINO:2021efb} or from the fact that the prompt lepton will be detected forward of the delayed lepton from the beta decay, building on ideas in Refs.~\cite{Vogel:1999zy, CHOOZ:1999hgz, Beacom:2003nk} and taking advantage of the excellent position resolution of JUNO. It will also be important to identify additional distinctive nuclear signatures in JUNO and other detectors; the prospects are encouraging because the channels considered here are only a small fraction of the total neutrino-nucleus cross section.  Overall, we expect that a comprehensive program on low-energy atmospheric neutrinos will lead to varied and valuable insights.

%%%%%%%%%%%%%%%%%%%%%%%%%%%%%%%%%%%%%%%%%%%%%%%%%%%%%%%%%%%%%%%%%%%%%%%%%%%%%%%%%%%%%%%%

\begin{acknowledgements}
We are grateful for helpful discussions with Baha Balantekin, Bhavesh Chauhan, Basudeb Dasgupta, Wick Haxton, Yufeng Li, Stephan Meighen-Berger, Irene Tamborra, Petr Vogel, Bei Zhou, and Shun Zhou. We also thank Yi Zhuang for providing us the atmospheric neutrino fluxes at the JUNO location.

A.M.S. was supported by the Network for Neutrinos Nuclear Astrophysics and Symmetries (N3AS), which is supported by National Science Foundation Physics Frontier Center  Grant No.\ PHY-2020275.  J.F.B. was supported by National Science Foundation Grant No.\ PHY-2012955.

In addition, A.M.S. thanks, for their kind hospitality and stimulating research environments, the Institute for Nuclear Theory (INT) at the University of Washington and the Mainz Institute for Theoretical Physics (MITP) of the Cluster of Excellence PRISMA$^{+}$ (Project ID 39083149). INT is supported by Department of Energy Grant No.\ DE-FG02-00ER41132.
\end{acknowledgements}
%%%%%%%%%%%%%%%%%%%%%%%%%%%%%%%%%%%%%%%%%%%%%%%%%%%%%%%%%%%%%%%%%%%%%%%%%%%%%%%%%%%%%%%%

\appendix

\section{Atmospheric-neutrino oscillations}
\label{app:atm-osc}

\begin{figure}[b]
\centering
\includegraphics[width=0.99\columnwidth]{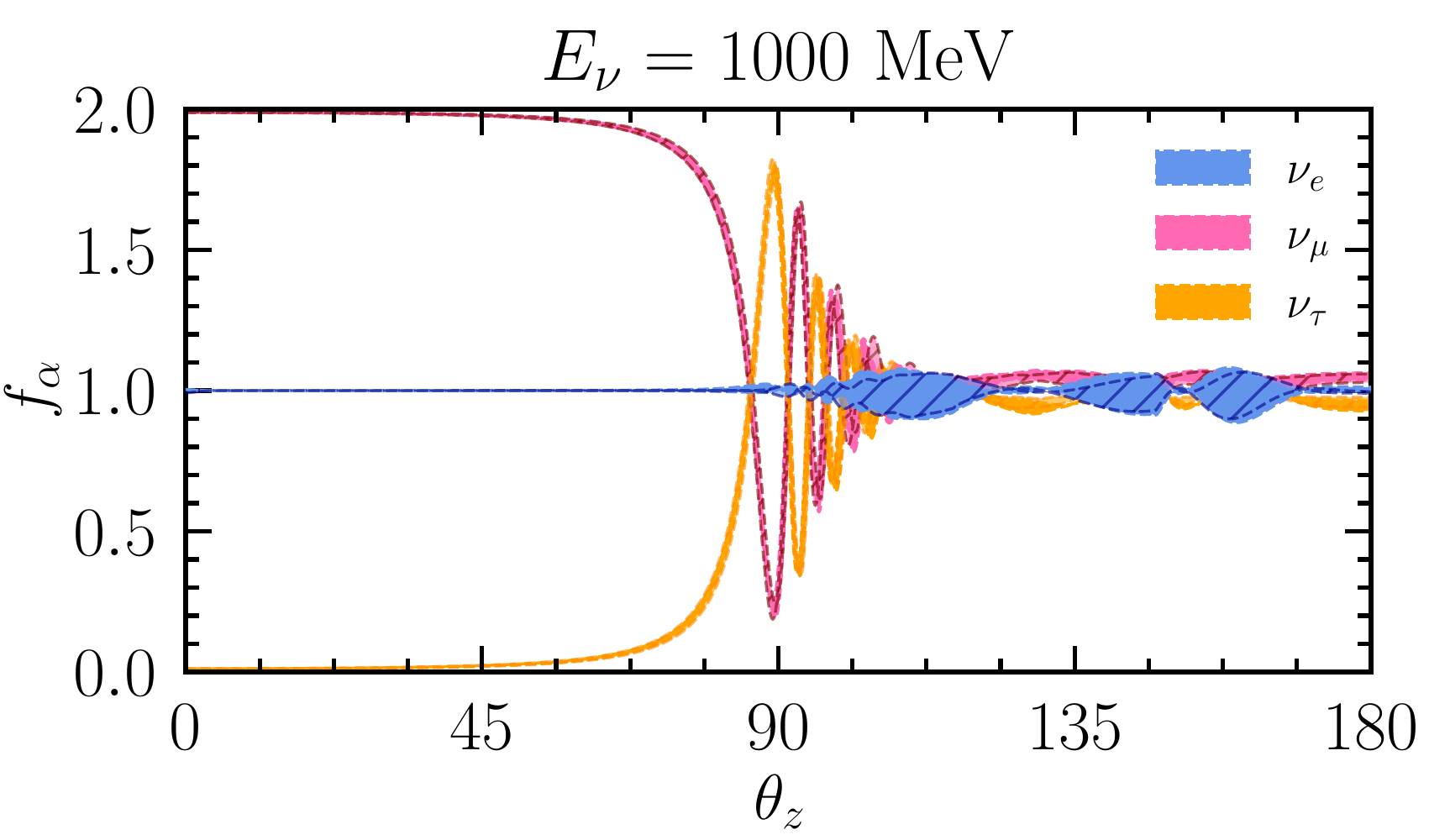}
\includegraphics[width=0.99\columnwidth]{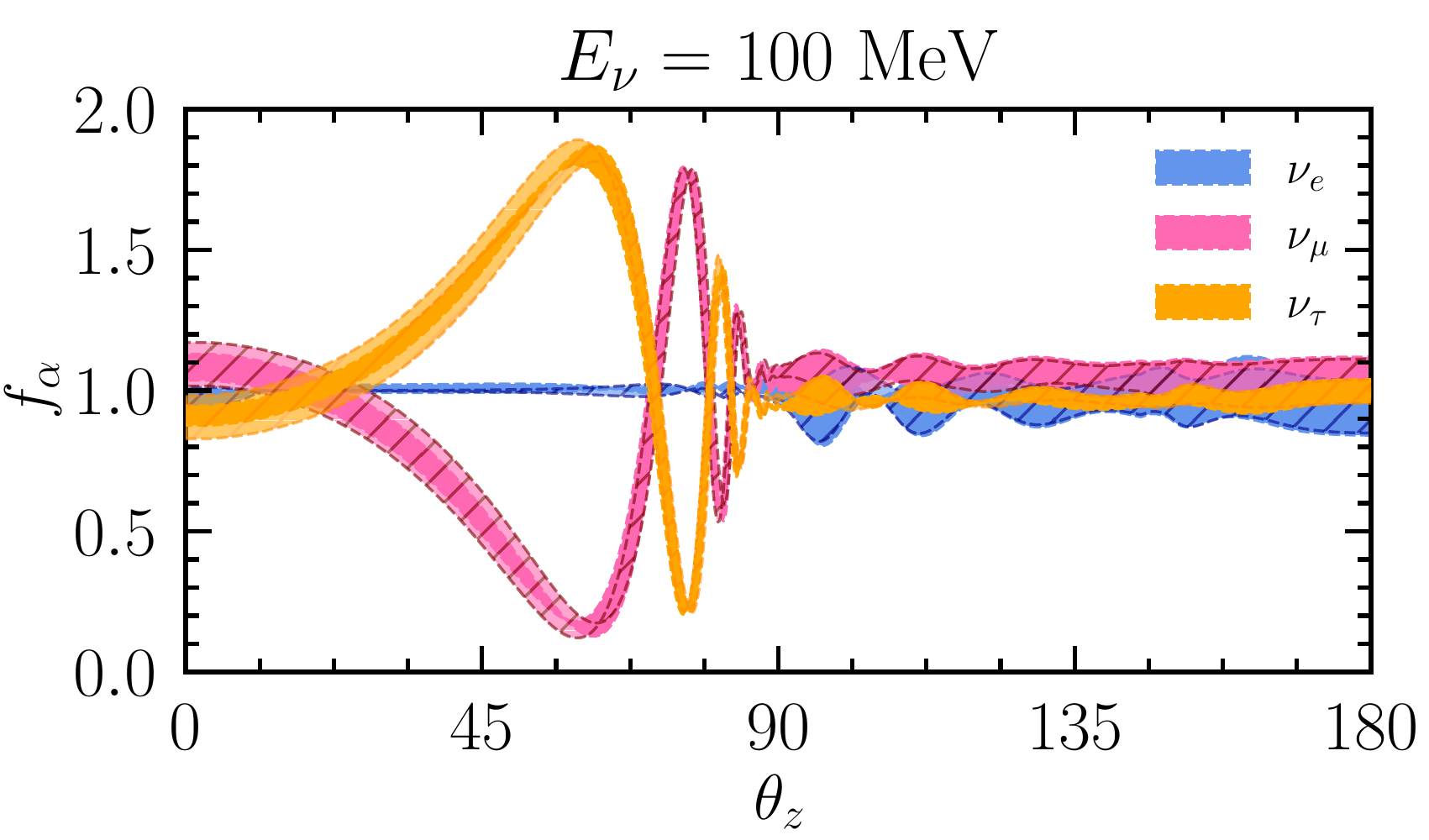}
\includegraphics[width=0.99\columnwidth]{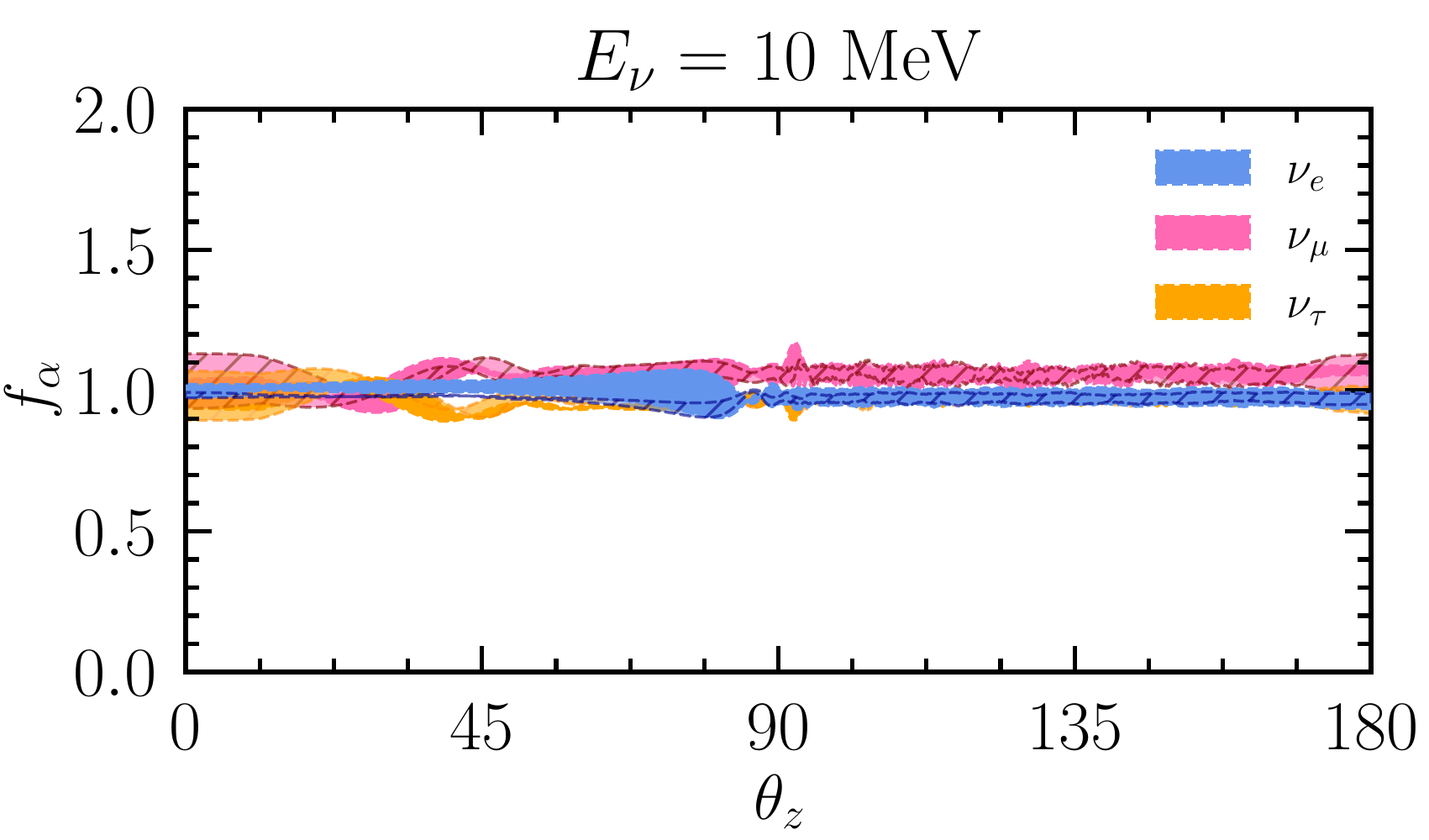}
\caption{
Atmospheric-neutrino flux ratios as a function of zenith angle after 10\% energy averaging.  The solid versus the hatched regions show the slight impact of changing of the mass ordering; see also Fig.~\ref{fig:flux}. 
}
\label{fig:av-survival-P}
\end{figure}

Figure~\ref{fig:av-survival-P} shows the flux ratios of atmospheric neutrinos arriving at JUNO as a function of zenith angles for three different energies (1000, 100, and 10~MeV) and a benchmark atmospheric height of 22~km.  To reduce the appearance of unmeasurable variations in the mixing probabilities, we include energy averaging via Gaussian smearing with width 10\%, which is optimistic.  Realistic averaging over energy, angle, and production height would further suppress the displayed variations.

At 1000~MeV, the mixing with the ``atmospheric" $\Delta m^2$ is unimportant for downgoing neutrinos, becomes important near the horizon, and then averages out for upgoing neutrinos.  At 100~MeV, the oscillation length being 10 times shorter means that a larger fraction of downgoing neutrinos undergo mixing, and all the more so at 10~MeV.

%%%%%%%%%%%%%%%%%%%%%%%%%%%%%%%%%%%%%%%%%%%%%
%%%%%%%%%%%%%%%%%%%%%%%%%%%%%%%%%%%%%%%%%%%%%

\clearpage

\bibliography{Atmospherics.bib}

\providecommand{\noopsort}[1]{}\providecommand{\singleletter}[1]{#1}%
\begin{thebibliography}{152}%
\makeatletter
\providecommand \@ifxundefined [1]{%
 \@ifx{#1\undefined}
}%
\providecommand \@ifnum [1]{%
 \ifnum #1\expandafter \@firstoftwo
 \else \expandafter \@secondoftwo
 \fi
}%
\providecommand \@ifx [1]{%
 \ifx #1\expandafter \@firstoftwo
 \else \expandafter \@secondoftwo
 \fi
}%
\providecommand \natexlab [1]{#1}%
\providecommand \enquote  [1]{``#1''}%
\providecommand \bibnamefont  [1]{#1}%
\providecommand \bibfnamefont [1]{#1}%
\providecommand \citenamefont [1]{#1}%
\providecommand \href@noop [0]{\@secondoftwo}%
\providecommand \href [0]{\begingroup \@sanitize@url \@href}%
\providecommand \@href[1]{\@@startlink{#1}\@@href}%
\providecommand \@@href[1]{\endgroup#1\@@endlink}%
\providecommand \@sanitize@url [0]{\catcode `\\12\catcode `\$12\catcode
  `\&12\catcode `\#12\catcode `\^12\catcode `\_12\catcode `\%12\relax}%
\providecommand \@@startlink[1]{}%
\providecommand \@@endlink[0]{}%
\providecommand \url  [0]{\begingroup\@sanitize@url \@url }%
\providecommand \@url [1]{\endgroup\@href {#1}{\urlprefix }}%
\providecommand \urlprefix  [0]{URL }%
\providecommand \Eprint [0]{\href }%
\providecommand \doibase [0]{http://dx.doi.org/}%
\providecommand \selectlanguage [0]{\@gobble}%
\providecommand \bibinfo  [0]{\@secondoftwo}%
\providecommand \bibfield  [0]{\@secondoftwo}%
\providecommand \translation [1]{[#1]}%
\providecommand \BibitemOpen [0]{}%
\providecommand \bibitemStop [0]{}%
\providecommand \bibitemNoStop [0]{.\EOS\space}%
\providecommand \EOS [0]{\spacefactor3000\relax}%
\providecommand \BibitemShut  [1]{\csname bibitem#1\endcsname}%
\let\auto@bib@innerbib\@empty
%</preamble>
\bibitem [{\citenamefont {Haxton}\ \emph {et~al.}(2013)\citenamefont {Haxton},
  \citenamefont {Hamish~Robertson},\ and\ \citenamefont
  {Serenelli}}]{Haxton:2012wfz}%
  \BibitemOpen
  \bibfield  {author} {\bibinfo {author} {\bibfnamefont {W.~C.}\ \bibnamefont
  {Haxton}}, \bibinfo {author} {\bibfnamefont {R.~G.}\ \bibnamefont
  {Hamish~Robertson}}, \ and\ \bibinfo {author} {\bibfnamefont {A.~M.}\
  \bibnamefont {Serenelli}},\ }\href {\doibase
  10.1146/annurev-astro-081811-125539} {\bibfield  {journal} {\bibinfo
  {journal} {Ann. Rev. Astron. Astrophys.}\ }\textbf {\bibinfo {volume} {51}},\
  \bibinfo {pages} {21} (\bibinfo {year} {2013})},\ \Eprint
  {http://arxiv.org/abs/1208.5723} {arXiv:1208.5723 [astro-ph.SR]} \BibitemShut
  {NoStop}%
\bibitem [{\citenamefont {Kajita}(2014)}]{Kajita:2014koa}%
  \BibitemOpen
  \bibfield  {author} {\bibinfo {author} {\bibfnamefont {T.}~\bibnamefont
  {Kajita}},\ }\href {\doibase 10.1146/annurev-nucl-102313-025402} {\bibfield
  {journal} {\bibinfo  {journal} {Ann. Rev. Nucl. Part. Sci.}\ }\textbf
  {\bibinfo {volume} {64}},\ \bibinfo {pages} {343} (\bibinfo {year}
  {2014})}\BibitemShut {NoStop}%
\bibitem [{\citenamefont {Vitagliano}\ \emph {et~al.}(2020)\citenamefont
  {Vitagliano}, \citenamefont {Tamborra},\ and\ \citenamefont
  {Raffelt}}]{Vitagliano:2019yzm}%
  \BibitemOpen
  \bibfield  {author} {\bibinfo {author} {\bibfnamefont {E.}~\bibnamefont
  {Vitagliano}}, \bibinfo {author} {\bibfnamefont {I.}~\bibnamefont
  {Tamborra}}, \ and\ \bibinfo {author} {\bibfnamefont {G.}~\bibnamefont
  {Raffelt}},\ }\href {\doibase 10.1103/RevModPhys.92.045006} {\bibfield
  {journal} {\bibinfo  {journal} {Rev. Mod. Phys.}\ }\textbf {\bibinfo {volume}
  {92}},\ \bibinfo {pages} {45006} (\bibinfo {year} {2020})},\ \Eprint
  {http://arxiv.org/abs/1910.11878} {arXiv:1910.11878 [astro-ph.HE]}
  \BibitemShut {NoStop}%
\bibitem [{\citenamefont {Workman}\ \emph {et~al.}(2022)\citenamefont {Workman}
  \emph {et~al.}}]{ParticleDataGroup:2022pth}%
  \BibitemOpen
  \bibfield  {author} {\bibinfo {author} {\bibfnamefont {R.~L.}\ \bibnamefont
  {Workman}} \emph {et~al.} (\bibinfo {collaboration} {Particle Data Group}),\
  }\href {\doibase 10.1093/ptep/ptac097} {\bibfield  {journal} {\bibinfo
  {journal} {PTEP}\ }\textbf {\bibinfo {volume} {2022}},\ \bibinfo {pages}
  {083C01} (\bibinfo {year} {2022})}\BibitemShut {NoStop}%
\bibitem [{\citenamefont {de~Gouv\^ea}\ \emph {et~al.}(2022)\citenamefont
  {de~Gouv\^ea} \emph {et~al.}}]{deGouvea:2022gut}%
  \BibitemOpen
  \bibfield  {author} {\bibinfo {author} {\bibfnamefont {A.}~\bibnamefont
  {de~Gouv\^ea}} \emph {et~al.},\ }\href@noop {} {\  (\bibinfo {year}
  {2022})},\ \Eprint {http://arxiv.org/abs/2209.07983} {arXiv:2209.07983
  [hep-ph]} \BibitemShut {NoStop}%
\bibitem [{\citenamefont {Aartsen}\ \emph {et~al.}(2021)\citenamefont {Aartsen}
  \emph {et~al.}}]{IceCube-Gen2:2020qha}%
  \BibitemOpen
  \bibfield  {author} {\bibinfo {author} {\bibfnamefont {M.~G.}\ \bibnamefont
  {Aartsen}} \emph {et~al.} (\bibinfo {collaboration} {IceCube-Gen2}),\ }\href
  {\doibase 10.1088/1361-6471/abbd48} {\bibfield  {journal} {\bibinfo
  {journal} {J. Phys. G}\ }\textbf {\bibinfo {volume} {48}},\ \bibinfo {pages}
  {060501} (\bibinfo {year} {2021})},\ \Eprint
  {http://arxiv.org/abs/2008.04323} {arXiv:2008.04323 [astro-ph.HE]}
  \BibitemShut {NoStop}%
\bibitem [{\citenamefont {Aiello}\ \emph {et~al.}(2022)\citenamefont {Aiello}
  \emph {et~al.}}]{KM3NeT:2021ozk}%
  \BibitemOpen
  \bibfield  {author} {\bibinfo {author} {\bibfnamefont {S.}~\bibnamefont
  {Aiello}} \emph {et~al.} (\bibinfo {collaboration} {KM3NeT}),\ }\href
  {\doibase 10.1140/epjc/s10052-021-09893-0} {\bibfield  {journal} {\bibinfo
  {journal} {Eur. Phys. J. C}\ }\textbf {\bibinfo {volume} {82}},\ \bibinfo
  {pages} {26} (\bibinfo {year} {2022})},\ \Eprint
  {http://arxiv.org/abs/2103.09885} {arXiv:2103.09885 [hep-ex]} \BibitemShut
  {NoStop}%
\bibitem [{\citenamefont {Abed~Abud}\ \emph {et~al.}(2022)\citenamefont
  {Abed~Abud} \emph {et~al.}}]{DUNE:2022aul}%
  \BibitemOpen
  \bibfield  {author} {\bibinfo {author} {\bibfnamefont {A.}~\bibnamefont
  {Abed~Abud}} \emph {et~al.} (\bibinfo {collaboration} {DUNE}),\ }\href@noop
  {} {\  (\bibinfo {year} {2022})},\ \Eprint {http://arxiv.org/abs/2203.06100}
  {arXiv:2203.06100 [hep-ex]} \BibitemShut {NoStop}%
\bibitem [{\citenamefont {Abusleme}\ \emph
  {et~al.}(2022{\natexlab{a}})\citenamefont {Abusleme} \emph
  {et~al.}}]{JUNO:2022mxj}%
  \BibitemOpen
  \bibfield  {author} {\bibinfo {author} {\bibfnamefont {A.}~\bibnamefont
  {Abusleme}} \emph {et~al.} (\bibinfo {collaboration} {JUNO}),\ }\href
  {\doibase 10.1088/1674-1137/ac8bc9} {\bibfield  {journal} {\bibinfo
  {journal} {Chin. Phys. C}\ }\textbf {\bibinfo {volume} {46}},\ \bibinfo
  {pages} {123001} (\bibinfo {year} {2022}{\natexlab{a}})},\ \Eprint
  {http://arxiv.org/abs/2204.13249} {arXiv:2204.13249 [hep-ex]} \BibitemShut
  {NoStop}%
\bibitem [{\citenamefont {Bian}\ \emph {et~al.}(2022)\citenamefont {Bian} \emph
  {et~al.}}]{Hyper-Kamiokande:2022smq}%
  \BibitemOpen
  \bibfield  {author} {\bibinfo {author} {\bibfnamefont {J.}~\bibnamefont
  {Bian}} \emph {et~al.} (\bibinfo {collaboration} {Hyper-Kamiokande}),\ }in\
  \href@noop {} {\emph {\bibinfo {booktitle} {{Snowmass 2021}}}}\ (\bibinfo
  {year} {2022})\ \Eprint {http://arxiv.org/abs/2203.02029} {arXiv:2203.02029
  [hep-ex]} \BibitemShut {NoStop}%
\bibitem [{\citenamefont {Denton}\ \emph {et~al.}(2022)\citenamefont {Denton},
  \citenamefont {Friend}, \citenamefont {Messier}, \citenamefont {Tanaka},
  \citenamefont {B\"oser}, \citenamefont {Coelho}, \citenamefont
  {Perrin-Terrin},\ and\ \citenamefont {Stuttard}}]{Denton:2022een}%
  \BibitemOpen
  \bibfield  {author} {\bibinfo {author} {\bibfnamefont {P.~B.}\ \bibnamefont
  {Denton}}, \bibinfo {author} {\bibfnamefont {M.}~\bibnamefont {Friend}},
  \bibinfo {author} {\bibfnamefont {M.~D.}\ \bibnamefont {Messier}}, \bibinfo
  {author} {\bibfnamefont {H.~A.}\ \bibnamefont {Tanaka}}, \bibinfo {author}
  {\bibfnamefont {S.}~\bibnamefont {B\"oser}}, \bibinfo {author} {\bibfnamefont
  {J.~a. A.~B.}\ \bibnamefont {Coelho}}, \bibinfo {author} {\bibfnamefont
  {M.}~\bibnamefont {Perrin-Terrin}}, \ and\ \bibinfo {author} {\bibfnamefont
  {T.}~\bibnamefont {Stuttard}},\ }\href@noop {} {\  (\bibinfo {year}
  {2022})},\ \Eprint {http://arxiv.org/abs/2212.00809} {arXiv:2212.00809
  [hep-ph]} \BibitemShut {NoStop}%
\bibitem [{\citenamefont {Huber}\ \emph {et~al.}(2022)\citenamefont {Huber}
  \emph {et~al.}}]{Huber:2022lpm}%
  \BibitemOpen
  \bibfield  {author} {\bibinfo {author} {\bibfnamefont {P.}~\bibnamefont
  {Huber}} \emph {et~al.},\ }in\ \href@noop {} {\emph {\bibinfo {booktitle}
  {{2022 Snowmass Summer Study}}}}\ (\bibinfo {year} {2022})\ \Eprint
  {http://arxiv.org/abs/2211.08641} {arXiv:2211.08641 [hep-ex]} \BibitemShut
  {NoStop}%
\bibitem [{\citenamefont {Abdullahi}\ \emph {et~al.}(2023)\citenamefont
  {Abdullahi} \emph {et~al.}}]{Abdullahi:2022jlv}%
  \BibitemOpen
  \bibfield  {author} {\bibinfo {author} {\bibfnamefont {A.~M.}\ \bibnamefont
  {Abdullahi}} \emph {et~al.},\ }\href {\doibase 10.1088/1361-6471/ac98f9}
  {\bibfield  {journal} {\bibinfo  {journal} {J. Phys. G}\ }\textbf {\bibinfo
  {volume} {50}},\ \bibinfo {pages} {020501} (\bibinfo {year} {2023})},\
  \Eprint {http://arxiv.org/abs/2203.08039} {arXiv:2203.08039 [hep-ph]}
  \BibitemShut {NoStop}%
\bibitem [{\citenamefont {Arg\"uelles}\ \emph {et~al.}(2023)\citenamefont
  {Arg\"uelles} \emph {et~al.}}]{Arguelles:2022tki}%
  \BibitemOpen
  \bibfield  {author} {\bibinfo {author} {\bibfnamefont {C.~A.}\ \bibnamefont
  {Arg\"uelles}} \emph {et~al.},\ }\href {\doibase
  10.1140/epjc/s10052-022-11049-7} {\bibfield  {journal} {\bibinfo  {journal}
  {Eur. Phys. J. C}\ }\textbf {\bibinfo {volume} {83}},\ \bibinfo {pages} {15}
  (\bibinfo {year} {2023})},\ \Eprint {http://arxiv.org/abs/2203.10811}
  {arXiv:2203.10811 [hep-ph]} \BibitemShut {NoStop}%
\bibitem [{\citenamefont {Berryman}\ \emph {et~al.}(2022)\citenamefont
  {Berryman} \emph {et~al.}}]{Berryman:2022hds}%
  \BibitemOpen
  \bibfield  {author} {\bibinfo {author} {\bibfnamefont {J.~M.}\ \bibnamefont
  {Berryman}} \emph {et~al.},\ }in\ \href@noop {} {\emph {\bibinfo {booktitle}
  {{Snowmass 2021}}}}\ (\bibinfo {year} {2022})\ \Eprint
  {http://arxiv.org/abs/2203.01955} {arXiv:2203.01955 [hep-ph]} \BibitemShut
  {NoStop}%
\bibitem [{\citenamefont {Gerbino}\ \emph {et~al.}(2022)\citenamefont {Gerbino}
  \emph {et~al.}}]{Gerbino:2022nvz}%
  \BibitemOpen
  \bibfield  {author} {\bibinfo {author} {\bibfnamefont {M.}~\bibnamefont
  {Gerbino}} \emph {et~al.},\ }\href@noop {} {\  (\bibinfo {year} {2022})},\
  \Eprint {http://arxiv.org/abs/2203.07377} {arXiv:2203.07377 [hep-ph]}
  \BibitemShut {NoStop}%
\bibitem [{\citenamefont {Acero}\ \emph {et~al.}(2022)\citenamefont {Acero}
  \emph {et~al.}}]{Acero:2022wqg}%
  \BibitemOpen
  \bibfield  {author} {\bibinfo {author} {\bibfnamefont {M.~A.}\ \bibnamefont
  {Acero}} \emph {et~al.},\ }\href@noop {} {\  (\bibinfo {year} {2022})},\
  \Eprint {http://arxiv.org/abs/2203.07323} {arXiv:2203.07323 [hep-ex]}
  \BibitemShut {NoStop}%
\bibitem [{\citenamefont {Duan}\ \emph {et~al.}(2010)\citenamefont {Duan},
  \citenamefont {Fuller},\ and\ \citenamefont {Qian}}]{Duan:2010bg}%
  \BibitemOpen
  \bibfield  {author} {\bibinfo {author} {\bibfnamefont {H.}~\bibnamefont
  {Duan}}, \bibinfo {author} {\bibfnamefont {G.~M.}\ \bibnamefont {Fuller}}, \
  and\ \bibinfo {author} {\bibfnamefont {Y.-Z.}\ \bibnamefont {Qian}},\ }\href
  {\doibase 10.1146/annurev.nucl.012809.104524} {\bibfield  {journal} {\bibinfo
   {journal} {Ann. Rev. Nucl. Part. Sci.}\ }\textbf {\bibinfo {volume} {60}},\
  \bibinfo {pages} {569} (\bibinfo {year} {2010})},\ \Eprint
  {http://arxiv.org/abs/1001.2799} {arXiv:1001.2799 [hep-ph]} \BibitemShut
  {NoStop}%
\bibitem [{\citenamefont {Chakraborty}\ \emph {et~al.}(2016)\citenamefont
  {Chakraborty}, \citenamefont {Hansen}, \citenamefont {Izaguirre},\ and\
  \citenamefont {Raffelt}}]{Chakraborty:2016yeg}%
  \BibitemOpen
  \bibfield  {author} {\bibinfo {author} {\bibfnamefont {S.}~\bibnamefont
  {Chakraborty}}, \bibinfo {author} {\bibfnamefont {R.}~\bibnamefont {Hansen}},
  \bibinfo {author} {\bibfnamefont {I.}~\bibnamefont {Izaguirre}}, \ and\
  \bibinfo {author} {\bibfnamefont {G.}~\bibnamefont {Raffelt}},\ }\href
  {\doibase 10.1016/j.nuclphysb.2016.02.012} {\bibfield  {journal} {\bibinfo
  {journal} {Nucl. Phys. B}\ }\textbf {\bibinfo {volume} {908}},\ \bibinfo
  {pages} {366} (\bibinfo {year} {2016})},\ \Eprint
  {http://arxiv.org/abs/1602.02766} {arXiv:1602.02766 [hep-ph]} \BibitemShut
  {NoStop}%
\bibitem [{\citenamefont {Tamborra}\ and\ \citenamefont
  {Shalgar}(2021)}]{Tamborra:2020cul}%
  \BibitemOpen
  \bibfield  {author} {\bibinfo {author} {\bibfnamefont {I.}~\bibnamefont
  {Tamborra}}\ and\ \bibinfo {author} {\bibfnamefont {S.}~\bibnamefont
  {Shalgar}},\ }\href {\doibase 10.1146/annurev-nucl-102920-050505} {\bibfield
  {journal} {\bibinfo  {journal} {Ann. Rev. Nucl. Part. Sci.}\ }\textbf
  {\bibinfo {volume} {71}},\ \bibinfo {pages} {165} (\bibinfo {year} {2021})},\
  \Eprint {http://arxiv.org/abs/2011.01948} {arXiv:2011.01948 [astro-ph.HE]}
  \BibitemShut {NoStop}%
\bibitem [{\citenamefont {Richers}\ and\ \citenamefont
  {Sen}(2022)}]{Richers:2022zug}%
  \BibitemOpen
  \bibfield  {author} {\bibinfo {author} {\bibfnamefont {S.}~\bibnamefont
  {Richers}}\ and\ \bibinfo {author} {\bibfnamefont {M.}~\bibnamefont {Sen}},\
  }\href@noop {} {\  (\bibinfo {year} {2022})},\ \Eprint
  {http://arxiv.org/abs/2207.03561} {arXiv:2207.03561 [astro-ph.HE]}
  \BibitemShut {NoStop}%
\bibitem [{\citenamefont {Patwardhan}\ \emph {et~al.}(2022)\citenamefont
  {Patwardhan}, \citenamefont {Cervia}, \citenamefont {Rrapaj}, \citenamefont
  {Siwach},\ and\ \citenamefont {Balantekin}}]{Patwardhan:2022mxg}%
  \BibitemOpen
  \bibfield  {author} {\bibinfo {author} {\bibfnamefont {A.~V.}\ \bibnamefont
  {Patwardhan}}, \bibinfo {author} {\bibfnamefont {M.~J.}\ \bibnamefont
  {Cervia}}, \bibinfo {author} {\bibfnamefont {E.}~\bibnamefont {Rrapaj}},
  \bibinfo {author} {\bibfnamefont {P.}~\bibnamefont {Siwach}}, \ and\ \bibinfo
  {author} {\bibfnamefont {A.~B.}\ \bibnamefont {Balantekin}},\ }\href
  {\doibase 10.1007/978-981-15-8818-1{\_}126-1} {\  (\bibinfo {year} {2022}),\
  10.1007/978-981-15-8818-1{\_}126-1},\ \Eprint
  {http://arxiv.org/abs/2301.00342} {arXiv:2301.00342 [hep-ph]} \BibitemShut
  {NoStop}%
\bibitem [{\citenamefont {Volpe}(2023)}]{Volpe:2023met}%
  \BibitemOpen
  \bibfield  {author} {\bibinfo {author} {\bibfnamefont {M.~C.}\ \bibnamefont
  {Volpe}},\ }\href@noop {} {\  (\bibinfo {year} {2023})},\ \Eprint
  {http://arxiv.org/abs/2301.11814} {arXiv:2301.11814 [hep-ph]} \BibitemShut
  {NoStop}%
\bibitem [{\citenamefont {Gaisser}\ and\ \citenamefont
  {O'Connell}(1986)}]{Gaisser:1986bv}%
  \BibitemOpen
  \bibfield  {author} {\bibinfo {author} {\bibfnamefont {T.~K.}\ \bibnamefont
  {Gaisser}}\ and\ \bibinfo {author} {\bibfnamefont {J.~S.}\ \bibnamefont
  {O'Connell}},\ }\href {\doibase 10.1103/PhysRevD.34.822} {\bibfield
  {journal} {\bibinfo  {journal} {Phys. Rev. D}\ }\textbf {\bibinfo {volume}
  {34}},\ \bibinfo {pages} {822} (\bibinfo {year} {1986})}\BibitemShut
  {NoStop}%
\bibitem [{\citenamefont {Nussinov}\ and\ \citenamefont
  {Shrock}(2001)}]{Nussinov:2000qc}%
  \BibitemOpen
  \bibfield  {author} {\bibinfo {author} {\bibfnamefont {S.}~\bibnamefont
  {Nussinov}}\ and\ \bibinfo {author} {\bibfnamefont {R.}~\bibnamefont
  {Shrock}},\ }\href {\doibase 10.1103/PhysRevLett.86.2223} {\bibfield
  {journal} {\bibinfo  {journal} {Phys. Rev. Lett.}\ }\textbf {\bibinfo
  {volume} {86}},\ \bibinfo {pages} {2223} (\bibinfo {year} {2001})},\ \Eprint
  {http://arxiv.org/abs/hep-ph/0009334} {arXiv:hep-ph/0009334} \BibitemShut
  {NoStop}%
\bibitem [{\citenamefont {Kolbe}\ \emph {et~al.}(2002)\citenamefont {Kolbe},
  \citenamefont {Langanke},\ and\ \citenamefont {Vogel}}]{Kolbe:2002gk}%
  \BibitemOpen
  \bibfield  {author} {\bibinfo {author} {\bibfnamefont {E.}~\bibnamefont
  {Kolbe}}, \bibinfo {author} {\bibfnamefont {K.}~\bibnamefont {Langanke}}, \
  and\ \bibinfo {author} {\bibfnamefont {P.}~\bibnamefont {Vogel}},\ }\href
  {\doibase 10.1103/PhysRevD.66.013007} {\bibfield  {journal} {\bibinfo
  {journal} {Phys. Rev. D}\ }\textbf {\bibinfo {volume} {66}},\ \bibinfo
  {pages} {013007} (\bibinfo {year} {2002})}\BibitemShut {NoStop}%
\bibitem [{\citenamefont {Kelly}\ \emph {et~al.}(2019)\citenamefont {Kelly},
  \citenamefont {Machado}, \citenamefont {Martinez~Soler}, \citenamefont
  {Parke},\ and\ \citenamefont {Perez~Gonzalez}}]{Kelly:2019itm}%
  \BibitemOpen
  \bibfield  {author} {\bibinfo {author} {\bibfnamefont {K.~J.}\ \bibnamefont
  {Kelly}}, \bibinfo {author} {\bibfnamefont {P.~A.}\ \bibnamefont {Machado}},
  \bibinfo {author} {\bibfnamefont {I.}~\bibnamefont {Martinez~Soler}},
  \bibinfo {author} {\bibfnamefont {S.~J.}\ \bibnamefont {Parke}}, \ and\
  \bibinfo {author} {\bibfnamefont {Y.~F.}\ \bibnamefont {Perez~Gonzalez}},\
  }\href {\doibase 10.1103/PhysRevLett.123.081801} {\bibfield  {journal}
  {\bibinfo  {journal} {Phys. Rev. Lett.}\ }\textbf {\bibinfo {volume} {123}},\
  \bibinfo {pages} {081801} (\bibinfo {year} {2019})},\ \Eprint
  {http://arxiv.org/abs/1904.02751} {arXiv:1904.02751 [hep-ph]} \BibitemShut
  {NoStop}%
\bibitem [{\citenamefont {Newstead}\ \emph {et~al.}(2021)\citenamefont
  {Newstead}, \citenamefont {Lang},\ and\ \citenamefont
  {Strigari}}]{Newstead:2020fie}%
  \BibitemOpen
  \bibfield  {author} {\bibinfo {author} {\bibfnamefont {J.~L.}\ \bibnamefont
  {Newstead}}, \bibinfo {author} {\bibfnamefont {R.~F.}\ \bibnamefont {Lang}},
  \ and\ \bibinfo {author} {\bibfnamefont {L.~E.}\ \bibnamefont {Strigari}},\
  }\href {\doibase 10.1103/PhysRevD.104.115022} {\bibfield  {journal} {\bibinfo
   {journal} {Phys. Rev. D}\ }\textbf {\bibinfo {volume} {104}},\ \bibinfo
  {pages} {115022} (\bibinfo {year} {2021})},\ \Eprint
  {http://arxiv.org/abs/2002.08566} {arXiv:2002.08566 [astro-ph.CO]}
  \BibitemShut {NoStop}%
\bibitem [{\citenamefont {Denton}\ and\ \citenamefont
  {Gehrlein}(2022)}]{Denton:2021mso}%
  \BibitemOpen
  \bibfield  {author} {\bibinfo {author} {\bibfnamefont {P.~B.}\ \bibnamefont
  {Denton}}\ and\ \bibinfo {author} {\bibfnamefont {J.}~\bibnamefont
  {Gehrlein}},\ }\href {\doibase 10.1007/JHEP06(2022)135} {\bibfield  {journal}
  {\bibinfo  {journal} {JHEP}\ }\textbf {\bibinfo {volume} {06}},\ \bibinfo
  {pages} {135} (\bibinfo {year} {2022})},\ \Eprint
  {http://arxiv.org/abs/2109.14575} {arXiv:2109.14575 [hep-ph]} \BibitemShut
  {NoStop}%
\bibitem [{\citenamefont {Zhuang}\ \emph {et~al.}(2022)\citenamefont {Zhuang},
  \citenamefont {Strigari},\ and\ \citenamefont {Lang}}]{Zhuang:2021rsg}%
  \BibitemOpen
  \bibfield  {author} {\bibinfo {author} {\bibfnamefont {Y.}~\bibnamefont
  {Zhuang}}, \bibinfo {author} {\bibfnamefont {L.~E.}\ \bibnamefont
  {Strigari}}, \ and\ \bibinfo {author} {\bibfnamefont {R.~F.}\ \bibnamefont
  {Lang}},\ }\href {\doibase 10.1103/PhysRevD.105.043001} {\bibfield  {journal}
  {\bibinfo  {journal} {Phys. Rev. D}\ }\textbf {\bibinfo {volume} {105}},\
  \bibinfo {pages} {043001} (\bibinfo {year} {2022})},\ \Eprint
  {http://arxiv.org/abs/2110.14723} {arXiv:2110.14723 [hep-ph]} \BibitemShut
  {NoStop}%
\bibitem [{\citenamefont {Kelly}\ \emph {et~al.}(2023)\citenamefont {Kelly},
  \citenamefont {Machado}, \citenamefont {Mishra}, \citenamefont {Strigari},\
  and\ \citenamefont {Zhuang}}]{Kelly:2023ugn}%
  \BibitemOpen
  \bibfield  {author} {\bibinfo {author} {\bibfnamefont {K.~J.}\ \bibnamefont
  {Kelly}}, \bibinfo {author} {\bibfnamefont {P.~A.~N.}\ \bibnamefont
  {Machado}}, \bibinfo {author} {\bibfnamefont {N.}~\bibnamefont {Mishra}},
  \bibinfo {author} {\bibfnamefont {L.~E.}\ \bibnamefont {Strigari}}, \ and\
  \bibinfo {author} {\bibfnamefont {Y.}~\bibnamefont {Zhuang}},\ }\href@noop {}
  {\  (\bibinfo {year} {2023})},\ \Eprint {http://arxiv.org/abs/2304.04689}
  {arXiv:2304.04689 [hep-ph]} \BibitemShut {NoStop}%
\bibitem [{\citenamefont {Peres}\ and\ \citenamefont
  {Smirnov}(2004)}]{Peres:2003wd}%
  \BibitemOpen
  \bibfield  {author} {\bibinfo {author} {\bibfnamefont {O.~L.~G.}\
  \bibnamefont {Peres}}\ and\ \bibinfo {author} {\bibfnamefont {A.~Y.}\
  \bibnamefont {Smirnov}},\ }\href {\doibase 10.1016/j.nuclphysb.2003.12.017}
  {\bibfield  {journal} {\bibinfo  {journal} {Nucl. Phys. B}\ }\textbf
  {\bibinfo {volume} {680}},\ \bibinfo {pages} {479} (\bibinfo {year}
  {2004})},\ \Eprint {http://arxiv.org/abs/hep-ph/0309312}
  {arXiv:hep-ph/0309312} \BibitemShut {NoStop}%
\bibitem [{\citenamefont {Mena}\ \emph {et~al.}(2008)\citenamefont {Mena},
  \citenamefont {Mocioiu},\ and\ \citenamefont {Razzaque}}]{Mena:2008rh}%
  \BibitemOpen
  \bibfield  {author} {\bibinfo {author} {\bibfnamefont {O.}~\bibnamefont
  {Mena}}, \bibinfo {author} {\bibfnamefont {I.}~\bibnamefont {Mocioiu}}, \
  and\ \bibinfo {author} {\bibfnamefont {S.}~\bibnamefont {Razzaque}},\ }\href
  {\doibase 10.1103/PhysRevD.78.093003} {\bibfield  {journal} {\bibinfo
  {journal} {Phys. Rev. D}\ }\textbf {\bibinfo {volume} {78}},\ \bibinfo
  {pages} {093003} (\bibinfo {year} {2008})},\ \Eprint
  {http://arxiv.org/abs/0803.3044} {arXiv:0803.3044 [hep-ph]} \BibitemShut
  {NoStop}%
\bibitem [{\citenamefont {Akhmedov}\ \emph {et~al.}(2013)\citenamefont
  {Akhmedov}, \citenamefont {Razzaque},\ and\ \citenamefont
  {Smirnov}}]{Akhmedov:2012ah}%
  \BibitemOpen
  \bibfield  {author} {\bibinfo {author} {\bibfnamefont {E.~K.}\ \bibnamefont
  {Akhmedov}}, \bibinfo {author} {\bibfnamefont {S.}~\bibnamefont {Razzaque}},
  \ and\ \bibinfo {author} {\bibfnamefont {A.~Y.}\ \bibnamefont {Smirnov}},\
  }\href {\doibase 10.1007/JHEP02(2013)082} {\bibfield  {journal} {\bibinfo
  {journal} {JHEP}\ }\textbf {\bibinfo {volume} {02}},\ \bibinfo {pages} {082}
  (\bibinfo {year} {2013})},\ \bibinfo {note} {[Erratum: JHEP 07, 026
  (2013)]},\ \Eprint {http://arxiv.org/abs/1205.7071} {arXiv:1205.7071
  [hep-ph]} \BibitemShut {NoStop}%
\bibitem [{\citenamefont {Ioannisian}\ \emph {et~al.}(2020)\citenamefont
  {Ioannisian}, \citenamefont {Pokorski}, \citenamefont {Rosiek},\ and\
  \citenamefont {Ryczkowski}}]{Ioannisian:2020isl}%
  \BibitemOpen
  \bibfield  {author} {\bibinfo {author} {\bibfnamefont {A.}~\bibnamefont
  {Ioannisian}}, \bibinfo {author} {\bibfnamefont {S.}~\bibnamefont
  {Pokorski}}, \bibinfo {author} {\bibfnamefont {J.}~\bibnamefont {Rosiek}}, \
  and\ \bibinfo {author} {\bibfnamefont {M.}~\bibnamefont {Ryczkowski}},\
  }\href {\doibase 10.1007/JHEP10(2020)120} {\bibfield  {journal} {\bibinfo
  {journal} {JHEP}\ }\textbf {\bibinfo {volume} {10}},\ \bibinfo {pages} {120}
  (\bibinfo {year} {2020})},\ \Eprint {http://arxiv.org/abs/2005.07719}
  {arXiv:2005.07719 [hep-ph]} \BibitemShut {NoStop}%
\bibitem [{\citenamefont {An}\ \emph {et~al.}(2016)\citenamefont {An} \emph
  {et~al.}}]{JUNO:2015zny}%
  \BibitemOpen
  \bibfield  {author} {\bibinfo {author} {\bibfnamefont {F.}~\bibnamefont {An}}
  \emph {et~al.} (\bibinfo {collaboration} {JUNO}),\ }\href {\doibase
  10.1088/0954-3899/43/3/030401} {\bibfield  {journal} {\bibinfo  {journal} {J.
  Phys. G}\ }\textbf {\bibinfo {volume} {43}},\ \bibinfo {pages} {030401}
  (\bibinfo {year} {2016})},\ \Eprint {http://arxiv.org/abs/1507.05613}
  {arXiv:1507.05613 [physics.ins-det]} \BibitemShut {NoStop}%
\bibitem [{\citenamefont {Djurcic}\ \emph {et~al.}(2015)\citenamefont {Djurcic}
  \emph {et~al.}}]{JUNO:2015sjr}%
  \BibitemOpen
  \bibfield  {author} {\bibinfo {author} {\bibfnamefont {Z.}~\bibnamefont
  {Djurcic}} \emph {et~al.} (\bibinfo {collaboration} {JUNO}),\ }\href@noop {}
  {\  (\bibinfo {year} {2015})},\ \Eprint {http://arxiv.org/abs/1508.07166}
  {arXiv:1508.07166 [physics.ins-det]} \BibitemShut {NoStop}%
\bibitem [{\citenamefont {Abusleme}\ \emph
  {et~al.}(2022{\natexlab{b}})\citenamefont {Abusleme} \emph
  {et~al.}}]{JUNO:2021vlw}%
  \BibitemOpen
  \bibfield  {author} {\bibinfo {author} {\bibfnamefont {A.}~\bibnamefont
  {Abusleme}} \emph {et~al.} (\bibinfo {collaboration} {JUNO}),\ }\href
  {\doibase 10.1016/j.ppnp.2021.103927} {\bibfield  {journal} {\bibinfo
  {journal} {Prog. Part. Nucl. Phys.}\ }\textbf {\bibinfo {volume} {123}},\
  \bibinfo {pages} {103927} (\bibinfo {year} {2022}{\natexlab{b}})},\ \Eprint
  {http://arxiv.org/abs/2104.02565} {arXiv:2104.02565 [hep-ex]} \BibitemShut
  {NoStop}%
\bibitem [{\citenamefont {Cheng}\ \emph
  {et~al.}(2021{\natexlab{a}})\citenamefont {Cheng}, \citenamefont {Li},
  \citenamefont {Wen},\ and\ \citenamefont {Zhou}}]{Cheng:2020aaw}%
  \BibitemOpen
  \bibfield  {author} {\bibinfo {author} {\bibfnamefont {J.}~\bibnamefont
  {Cheng}}, \bibinfo {author} {\bibfnamefont {Y.-F.}\ \bibnamefont {Li}},
  \bibinfo {author} {\bibfnamefont {L.-J.}\ \bibnamefont {Wen}}, \ and\
  \bibinfo {author} {\bibfnamefont {S.}~\bibnamefont {Zhou}},\ }\href {\doibase
  10.1103/PhysRevD.103.053001} {\bibfield  {journal} {\bibinfo  {journal}
  {Phys. Rev. D}\ }\textbf {\bibinfo {volume} {103}},\ \bibinfo {pages}
  {053001} (\bibinfo {year} {2021}{\natexlab{a}})},\ \Eprint
  {http://arxiv.org/abs/2008.04633} {arXiv:2008.04633 [hep-ph]} \BibitemShut
  {NoStop}%
\bibitem [{\citenamefont {Cheng}\ \emph
  {et~al.}(2021{\natexlab{b}})\citenamefont {Cheng}, \citenamefont {Li},
  \citenamefont {Lu},\ and\ \citenamefont {Wen}}]{Cheng:2020oko}%
  \BibitemOpen
  \bibfield  {author} {\bibinfo {author} {\bibfnamefont {J.}~\bibnamefont
  {Cheng}}, \bibinfo {author} {\bibfnamefont {Y.-F.}\ \bibnamefont {Li}},
  \bibinfo {author} {\bibfnamefont {H.-Q.}\ \bibnamefont {Lu}}, \ and\ \bibinfo
  {author} {\bibfnamefont {L.-J.}\ \bibnamefont {Wen}},\ }\href {\doibase
  10.1103/PhysRevD.103.053002} {\bibfield  {journal} {\bibinfo  {journal}
  {Phys. Rev. D}\ }\textbf {\bibinfo {volume} {103}},\ \bibinfo {pages}
  {053002} (\bibinfo {year} {2021}{\natexlab{b}})},\ \Eprint
  {http://arxiv.org/abs/2009.04085} {arXiv:2009.04085 [hep-ex]} \BibitemShut
  {NoStop}%
\bibitem [{\citenamefont {Abusleme}\ \emph
  {et~al.}(2021{\natexlab{a}})\citenamefont {Abusleme} \emph
  {et~al.}}]{JUNO:2021tll}%
  \BibitemOpen
  \bibfield  {author} {\bibinfo {author} {\bibfnamefont {A.}~\bibnamefont
  {Abusleme}} \emph {et~al.} (\bibinfo {collaboration} {JUNO}),\ }\href
  {\doibase 10.1140/epjc/s10052-021-09565-z} {\bibfield  {journal} {\bibinfo
  {journal} {Eur. Phys. J. C}\ }\textbf {\bibinfo {volume} {81}},\ \bibinfo
  {pages} {10} (\bibinfo {year} {2021}{\natexlab{a}})},\ \Eprint
  {http://arxiv.org/abs/2103.09908} {arXiv:2103.09908 [hep-ex]} \BibitemShut
  {NoStop}%
\bibitem [{\citenamefont {Fukugita}\ \emph {et~al.}(1988)\citenamefont
  {Fukugita}, \citenamefont {Kohyama},\ and\ \citenamefont
  {Kubodera}}]{Fukugita:1988hg}%
  \BibitemOpen
  \bibfield  {author} {\bibinfo {author} {\bibfnamefont {M.}~\bibnamefont
  {Fukugita}}, \bibinfo {author} {\bibfnamefont {Y.}~\bibnamefont {Kohyama}}, \
  and\ \bibinfo {author} {\bibfnamefont {K.}~\bibnamefont {Kubodera}},\ }\href
  {\doibase 10.1016/0370-2693(88)90513-8} {\bibfield  {journal} {\bibinfo
  {journal} {Phys. Lett. B}\ }\textbf {\bibinfo {volume} {212}},\ \bibinfo
  {pages} {139} (\bibinfo {year} {1988})}\BibitemShut {NoStop}%
\bibitem [{\citenamefont {Engel}\ \emph {et~al.}(1996)\citenamefont {Engel},
  \citenamefont {Kolbe}, \citenamefont {Langanke},\ and\ \citenamefont
  {Vogel}}]{Engel:1996zt}%
  \BibitemOpen
  \bibfield  {author} {\bibinfo {author} {\bibfnamefont {J.}~\bibnamefont
  {Engel}}, \bibinfo {author} {\bibfnamefont {E.}~\bibnamefont {Kolbe}},
  \bibinfo {author} {\bibfnamefont {K.}~\bibnamefont {Langanke}}, \ and\
  \bibinfo {author} {\bibfnamefont {P.}~\bibnamefont {Vogel}},\ }\href
  {\doibase 10.1103/PhysRevC.54.2740} {\bibfield  {journal} {\bibinfo
  {journal} {Phys. Rev. C}\ }\textbf {\bibinfo {volume} {54}},\ \bibinfo
  {pages} {2740} (\bibinfo {year} {1996})},\ \Eprint
  {http://arxiv.org/abs/nucl-th/9606031} {arXiv:nucl-th/9606031} \BibitemShut
  {NoStop}%
\bibitem [{\citenamefont {Kubodera}\ and\ \citenamefont
  {Nozawa}(1994)}]{Kubodera:1993rk}%
  \BibitemOpen
  \bibfield  {author} {\bibinfo {author} {\bibfnamefont {K.}~\bibnamefont
  {Kubodera}}\ and\ \bibinfo {author} {\bibfnamefont {S.}~\bibnamefont
  {Nozawa}},\ }\href {\doibase 10.1142/S0218301394000048} {\bibfield  {journal}
  {\bibinfo  {journal} {Int. J. Mod. Phys. E}\ }\textbf {\bibinfo {volume}
  {3}},\ \bibinfo {pages} {101} (\bibinfo {year} {1994})},\ \Eprint
  {http://arxiv.org/abs/nucl-th/9310014} {arXiv:nucl-th/9310014} \BibitemShut
  {NoStop}%
\bibitem [{\citenamefont {Kolbe}(1996)}]{Kolbe:1996km}%
  \BibitemOpen
  \bibfield  {author} {\bibinfo {author} {\bibfnamefont {E.}~\bibnamefont
  {Kolbe}},\ }\href {\doibase 10.1103/PhysRevC.54.1741} {\bibfield  {journal}
  {\bibinfo  {journal} {Phys. Rev. C}\ }\textbf {\bibinfo {volume} {54}},\
  \bibinfo {pages} {1741} (\bibinfo {year} {1996})}\BibitemShut {NoStop}%
\bibitem [{\citenamefont {Kelley}\ \emph {et~al.}(2017)\citenamefont {Kelley},
  \citenamefont {Purcell},\ and\ \citenamefont {Sheu}}]{Kelley:2017qgh}%
  \BibitemOpen
  \bibfield  {author} {\bibinfo {author} {\bibfnamefont {J.~H.}\ \bibnamefont
  {Kelley}}, \bibinfo {author} {\bibfnamefont {J.~E.}\ \bibnamefont {Purcell}},
  \ and\ \bibinfo {author} {\bibfnamefont {C.~G.}\ \bibnamefont {Sheu}},\
  }\href {\doibase 10.1016/j.nuclphysa.2017.07.015} {\bibfield  {journal}
  {\bibinfo  {journal} {Nucl. Phys. A}\ }\textbf {\bibinfo {volume} {968}},\
  \bibinfo {pages} {71} (\bibinfo {year} {2017})}\BibitemShut {NoStop}%
\bibitem [{\citenamefont {Bellini}\ \emph {et~al.}(2012)\citenamefont {Bellini}
  \emph {et~al.}}]{Borexino:2011ufb}%
  \BibitemOpen
  \bibfield  {author} {\bibinfo {author} {\bibfnamefont {G.}~\bibnamefont
  {Bellini}} \emph {et~al.} (\bibinfo {collaboration} {Borexino}),\ }\href
  {\doibase 10.1103/PhysRevLett.108.051302} {\bibfield  {journal} {\bibinfo
  {journal} {Phys. Rev. Lett.}\ }\textbf {\bibinfo {volume} {108}},\ \bibinfo
  {pages} {051302} (\bibinfo {year} {2012})},\ \Eprint
  {http://arxiv.org/abs/1110.3230} {arXiv:1110.3230 [hep-ex]} \BibitemShut
  {NoStop}%
\bibitem [{\citenamefont {Bellini}\ \emph {et~al.}(2014)\citenamefont {Bellini}
  \emph {et~al.}}]{Borexino:2013zhu}%
  \BibitemOpen
  \bibfield  {author} {\bibinfo {author} {\bibfnamefont {G.}~\bibnamefont
  {Bellini}} \emph {et~al.} (\bibinfo {collaboration} {Borexino}),\ }\href
  {\doibase 10.1103/PhysRevD.89.112007} {\bibfield  {journal} {\bibinfo
  {journal} {Phys. Rev. D}\ }\textbf {\bibinfo {volume} {89}},\ \bibinfo
  {pages} {112007} (\bibinfo {year} {2014})},\ \Eprint
  {http://arxiv.org/abs/1308.0443} {arXiv:1308.0443 [hep-ex]} \BibitemShut
  {NoStop}%
\bibitem [{\citenamefont {Measday}(2001)}]{Measday:2001yr}%
  \BibitemOpen
  \bibfield  {author} {\bibinfo {author} {\bibfnamefont {D.~F.}\ \bibnamefont
  {Measday}},\ }\href {\doibase 10.1016/S0370-1573(01)00012-6} {\bibfield
  {journal} {\bibinfo  {journal} {Phys. Rept.}\ }\textbf {\bibinfo {volume}
  {354}},\ \bibinfo {pages} {243} (\bibinfo {year} {2001})}\BibitemShut
  {NoStop}%
\bibitem [{\citenamefont {Auerbach}\ \emph {et~al.}(2002)\citenamefont
  {Auerbach} \emph {et~al.}}]{LSND:2002oco}%
  \BibitemOpen
  \bibfield  {author} {\bibinfo {author} {\bibfnamefont {L.~B.}\ \bibnamefont
  {Auerbach}} \emph {et~al.} (\bibinfo {collaboration} {LSND}),\ }\href
  {\doibase 10.1103/PhysRevC.66.015501} {\bibfield  {journal} {\bibinfo
  {journal} {Phys. Rev. C}\ }\textbf {\bibinfo {volume} {66}},\ \bibinfo
  {pages} {015501} (\bibinfo {year} {2002})},\ \Eprint
  {http://arxiv.org/abs/nucl-ex/0203011} {arXiv:nucl-ex/0203011} \BibitemShut
  {NoStop}%
\bibitem [{\citenamefont {Abe}\ \emph {et~al.}(2016)\citenamefont {Abe} \emph
  {et~al.}}]{DoubleChooz:2015jlf}%
  \BibitemOpen
  \bibfield  {author} {\bibinfo {author} {\bibfnamefont {Y.}~\bibnamefont
  {Abe}} \emph {et~al.} (\bibinfo {collaboration} {Double Chooz}),\ }\href
  {\doibase 10.1103/PhysRevC.93.054608} {\bibfield  {journal} {\bibinfo
  {journal} {Phys. Rev. C}\ }\textbf {\bibinfo {volume} {93}},\ \bibinfo
  {pages} {054608} (\bibinfo {year} {2016})},\ \Eprint
  {http://arxiv.org/abs/1512.07562} {arXiv:1512.07562 [nucl-ex]} \BibitemShut
  {NoStop}%
\bibitem [{\citenamefont {{Greisen}}(1960)}]{Greisen1960}%
  \BibitemOpen
  \bibfield  {author} {\bibinfo {author} {\bibfnamefont {K.}~\bibnamefont
  {{Greisen}}},\ }\href {\doibase 10.1146/annurev.ns.10.120160.000431}
  {\bibfield  {journal} {\bibinfo  {journal} {Annual Review of Nuclear and
  Particle Science}\ }\textbf {\bibinfo {volume} {10}},\ \bibinfo {pages} {63}
  (\bibinfo {year} {1960})}\BibitemShut {NoStop}%
\bibitem [{\citenamefont {Gaisser}\ \emph {et~al.}(2016)\citenamefont
  {Gaisser}, \citenamefont {Engel},\ and\ \citenamefont
  {Resconi}}]{Gaisserbook}%
  \BibitemOpen
  \bibfield  {author} {\bibinfo {author} {\bibfnamefont {T.~K.}\ \bibnamefont
  {Gaisser}}, \bibinfo {author} {\bibfnamefont {R.}~\bibnamefont {Engel}}, \
  and\ \bibinfo {author} {\bibfnamefont {E.}~\bibnamefont {Resconi}},\
  }\enquote {\bibinfo {title} {Atmospheric muons and neutrinos},}\ in\ \href
  {\doibase 10.1017/CBO9781139192194.008} {\emph {\bibinfo {booktitle} {Cosmic
  Rays and Particle Physics}}}\ (\bibinfo  {publisher} {Cambridge University
  Press},\ \bibinfo {year} {2016})\ p.\ \bibinfo {pages} {126–148},\ \bibinfo
  {edition} {2nd}\ ed.\BibitemShut {Stop}%
\bibitem [{\citenamefont {Gaisser}\ and\ \citenamefont
  {Honda}(2002)}]{Gaisser:2002jj}%
  \BibitemOpen
  \bibfield  {author} {\bibinfo {author} {\bibfnamefont {T.~K.}\ \bibnamefont
  {Gaisser}}\ and\ \bibinfo {author} {\bibfnamefont {M.}~\bibnamefont
  {Honda}},\ }\href {\doibase 10.1146/annurev.nucl.52.050102.090645} {\bibfield
   {journal} {\bibinfo  {journal} {Ann. Rev. Nucl. Part. Sci.}\ }\textbf
  {\bibinfo {volume} {52}},\ \bibinfo {pages} {153} (\bibinfo {year} {2002})},\
  \Eprint {http://arxiv.org/abs/hep-ph/0203272} {arXiv:hep-ph/0203272}
  \BibitemShut {NoStop}%
\bibitem [{\citenamefont {Potgieter}(2013)}]{Potgieter:2013pdj}%
  \BibitemOpen
  \bibfield  {author} {\bibinfo {author} {\bibfnamefont {M.}~\bibnamefont
  {Potgieter}},\ }\href {\doibase 10.12942/lrsp-2013-3} {\bibfield  {journal}
  {\bibinfo  {journal} {Living Rev. Solar Phys.}\ }\textbf {\bibinfo {volume}
  {10}},\ \bibinfo {pages} {3} (\bibinfo {year} {2013})},\ \Eprint
  {http://arxiv.org/abs/1306.4421} {arXiv:1306.4421 [physics.space-ph]}
  \BibitemShut {NoStop}%
\bibitem [{\citenamefont {Moraal}(2013)}]{Moraal:2013jxa}%
  \BibitemOpen
  \bibfield  {author} {\bibinfo {author} {\bibfnamefont {H.}~\bibnamefont
  {Moraal}},\ }\href {\doibase 10.1007/s11214-011-9819-3} {\bibfield  {journal}
  {\bibinfo  {journal} {Space Sci. Rev.}\ }\textbf {\bibinfo {volume} {176}},\
  \bibinfo {pages} {299} (\bibinfo {year} {2013})}\BibitemShut {NoStop}%
\bibitem [{\citenamefont {Li}\ \emph {et~al.}(2022)\citenamefont {Li},
  \citenamefont {Beacom},\ and\ \citenamefont {Peter}}]{Li:2022zio}%
  \BibitemOpen
  \bibfield  {author} {\bibinfo {author} {\bibfnamefont {J.-T.}\ \bibnamefont
  {Li}}, \bibinfo {author} {\bibfnamefont {J.~F.}\ \bibnamefont {Beacom}}, \
  and\ \bibinfo {author} {\bibfnamefont {A.~H.~G.}\ \bibnamefont {Peter}},\
  }\href {\doibase 10.3847/1538-4357/ac8cf3} {\bibfield  {journal} {\bibinfo
  {journal} {Astrophys. J.}\ }\textbf {\bibinfo {volume} {937}},\ \bibinfo
  {pages} {27} (\bibinfo {year} {2022})},\ \Eprint
  {http://arxiv.org/abs/2206.14815} {arXiv:2206.14815 [astro-ph.SR]}
  \BibitemShut {NoStop}%
\bibitem [{\citenamefont {Honda}\ \emph {et~al.}(2001)\citenamefont {Honda},
  \citenamefont {Kajita}, \citenamefont {Kasahara},\ and\ \citenamefont
  {Midorikawa}}]{Honda:2001xy}%
  \BibitemOpen
  \bibfield  {author} {\bibinfo {author} {\bibfnamefont {M.}~\bibnamefont
  {Honda}}, \bibinfo {author} {\bibfnamefont {T.}~\bibnamefont {Kajita}},
  \bibinfo {author} {\bibfnamefont {K.}~\bibnamefont {Kasahara}}, \ and\
  \bibinfo {author} {\bibfnamefont {S.}~\bibnamefont {Midorikawa}},\ }\href
  {\doibase 10.1103/PhysRevD.64.053011} {\bibfield  {journal} {\bibinfo
  {journal} {Phys. Rev. D}\ }\textbf {\bibinfo {volume} {64}},\ \bibinfo
  {pages} {053011} (\bibinfo {year} {2001})},\ \Eprint
  {http://arxiv.org/abs/hep-ph/0103328} {arXiv:hep-ph/0103328} \BibitemShut
  {NoStop}%
\bibitem [{\citenamefont {Barr}\ \emph {et~al.}(2004)\citenamefont {Barr},
  \citenamefont {Gaisser}, \citenamefont {Lipari}, \citenamefont {Robbins},\
  and\ \citenamefont {Stanev}}]{Barr:2004br}%
  \BibitemOpen
  \bibfield  {author} {\bibinfo {author} {\bibfnamefont {G.~D.}\ \bibnamefont
  {Barr}}, \bibinfo {author} {\bibfnamefont {T.~K.}\ \bibnamefont {Gaisser}},
  \bibinfo {author} {\bibfnamefont {P.}~\bibnamefont {Lipari}}, \bibinfo
  {author} {\bibfnamefont {S.}~\bibnamefont {Robbins}}, \ and\ \bibinfo
  {author} {\bibfnamefont {T.}~\bibnamefont {Stanev}},\ }\href {\doibase
  10.1103/PhysRevD.70.023006} {\bibfield  {journal} {\bibinfo  {journal} {Phys.
  Rev. D}\ }\textbf {\bibinfo {volume} {70}},\ \bibinfo {pages} {023006}
  (\bibinfo {year} {2004})},\ \Eprint {http://arxiv.org/abs/astro-ph/0403630}
  {arXiv:astro-ph/0403630} \BibitemShut {NoStop}%
\bibitem [{\citenamefont {Liu}\ \emph {et~al.}(2003)\citenamefont {Liu},
  \citenamefont {Derome},\ and\ \citenamefont {Buenerd}}]{Liu:2002sq}%
  \BibitemOpen
  \bibfield  {author} {\bibinfo {author} {\bibfnamefont {Y.}~\bibnamefont
  {Liu}}, \bibinfo {author} {\bibfnamefont {L.}~\bibnamefont {Derome}}, \ and\
  \bibinfo {author} {\bibfnamefont {M.}~\bibnamefont {Buenerd}},\ }\href
  {\doibase 10.1103/PhysRevD.67.073022} {\bibfield  {journal} {\bibinfo
  {journal} {Phys. Rev. D}\ }\textbf {\bibinfo {volume} {67}},\ \bibinfo
  {pages} {073022} (\bibinfo {year} {2003})},\ \Eprint
  {http://arxiv.org/abs/astro-ph/0211632} {arXiv:astro-ph/0211632} \BibitemShut
  {NoStop}%
\bibitem [{\citenamefont {Battistoni}\ \emph {et~al.}(2005)\citenamefont
  {Battistoni}, \citenamefont {Ferrari}, \citenamefont {Montaruli},\ and\
  \citenamefont {Sala}}]{Battistoni:2005pd}%
  \BibitemOpen
  \bibfield  {author} {\bibinfo {author} {\bibfnamefont {G.}~\bibnamefont
  {Battistoni}}, \bibinfo {author} {\bibfnamefont {A.}~\bibnamefont {Ferrari}},
  \bibinfo {author} {\bibfnamefont {T.}~\bibnamefont {Montaruli}}, \ and\
  \bibinfo {author} {\bibfnamefont {P.~R.}\ \bibnamefont {Sala}},\ }\href
  {\doibase 10.1016/j.astropartphys.2005.03.006} {\bibfield  {journal}
  {\bibinfo  {journal} {Astropart. Phys.}\ }\textbf {\bibinfo {volume} {23}},\
  \bibinfo {pages} {526} (\bibinfo {year} {2005})}\BibitemShut {NoStop}%
\bibitem [{\citenamefont {Wentz}\ \emph {et~al.}(2003)\citenamefont {Wentz},
  \citenamefont {Brancus}, \citenamefont {Bercuci}, \citenamefont {Heck},
  \citenamefont {Oehlschlager}, \citenamefont {Rebel},\ and\ \citenamefont
  {Vulpescu}}]{Wentz:2003bp}%
  \BibitemOpen
  \bibfield  {author} {\bibinfo {author} {\bibfnamefont {J.}~\bibnamefont
  {Wentz}}, \bibinfo {author} {\bibfnamefont {I.~M.}\ \bibnamefont {Brancus}},
  \bibinfo {author} {\bibfnamefont {A.}~\bibnamefont {Bercuci}}, \bibinfo
  {author} {\bibfnamefont {D.}~\bibnamefont {Heck}}, \bibinfo {author}
  {\bibfnamefont {J.}~\bibnamefont {Oehlschlager}}, \bibinfo {author}
  {\bibfnamefont {H.}~\bibnamefont {Rebel}}, \ and\ \bibinfo {author}
  {\bibfnamefont {B.}~\bibnamefont {Vulpescu}},\ }\href {\doibase
  10.1103/PhysRevD.67.073020} {\bibfield  {journal} {\bibinfo  {journal} {Phys.
  Rev. D}\ }\textbf {\bibinfo {volume} {67}},\ \bibinfo {pages} {073020}
  (\bibinfo {year} {2003})},\ \Eprint {http://arxiv.org/abs/hep-ph/0301199}
  {arXiv:hep-ph/0301199} \BibitemShut {NoStop}%
\bibitem [{\citenamefont {Zhuang}\ and\ \citenamefont
  {Strigari}(2023)}]{Yi-Zhuang:2023}%
  \BibitemOpen
  \bibfield  {author} {\bibinfo {author} {\bibfnamefont {Y.}~\bibnamefont
  {Zhuang}}\ and\ \bibinfo {author} {\bibfnamefont {L.~E.}\ \bibnamefont
  {Strigari}},\ }\href@noop {} {\enquote {\bibinfo {title} {private
  communication},}\ } (\bibinfo {year} {2023})\BibitemShut {NoStop}%
\bibitem [{\citenamefont {{Stecker}}(1971)}]{Steckerbook}%
  \BibitemOpen
  \bibfield  {author} {\bibinfo {author} {\bibfnamefont {F.~W.}\ \bibnamefont
  {{Stecker}}},\ }\href@noop {} {\emph {\bibinfo {title} {{Cosmic gamma
  rays}}}},\ Vol.\ \bibinfo {volume} {249}\ (\bibinfo  {publisher} {Baltimore:
  Mono Book Corp.},\ \bibinfo {year} {1971})\BibitemShut {NoStop}%
\bibitem [{\citenamefont {Kelner}\ \emph {et~al.}(2006)\citenamefont {Kelner},
  \citenamefont {Aharonian},\ and\ \citenamefont {Bugayov}}]{Kelner:2006tc}%
  \BibitemOpen
  \bibfield  {author} {\bibinfo {author} {\bibfnamefont {S.~R.}\ \bibnamefont
  {Kelner}}, \bibinfo {author} {\bibfnamefont {F.~A.}\ \bibnamefont
  {Aharonian}}, \ and\ \bibinfo {author} {\bibfnamefont {V.~V.}\ \bibnamefont
  {Bugayov}},\ }\href {\doibase 10.1103/PhysRevD.74.034018} {\bibfield
  {journal} {\bibinfo  {journal} {Phys. Rev. D}\ }\textbf {\bibinfo {volume}
  {74}},\ \bibinfo {pages} {034018} (\bibinfo {year} {2006})},\ \bibinfo {note}
  {[Erratum: Phys.Rev.D 79, 039901 (2009)]},\ \Eprint
  {http://arxiv.org/abs/astro-ph/0606058} {arXiv:astro-ph/0606058} \BibitemShut
  {NoStop}%
\bibitem [{\citenamefont {Wallraff}\ and\ \citenamefont
  {Wiebusch}(2015)}]{Wallraff:2014qka}%
  \BibitemOpen
  \bibfield  {author} {\bibinfo {author} {\bibfnamefont {M.}~\bibnamefont
  {Wallraff}}\ and\ \bibinfo {author} {\bibfnamefont {C.}~\bibnamefont
  {Wiebusch}},\ }\href {\doibase 10.1016/j.cpc.2015.07.010} {\bibfield
  {journal} {\bibinfo  {journal} {Comput. Phys. Commun.}\ }\textbf {\bibinfo
  {volume} {197}},\ \bibinfo {pages} {185} (\bibinfo {year} {2015})},\ \Eprint
  {http://arxiv.org/abs/1409.1387} {arXiv:1409.1387 [astro-ph.IM]} \BibitemShut
  {NoStop}%
\bibitem [{\citenamefont {Wolfenstein}(1978)}]{Wolfenstein:1977ue}%
  \BibitemOpen
  \bibfield  {author} {\bibinfo {author} {\bibfnamefont {L.}~\bibnamefont
  {Wolfenstein}},\ }\href {\doibase 10.1103/PhysRevD.17.2369} {\bibfield
  {journal} {\bibinfo  {journal} {Phys. Rev. D}\ }\textbf {\bibinfo {volume}
  {17}},\ \bibinfo {pages} {2369} (\bibinfo {year} {1978})}\BibitemShut
  {NoStop}%
\bibitem [{\citenamefont {Mikheyev}\ and\ \citenamefont
  {Smirnov}(1985)}]{Mikheyev:1985zog}%
  \BibitemOpen
  \bibfield  {author} {\bibinfo {author} {\bibfnamefont {S.~P.}\ \bibnamefont
  {Mikheyev}}\ and\ \bibinfo {author} {\bibfnamefont {A.~Y.}\ \bibnamefont
  {Smirnov}},\ }\href@noop {} {\bibfield  {journal} {\bibinfo  {journal} {Sov.
  J. Nucl. Phys.}\ }\textbf {\bibinfo {volume} {42}},\ \bibinfo {pages} {913}
  (\bibinfo {year} {1985})}\BibitemShut {NoStop}%
\bibitem [{\citenamefont {Akhmedov}(1988)}]{Akhmedov:1988kd}%
  \BibitemOpen
  \bibfield  {author} {\bibinfo {author} {\bibfnamefont {E.~K.}\ \bibnamefont
  {Akhmedov}},\ }\href@noop {} {\bibfield  {journal} {\bibinfo  {journal} {Sov.
  J. Nucl. Phys.}\ }\textbf {\bibinfo {volume} {47}},\ \bibinfo {pages} {301}
  (\bibinfo {year} {1988})}\BibitemShut {NoStop}%
\bibitem [{\citenamefont {Krastev}\ and\ \citenamefont
  {Smirnov}(1989)}]{Krastev:1989ix}%
  \BibitemOpen
  \bibfield  {author} {\bibinfo {author} {\bibfnamefont {P.~I.}\ \bibnamefont
  {Krastev}}\ and\ \bibinfo {author} {\bibfnamefont {A.~Y.}\ \bibnamefont
  {Smirnov}},\ }\href {\doibase 10.1016/0370-2693(89)91206-9} {\bibfield
  {journal} {\bibinfo  {journal} {Phys. Lett. B}\ }\textbf {\bibinfo {volume}
  {226}},\ \bibinfo {pages} {341} (\bibinfo {year} {1989})}\BibitemShut
  {NoStop}%
\bibitem [{\citenamefont {Armbruster}\ \emph {et~al.}(1998)\citenamefont
  {Armbruster} \emph {et~al.}}]{KARMEN:1998xmo}%
  \BibitemOpen
  \bibfield  {author} {\bibinfo {author} {\bibfnamefont {B.}~\bibnamefont
  {Armbruster}} \emph {et~al.} (\bibinfo {collaboration} {KARMEN}),\ }\href
  {\doibase 10.1016/S0370-2693(98)00087-2} {\bibfield  {journal} {\bibinfo
  {journal} {Phys. Lett. B}\ }\textbf {\bibinfo {volume} {423}},\ \bibinfo
  {pages} {15} (\bibinfo {year} {1998})}\BibitemShut {NoStop}%
\bibitem [{\citenamefont {Zeitnitz}(1994)}]{Zeitnitz:1994kz}%
  \BibitemOpen
  \bibfield  {author} {\bibinfo {author} {\bibfnamefont {B.}~\bibnamefont
  {Zeitnitz}} (\bibinfo {collaboration} {KARMEN}),\ }\href {\doibase
  10.1016/0146-6410(94)90034-5} {\bibfield  {journal} {\bibinfo  {journal}
  {Prog. Part. Nucl. Phys.}\ }\textbf {\bibinfo {volume} {32}},\ \bibinfo
  {pages} {351} (\bibinfo {year} {1994})}\BibitemShut {NoStop}%
\bibitem [{\citenamefont {Athanassopoulos}\ \emph {et~al.}(1997)\citenamefont
  {Athanassopoulos} \emph {et~al.}}]{LSND:1997lta}%
  \BibitemOpen
  \bibfield  {author} {\bibinfo {author} {\bibfnamefont {C.}~\bibnamefont
  {Athanassopoulos}} \emph {et~al.} (\bibinfo {collaboration} {LSND}),\ }\href
  {\doibase 10.1103/PhysRevC.55.2078} {\bibfield  {journal} {\bibinfo
  {journal} {Phys. Rev. C}\ }\textbf {\bibinfo {volume} {55}},\ \bibinfo
  {pages} {2078} (\bibinfo {year} {1997})},\ \Eprint
  {http://arxiv.org/abs/nucl-ex/9705001} {arXiv:nucl-ex/9705001} \BibitemShut
  {NoStop}%
\bibitem [{\citenamefont {Auerbach}\ \emph
  {et~al.}(2001{\natexlab{a}})\citenamefont {Auerbach} \emph
  {et~al.}}]{LSND:2001fbw}%
  \BibitemOpen
  \bibfield  {author} {\bibinfo {author} {\bibfnamefont {L.~B.}\ \bibnamefont
  {Auerbach}} \emph {et~al.} (\bibinfo {collaboration} {LSND}),\ }\href
  {\doibase 10.1103/PhysRevC.64.065501} {\bibfield  {journal} {\bibinfo
  {journal} {Phys. Rev. C}\ }\textbf {\bibinfo {volume} {64}},\ \bibinfo
  {pages} {065501} (\bibinfo {year} {2001}{\natexlab{a}})},\ \Eprint
  {http://arxiv.org/abs/hep-ex/0105068} {arXiv:hep-ex/0105068} \BibitemShut
  {NoStop}%
\bibitem [{\citenamefont {Fujii}\ and\ \citenamefont
  {Yamaguchi}(1964)}]{Akihiko:1964}%
  \BibitemOpen
  \bibfield  {author} {\bibinfo {author} {\bibfnamefont {A.}~\bibnamefont
  {Fujii}}\ and\ \bibinfo {author} {\bibfnamefont {Y.}~\bibnamefont
  {Yamaguchi}},\ }\href {\doibase 10.1143/PTP.31.107} {\bibfield  {journal}
  {\bibinfo  {journal} {Progress of Theoretical Physics}\ }\textbf {\bibinfo
  {volume} {31}},\ \bibinfo {pages} {107} (\bibinfo {year} {1964})},\ \Eprint
  {http://arxiv.org/abs/https://academic.oup.com/ptp/article-pdf/31/1/107/5174549/31-1-107.pdf}
  {https://academic.oup.com/ptp/article-pdf/31/1/107/5174549/31-1-107.pdf}
  \BibitemShut {NoStop}%
\bibitem [{\citenamefont {Kim}\ and\ \citenamefont
  {Primakoff}(1965)}]{Kim:1965zzc}%
  \BibitemOpen
  \bibfield  {author} {\bibinfo {author} {\bibfnamefont {C.~W.}\ \bibnamefont
  {Kim}}\ and\ \bibinfo {author} {\bibfnamefont {H.}~\bibnamefont
  {Primakoff}},\ }\href {\doibase 10.1103/PhysRev.140.B566} {\bibfield
  {journal} {\bibinfo  {journal} {Phys. Rev.}\ }\textbf {\bibinfo {volume}
  {140}},\ \bibinfo {pages} {B566} (\bibinfo {year} {1965})}\BibitemShut
  {NoStop}%
\bibitem [{\citenamefont {Kubodera}\ and\ \citenamefont
  {Kim}(1973)}]{Kubodera:1973kv}%
  \BibitemOpen
  \bibfield  {author} {\bibinfo {author} {\bibfnamefont {K.}~\bibnamefont
  {Kubodera}}\ and\ \bibinfo {author} {\bibfnamefont {C.~W.}\ \bibnamefont
  {Kim}},\ }\href {\doibase 10.1016/0370-2693(73)90438-3} {\bibfield  {journal}
  {\bibinfo  {journal} {Phys. Lett. B}\ }\textbf {\bibinfo {volume} {43}},\
  \bibinfo {pages} {275} (\bibinfo {year} {1973})}\BibitemShut {NoStop}%
\bibitem [{\citenamefont {Hwang}\ and\ \citenamefont
  {Primakoff}(1977)}]{Hwang:1977fi}%
  \BibitemOpen
  \bibfield  {author} {\bibinfo {author} {\bibfnamefont {W.~Y.~P.}\
  \bibnamefont {Hwang}}\ and\ \bibinfo {author} {\bibfnamefont
  {H.}~\bibnamefont {Primakoff}},\ }\href {\doibase 10.1103/PhysRevC.16.397}
  {\bibfield  {journal} {\bibinfo  {journal} {Phys. Rev. C}\ }\textbf {\bibinfo
  {volume} {16}},\ \bibinfo {pages} {397} (\bibinfo {year} {1977})}\BibitemShut
  {NoStop}%
\bibitem [{\citenamefont {Nozawa}\ \emph {et~al.}(1983)\citenamefont {Nozawa},
  \citenamefont {Kohyama},\ and\ \citenamefont {Kubodera}}]{Nozawa:1983imd}%
  \BibitemOpen
  \bibfield  {author} {\bibinfo {author} {\bibfnamefont {S.}~\bibnamefont
  {Nozawa}}, \bibinfo {author} {\bibfnamefont {Y.}~\bibnamefont {Kohyama}}, \
  and\ \bibinfo {author} {\bibfnamefont {K.}~\bibnamefont {Kubodera}},\ }\href
  {\doibase 10.1143/PTP.70.892} {\bibfield  {journal} {\bibinfo  {journal}
  {Prog. Theor. Phys.}\ }\textbf {\bibinfo {volume} {70}},\ \bibinfo {pages}
  {892} (\bibinfo {year} {1983})}\BibitemShut {NoStop}%
\bibitem [{\citenamefont {Leitner}\ \emph {et~al.}(2006)\citenamefont
  {Leitner}, \citenamefont {Alvarez-Ruso},\ and\ \citenamefont
  {Mosel}}]{Leitner:2006ww}%
  \BibitemOpen
  \bibfield  {author} {\bibinfo {author} {\bibfnamefont {T.}~\bibnamefont
  {Leitner}}, \bibinfo {author} {\bibfnamefont {L.}~\bibnamefont
  {Alvarez-Ruso}}, \ and\ \bibinfo {author} {\bibfnamefont {U.}~\bibnamefont
  {Mosel}},\ }\href {\doibase 10.1103/PhysRevC.73.065502} {\bibfield  {journal}
  {\bibinfo  {journal} {Phys. Rev. C}\ }\textbf {\bibinfo {volume} {73}},\
  \bibinfo {pages} {065502} (\bibinfo {year} {2006})},\ \Eprint
  {http://arxiv.org/abs/nucl-th/0601103} {arXiv:nucl-th/0601103} \BibitemShut
  {NoStop}%
\bibitem [{\citenamefont {Benhar}\ and\ \citenamefont
  {Rocco}(2013)}]{Benhar:2013bwa}%
  \BibitemOpen
  \bibfield  {author} {\bibinfo {author} {\bibfnamefont {O.}~\bibnamefont
  {Benhar}}\ and\ \bibinfo {author} {\bibfnamefont {N.}~\bibnamefont {Rocco}},\
  }\href {\doibase 10.1155/2013/912702} {\bibfield  {journal} {\bibinfo
  {journal} {Adv. High Energy Phys.}\ }\textbf {\bibinfo {volume} {2013}},\
  \bibinfo {pages} {912702} (\bibinfo {year} {2013})},\ \Eprint
  {http://arxiv.org/abs/1310.3869} {arXiv:1310.3869 [nucl-th]} \BibitemShut
  {NoStop}%
\bibitem [{\citenamefont {Ruso}\ \emph {et~al.}(2022)\citenamefont {Ruso} \emph
  {et~al.}}]{Ruso:2022qes}%
  \BibitemOpen
  \bibfield  {author} {\bibinfo {author} {\bibfnamefont {L.~A.}\ \bibnamefont
  {Ruso}} \emph {et~al.},\ }\href@noop {} {\  (\bibinfo {year} {2022})},\
  \Eprint {http://arxiv.org/abs/2203.09030} {arXiv:2203.09030 [hep-ph]}
  \BibitemShut {NoStop}%
\bibitem [{\citenamefont {Pourkaviani}\ and\ \citenamefont
  {Mintz}(1990)}]{Pourkaviani:1990et}%
  \BibitemOpen
  \bibfield  {author} {\bibinfo {author} {\bibfnamefont {M.}~\bibnamefont
  {Pourkaviani}}\ and\ \bibinfo {author} {\bibfnamefont {S.~L.}\ \bibnamefont
  {Mintz}},\ }\href {\doibase 10.1088/0954-3899/16/4/007} {\bibfield  {journal}
  {\bibinfo  {journal} {J. Phys. G}\ }\textbf {\bibinfo {volume} {16}},\
  \bibinfo {pages} {569} (\bibinfo {year} {1990})}\BibitemShut {NoStop}%
\bibitem [{\citenamefont {Chertok}\ \emph {et~al.}(1973)\citenamefont
  {Chertok}, \citenamefont {Sheffield}, \citenamefont {Lightbody},
  \citenamefont {Penner},\ and\ \citenamefont {Blum}}]{Chertok:1973}%
  \BibitemOpen
  \bibfield  {author} {\bibinfo {author} {\bibfnamefont {B.~T.}\ \bibnamefont
  {Chertok}}, \bibinfo {author} {\bibfnamefont {C.}~\bibnamefont {Sheffield}},
  \bibinfo {author} {\bibfnamefont {J.~W.}\ \bibnamefont {Lightbody}}, \bibinfo
  {author} {\bibfnamefont {S.}~\bibnamefont {Penner}}, \ and\ \bibinfo {author}
  {\bibfnamefont {D.}~\bibnamefont {Blum}},\ }\href {\doibase
  10.1103/PhysRevC.8.23} {\bibfield  {journal} {\bibinfo  {journal} {Phys. Rev.
  C}\ }\textbf {\bibinfo {volume} {8}},\ \bibinfo {pages} {23} (\bibinfo {year}
  {1973})}\BibitemShut {NoStop}%
\bibitem [{\citenamefont {Haxton}(1978)}]{Haxton:1977jr}%
  \BibitemOpen
  \bibfield  {author} {\bibinfo {author} {\bibfnamefont {W.~C.}\ \bibnamefont
  {Haxton}},\ }\href {\doibase 10.1016/0370-2693(78)90266-6} {\bibfield
  {journal} {\bibinfo  {journal} {Phys. Lett. B}\ }\textbf {\bibinfo {volume}
  {76}},\ \bibinfo {pages} {165} (\bibinfo {year} {1978})}\BibitemShut
  {NoStop}%
\bibitem [{\citenamefont {Fukugita}\ and\ \citenamefont
  {Yanagida}(1993)}]{Fukugita:1993fr}%
  \BibitemOpen
  \bibfield  {author} {\bibinfo {author} {\bibfnamefont {M.}~\bibnamefont
  {Fukugita}}\ and\ \bibinfo {author} {\bibfnamefont {T.}~\bibnamefont
  {Yanagida}},\ }\enquote {\bibinfo {title} {{Physics of neutrinos}},}\ in\
  \href {\doibase 10.1007/978-4-431-67029-2_1} {\emph {\bibinfo {booktitle}
  {{Physics and astrophysics of neutrinos}}}},\ \bibinfo {editor} {edited by\
  \bibinfo {editor} {\bibfnamefont {M.}~\bibnamefont {Fukugita}}\ and\ \bibinfo
  {editor} {\bibfnamefont {A.}~\bibnamefont {Suzuki}}}\ (\bibinfo  {publisher}
  {Springer, Tokyo},\ \bibinfo {year} {1993})\ pp.\ \bibinfo {pages}
  {1--248}\BibitemShut {NoStop}%
\bibitem [{\citenamefont {{Fermi}}(1934)}]{1934ZPhy...88..161F}%
  \BibitemOpen
  \bibfield  {author} {\bibinfo {author} {\bibfnamefont {E.}~\bibnamefont
  {{Fermi}}},\ }\href {\doibase 10.1007/BF01351864} {\bibfield  {journal}
  {\bibinfo  {journal} {Zeitschrift fur Physik}\ }\textbf {\bibinfo {volume}
  {88}},\ \bibinfo {pages} {161} (\bibinfo {year} {1934})}\BibitemShut
  {NoStop}%
\bibitem [{\citenamefont {Schenter}\ and\ \citenamefont
  {Vogel}(1983)}]{doi:10.13182/NSE83-A17574}%
  \BibitemOpen
  \bibfield  {author} {\bibinfo {author} {\bibfnamefont {G.~K.}\ \bibnamefont
  {Schenter}}\ and\ \bibinfo {author} {\bibfnamefont {P.}~\bibnamefont
  {Vogel}},\ }\href {\doibase 10.13182/NSE83-A17574} {\bibfield  {journal}
  {\bibinfo  {journal} {Nuclear Science and Engineering}\ }\textbf {\bibinfo
  {volume} {83}},\ \bibinfo {pages} {393} (\bibinfo {year} {1983})},\ \Eprint
  {http://arxiv.org/abs/https://doi.org/10.13182/NSE83-A17574}
  {https://doi.org/10.13182/NSE83-A17574} \BibitemShut {NoStop}%
\bibitem [{\citenamefont {Venkataramaiah}\ \emph {et~al.}(1985)\citenamefont
  {Venkataramaiah}, \citenamefont {Gopala}, \citenamefont {Basavaraju},
  \citenamefont {Suryanarayana},\ and\ \citenamefont
  {Sanjeeviah}}]{Venkataramaiah_1985}%
  \BibitemOpen
  \bibfield  {author} {\bibinfo {author} {\bibfnamefont {P.}~\bibnamefont
  {Venkataramaiah}}, \bibinfo {author} {\bibfnamefont {K.}~\bibnamefont
  {Gopala}}, \bibinfo {author} {\bibfnamefont {A.}~\bibnamefont {Basavaraju}},
  \bibinfo {author} {\bibfnamefont {S.~S.}\ \bibnamefont {Suryanarayana}}, \
  and\ \bibinfo {author} {\bibfnamefont {H.}~\bibnamefont {Sanjeeviah}},\
  }\href {\doibase 10.1088/0305-4616/11/3/014} {\bibfield  {journal} {\bibinfo
  {journal} {Journal of Physics G: Nuclear Physics}\ }\textbf {\bibinfo
  {volume} {11}},\ \bibinfo {pages} {359} (\bibinfo {year} {1985})}\BibitemShut
  {NoStop}%
\bibitem [{\citenamefont {Auerbach}\ \emph
  {et~al.}(2001{\natexlab{b}})\citenamefont {Auerbach} \emph
  {et~al.}}]{LSND:2001akn}%
  \BibitemOpen
  \bibfield  {author} {\bibinfo {author} {\bibfnamefont {L.~B.}\ \bibnamefont
  {Auerbach}} \emph {et~al.} (\bibinfo {collaboration} {LSND}),\ }\href
  {\doibase 10.1103/PhysRevD.63.112001} {\bibfield  {journal} {\bibinfo
  {journal} {Phys. Rev. D}\ }\textbf {\bibinfo {volume} {63}},\ \bibinfo
  {pages} {112001} (\bibinfo {year} {2001}{\natexlab{b}})},\ \Eprint
  {http://arxiv.org/abs/hep-ex/0101039} {arXiv:hep-ex/0101039} \BibitemShut
  {NoStop}%
\bibitem [{\citenamefont {Aguilar-Arevalo}\ \emph {et~al.}(2008)\citenamefont
  {Aguilar-Arevalo} \emph {et~al.}}]{MiniBooNE:2007iti}%
  \BibitemOpen
  \bibfield  {author} {\bibinfo {author} {\bibfnamefont {A.~A.}\ \bibnamefont
  {Aguilar-Arevalo}} \emph {et~al.} (\bibinfo {collaboration} {MiniBooNE}),\
  }\href {\doibase 10.1103/PhysRevLett.100.032301} {\bibfield  {journal}
  {\bibinfo  {journal} {Phys. Rev. Lett.}\ }\textbf {\bibinfo {volume} {100}},\
  \bibinfo {pages} {032301} (\bibinfo {year} {2008})},\ \Eprint
  {http://arxiv.org/abs/0706.0926} {arXiv:0706.0926 [hep-ex]} \BibitemShut
  {NoStop}%
\bibitem [{\citenamefont {Aguilar-Arevalo}\ \emph {et~al.}(2010)\citenamefont
  {Aguilar-Arevalo} \emph {et~al.}}]{MiniBooNE:2010bsu}%
  \BibitemOpen
  \bibfield  {author} {\bibinfo {author} {\bibfnamefont {A.~A.}\ \bibnamefont
  {Aguilar-Arevalo}} \emph {et~al.} (\bibinfo {collaboration} {MiniBooNE}),\
  }\href {\doibase 10.1103/PhysRevD.81.092005} {\bibfield  {journal} {\bibinfo
  {journal} {Phys. Rev. D}\ }\textbf {\bibinfo {volume} {81}},\ \bibinfo
  {pages} {092005} (\bibinfo {year} {2010})},\ \Eprint
  {http://arxiv.org/abs/1002.2680} {arXiv:1002.2680 [hep-ex]} \BibitemShut
  {NoStop}%
\bibitem [{\citenamefont {Aguilar-Arevalo}\ \emph {et~al.}(2013)\citenamefont
  {Aguilar-Arevalo} \emph {et~al.}}]{MiniBooNE:2013qnd}%
  \BibitemOpen
  \bibfield  {author} {\bibinfo {author} {\bibfnamefont {A.~A.}\ \bibnamefont
  {Aguilar-Arevalo}} \emph {et~al.} (\bibinfo {collaboration} {MiniBooNE}),\
  }\href {\doibase 10.1103/PhysRevD.88.032001} {\bibfield  {journal} {\bibinfo
  {journal} {Phys. Rev. D}\ }\textbf {\bibinfo {volume} {88}},\ \bibinfo
  {pages} {032001} (\bibinfo {year} {2013})},\ \Eprint
  {http://arxiv.org/abs/1301.7067} {arXiv:1301.7067 [hep-ex]} \BibitemShut
  {NoStop}%
\bibitem [{\citenamefont {Kolbe}\ \emph {et~al.}(1994)\citenamefont {Kolbe},
  \citenamefont {Langanke},\ and\ \citenamefont {Krewald}}]{Kolbe:1994xb}%
  \BibitemOpen
  \bibfield  {author} {\bibinfo {author} {\bibfnamefont {E.}~\bibnamefont
  {Kolbe}}, \bibinfo {author} {\bibfnamefont {K.}~\bibnamefont {Langanke}}, \
  and\ \bibinfo {author} {\bibfnamefont {S.}~\bibnamefont {Krewald}},\ }\href
  {\doibase 10.1103/PhysRevC.49.1122} {\bibfield  {journal} {\bibinfo
  {journal} {Phys. Rev. C}\ }\textbf {\bibinfo {volume} {49}},\ \bibinfo
  {pages} {1122} (\bibinfo {year} {1994})}\BibitemShut {NoStop}%
\bibitem [{\citenamefont {Auerbach}\ \emph {et~al.}(1997)\citenamefont
  {Auerbach}, \citenamefont {Van~Giai},\ and\ \citenamefont
  {Vorov}}]{Auerbach:1997ay}%
  \BibitemOpen
  \bibfield  {author} {\bibinfo {author} {\bibfnamefont {N.}~\bibnamefont
  {Auerbach}}, \bibinfo {author} {\bibfnamefont {N.}~\bibnamefont {Van~Giai}},
  \ and\ \bibinfo {author} {\bibfnamefont {O.~K.}\ \bibnamefont {Vorov}},\
  }\href {\doibase 10.1103/PhysRevC.56.R2368} {\bibfield  {journal} {\bibinfo
  {journal} {Phys. Rev. C}\ }\textbf {\bibinfo {volume} {56}},\ \bibinfo
  {pages} {R2368} (\bibinfo {year} {1997})},\ \Eprint
  {http://arxiv.org/abs/nucl-th/9705003} {arXiv:nucl-th/9705003} \BibitemShut
  {NoStop}%
\bibitem [{\citenamefont {Vogel}(2006)}]{Vogel:2006sg}%
  \BibitemOpen
  \bibfield  {author} {\bibinfo {author} {\bibfnamefont {P.}~\bibnamefont
  {Vogel}},\ }\href {\doibase 10.1016/j.nuclphysa.2005.12.002} {\bibfield
  {journal} {\bibinfo  {journal} {Nucl. Phys. A}\ }\textbf {\bibinfo {volume}
  {777}},\ \bibinfo {pages} {340} (\bibinfo {year} {2006})}\BibitemShut
  {NoStop}%
\bibitem [{\citenamefont {Samana}\ \emph {et~al.}(2011)\citenamefont {Samana},
  \citenamefont {Krmpotic}, \citenamefont {Paar},\ and\ \citenamefont
  {Bertulani}}]{Samana:2010up}%
  \BibitemOpen
  \bibfield  {author} {\bibinfo {author} {\bibfnamefont {A.~R.}\ \bibnamefont
  {Samana}}, \bibinfo {author} {\bibfnamefont {F.}~\bibnamefont {Krmpotic}},
  \bibinfo {author} {\bibfnamefont {N.}~\bibnamefont {Paar}}, \ and\ \bibinfo
  {author} {\bibfnamefont {C.~A.}\ \bibnamefont {Bertulani}},\ }\href {\doibase
  10.1103/PhysRevC.83.024303} {\bibfield  {journal} {\bibinfo  {journal} {Phys.
  Rev. C}\ }\textbf {\bibinfo {volume} {83}},\ \bibinfo {pages} {024303}
  (\bibinfo {year} {2011})},\ \Eprint {http://arxiv.org/abs/1005.2134}
  {arXiv:1005.2134 [nucl-th]} \BibitemShut {NoStop}%
\bibitem [{\citenamefont {Drexlin}\ \emph {et~al.}(1991)\citenamefont {Drexlin}
  \emph {et~al.}}]{KARMEN:1991vkr}%
  \BibitemOpen
  \bibfield  {author} {\bibinfo {author} {\bibfnamefont {G.}~\bibnamefont
  {Drexlin}} \emph {et~al.} (\bibinfo {collaboration} {KARMEN}),\ }\href
  {\doibase 10.1016/0370-2693(91)90939-N} {\bibfield  {journal} {\bibinfo
  {journal} {Phys. Lett. B}\ }\textbf {\bibinfo {volume} {267}},\ \bibinfo
  {pages} {321} (\bibinfo {year} {1991})}\BibitemShut {NoStop}%
\bibitem [{\citenamefont {Bodmann}\ \emph {et~al.}(1994)\citenamefont {Bodmann}
  \emph {et~al.}}]{KARMEN:1994xse}%
  \BibitemOpen
  \bibfield  {author} {\bibinfo {author} {\bibfnamefont {B.~E.}\ \bibnamefont
  {Bodmann}} \emph {et~al.} (\bibinfo {collaboration} {KARMEN}),\ }\href
  {\doibase 10.1016/0370-2693(94)91250-5} {\bibfield  {journal} {\bibinfo
  {journal} {Phys. Lett. B}\ }\textbf {\bibinfo {volume} {332}},\ \bibinfo
  {pages} {251} (\bibinfo {year} {1994})}\BibitemShut {NoStop}%
\bibitem [{\citenamefont {Czarnecki}\ \emph {et~al.}(2011)\citenamefont
  {Czarnecki}, \citenamefont {Garcia~i Tormo},\ and\ \citenamefont
  {Marciano}}]{Czarnecki:2011mx}%
  \BibitemOpen
  \bibfield  {author} {\bibinfo {author} {\bibfnamefont {A.}~\bibnamefont
  {Czarnecki}}, \bibinfo {author} {\bibfnamefont {X.}~\bibnamefont {Garcia~i
  Tormo}}, \ and\ \bibinfo {author} {\bibfnamefont {W.~J.}\ \bibnamefont
  {Marciano}},\ }\href {\doibase 10.1103/PhysRevD.84.013006} {\bibfield
  {journal} {\bibinfo  {journal} {Phys. Rev. D}\ }\textbf {\bibinfo {volume}
  {84}},\ \bibinfo {pages} {013006} (\bibinfo {year} {2011})},\ \Eprint
  {http://arxiv.org/abs/1106.4756} {arXiv:1106.4756 [hep-ph]} \BibitemShut
  {NoStop}%
\bibitem [{\citenamefont {Zhao}(2022)}]{jie_zhao_2022_6683749}%
  \BibitemOpen
  \bibfield  {author} {\bibinfo {author} {\bibfnamefont {J.}~\bibnamefont
  {Zhao}},\ }\href {\doibase 10.5281/zenodo.6683749} {\enquote {\bibinfo
  {title} {Reactor neutrino ii \ juno status \& prospects},}\ } (\bibinfo
  {year} {Neutrino 2022, Seoul, South Korea, 2022})\BibitemShut {NoStop}%
\bibitem [{\citenamefont {Lin}\ \emph {et~al.}(2022)\citenamefont {Lin} \emph
  {et~al.}}]{Lin:2022htc}%
  \BibitemOpen
  \bibfield  {author} {\bibinfo {author} {\bibfnamefont {T.}~\bibnamefont
  {Lin}} \emph {et~al.},\ }\href@noop {} {\  (\bibinfo {year} {2022})},\
  \Eprint {http://arxiv.org/abs/2212.10741} {arXiv:2212.10741 [hep-ex]}
  \BibitemShut {NoStop}%
\bibitem [{\citenamefont {Sisti}(2022)}]{JUNO-NOW}%
  \BibitemOpen
  \bibfield  {author} {\bibinfo {author} {\bibfnamefont {M.}~\bibnamefont
  {Sisti}},\ }\href {https://agenda.infn.it/event/30418/contributions/170630/}
  {\enquote {\bibinfo {title} {Jiangmen underground neutrino observatory
  (juno): on the way to physics data},}\ } (\bibinfo {year} {Neutrino
  Oscillation Workshop, Ostuni, Italy, 2022})\BibitemShut {NoStop}%
\bibitem [{\citenamefont {Qian}\ \emph {et~al.}(2021)\citenamefont {Qian} \emph
  {et~al.}}]{Qian:2021vnh}%
  \BibitemOpen
  \bibfield  {author} {\bibinfo {author} {\bibfnamefont {Z.}~\bibnamefont
  {Qian}} \emph {et~al.},\ }\href {\doibase 10.1016/j.nima.2021.165527}
  {\bibfield  {journal} {\bibinfo  {journal} {Nucl. Instrum. Meth. A}\ }\textbf
  {\bibinfo {volume} {1010}},\ \bibinfo {pages} {165527} (\bibinfo {year}
  {2021})},\ \Eprint {http://arxiv.org/abs/2101.04839} {arXiv:2101.04839
  [physics.ins-det]} \BibitemShut {NoStop}%
\bibitem [{\citenamefont {Fang}\ \emph {et~al.}(2020)\citenamefont {Fang},
  \citenamefont {Zhang}, \citenamefont {Gong}, \citenamefont {Cao},
  \citenamefont {Lin}, \citenamefont {Yang},\ and\ \citenamefont
  {Li}}]{Fang:2019lej}%
  \BibitemOpen
  \bibfield  {author} {\bibinfo {author} {\bibfnamefont {X.}~\bibnamefont
  {Fang}}, \bibinfo {author} {\bibfnamefont {Y.}~\bibnamefont {Zhang}},
  \bibinfo {author} {\bibfnamefont {G.~H.}\ \bibnamefont {Gong}}, \bibinfo
  {author} {\bibfnamefont {G.~F.}\ \bibnamefont {Cao}}, \bibinfo {author}
  {\bibfnamefont {T.}~\bibnamefont {Lin}}, \bibinfo {author} {\bibfnamefont
  {C.~W.}\ \bibnamefont {Yang}}, \ and\ \bibinfo {author} {\bibfnamefont
  {W.~D.}\ \bibnamefont {Li}},\ }\href {\doibase
  10.1088/1748-0221/15/03/P03020} {\bibfield  {journal} {\bibinfo  {journal}
  {JINST}\ }\textbf {\bibinfo {volume} {15}},\ \bibinfo {pages} {P03020}
  (\bibinfo {year} {2020})},\ \Eprint {http://arxiv.org/abs/1912.01864}
  {arXiv:1912.01864 [physics.ins-det]} \BibitemShut {NoStop}%
\bibitem [{\citenamefont {Beacom}\ and\ \citenamefont
  {Palomares-Ruiz}(2003)}]{Beacom:2003zu}%
  \BibitemOpen
  \bibfield  {author} {\bibinfo {author} {\bibfnamefont {J.~F.}\ \bibnamefont
  {Beacom}}\ and\ \bibinfo {author} {\bibfnamefont {S.}~\bibnamefont
  {Palomares-Ruiz}},\ }\href {\doibase 10.1103/PhysRevD.67.093001} {\bibfield
  {journal} {\bibinfo  {journal} {Phys. Rev. D}\ }\textbf {\bibinfo {volume}
  {67}},\ \bibinfo {pages} {093001} (\bibinfo {year} {2003})},\ \Eprint
  {http://arxiv.org/abs/hep-ph/0301060} {arXiv:hep-ph/0301060} \BibitemShut
  {NoStop}%
\bibitem [{\citenamefont {Chauhan}\ \emph {et~al.}(2022)\citenamefont
  {Chauhan}, \citenamefont {Dasgupta},\ and\ \citenamefont
  {Dighe}}]{Chauhan:2021fzu}%
  \BibitemOpen
  \bibfield  {author} {\bibinfo {author} {\bibfnamefont {B.}~\bibnamefont
  {Chauhan}}, \bibinfo {author} {\bibfnamefont {B.}~\bibnamefont {Dasgupta}}, \
  and\ \bibinfo {author} {\bibfnamefont {A.}~\bibnamefont {Dighe}},\ }\href
  {\doibase 10.1103/PhysRevD.105.095035} {\bibfield  {journal} {\bibinfo
  {journal} {Phys. Rev. D}\ }\textbf {\bibinfo {volume} {105}},\ \bibinfo
  {pages} {095035} (\bibinfo {year} {2022})},\ \Eprint
  {http://arxiv.org/abs/2111.14586} {arXiv:2111.14586 [hep-ph]} \BibitemShut
  {NoStop}%
\bibitem [{\citenamefont {Mintz}\ and\ \citenamefont
  {Pourkaviani}(1996)}]{Mintz:1996qh}%
  \BibitemOpen
  \bibfield  {author} {\bibinfo {author} {\bibfnamefont {S.~L.}\ \bibnamefont
  {Mintz}}\ and\ \bibinfo {author} {\bibfnamefont {M.}~\bibnamefont
  {Pourkaviani}},\ }\href {\doibase 10.1016/S0375-9474(96)00266-7} {\bibfield
  {journal} {\bibinfo  {journal} {Nucl. Phys. A}\ }\textbf {\bibinfo {volume}
  {609}},\ \bibinfo {pages} {411} (\bibinfo {year} {1996})}\BibitemShut
  {NoStop}%
\bibitem [{\citenamefont {Abusleme}\ \emph
  {et~al.}(2021{\natexlab{b}})\citenamefont {Abusleme} \emph
  {et~al.}}]{JUNO:2020hqc}%
  \BibitemOpen
  \bibfield  {author} {\bibinfo {author} {\bibfnamefont {A.}~\bibnamefont
  {Abusleme}} \emph {et~al.} (\bibinfo {collaboration} {JUNO}),\ }\href
  {\doibase 10.1088/1674-1137/abd92a} {\bibfield  {journal} {\bibinfo
  {journal} {Chin. Phys. C}\ }\textbf {\bibinfo {volume} {45}},\ \bibinfo
  {pages} {023004} (\bibinfo {year} {2021}{\natexlab{b}})},\ \Eprint
  {http://arxiv.org/abs/2006.11760} {arXiv:2006.11760 [hep-ex]} \BibitemShut
  {NoStop}%
\bibitem [{\citenamefont {Li}\ and\ \citenamefont {Beacom}(2014)}]{Li:2014sea}%
  \BibitemOpen
  \bibfield  {author} {\bibinfo {author} {\bibfnamefont {S.~W.}\ \bibnamefont
  {Li}}\ and\ \bibinfo {author} {\bibfnamefont {J.~F.}\ \bibnamefont
  {Beacom}},\ }\href {\doibase 10.1103/PhysRevC.89.045801} {\bibfield
  {journal} {\bibinfo  {journal} {Phys. Rev. C}\ }\textbf {\bibinfo {volume}
  {89}},\ \bibinfo {pages} {045801} (\bibinfo {year} {2014})},\ \Eprint
  {http://arxiv.org/abs/1402.4687} {arXiv:1402.4687 [hep-ph]} \BibitemShut
  {NoStop}%
\bibitem [{\citenamefont {Li}\ and\ \citenamefont
  {Beacom}(2015{\natexlab{a}})}]{Li:2015kpa}%
  \BibitemOpen
  \bibfield  {author} {\bibinfo {author} {\bibfnamefont {S.~W.}\ \bibnamefont
  {Li}}\ and\ \bibinfo {author} {\bibfnamefont {J.~F.}\ \bibnamefont
  {Beacom}},\ }\href {\doibase 10.1103/PhysRevD.91.105005} {\bibfield
  {journal} {\bibinfo  {journal} {Phys. Rev. D}\ }\textbf {\bibinfo {volume}
  {91}},\ \bibinfo {pages} {105005} (\bibinfo {year} {2015}{\natexlab{a}})},\
  \Eprint {http://arxiv.org/abs/1503.04823} {arXiv:1503.04823 [hep-ph]}
  \BibitemShut {NoStop}%
\bibitem [{\citenamefont {Li}\ and\ \citenamefont
  {Beacom}(2015{\natexlab{b}})}]{Li:2015lxa}%
  \BibitemOpen
  \bibfield  {author} {\bibinfo {author} {\bibfnamefont {S.~W.}\ \bibnamefont
  {Li}}\ and\ \bibinfo {author} {\bibfnamefont {J.~F.}\ \bibnamefont
  {Beacom}},\ }\href {\doibase 10.1103/PhysRevD.92.105033} {\bibfield
  {journal} {\bibinfo  {journal} {Phys. Rev. D}\ }\textbf {\bibinfo {volume}
  {92}},\ \bibinfo {pages} {105033} (\bibinfo {year} {2015}{\natexlab{b}})},\
  \Eprint {http://arxiv.org/abs/1508.05389} {arXiv:1508.05389
  [physics.ins-det]} \BibitemShut {NoStop}%
\bibitem [{\citenamefont {Zhang}\ \emph {et~al.}(2016)\citenamefont {Zhang}
  \emph {et~al.}}]{Super-Kamiokande:2015xra}%
  \BibitemOpen
  \bibfield  {author} {\bibinfo {author} {\bibfnamefont {Y.}~\bibnamefont
  {Zhang}} \emph {et~al.} (\bibinfo {collaboration} {Super-Kamiokande}),\
  }\href {\doibase 10.1103/PhysRevD.93.012004} {\bibfield  {journal} {\bibinfo
  {journal} {Phys. Rev. D}\ }\textbf {\bibinfo {volume} {93}},\ \bibinfo
  {pages} {012004} (\bibinfo {year} {2016})},\ \Eprint
  {http://arxiv.org/abs/1509.08168} {arXiv:1509.08168 [hep-ex]} \BibitemShut
  {NoStop}%
\bibitem [{\citenamefont {Locke}\ \emph {et~al.}(2021)\citenamefont {Locke}
  \emph {et~al.}}]{Super-Kamiokande:2021snn}%
  \BibitemOpen
  \bibfield  {author} {\bibinfo {author} {\bibfnamefont {S.}~\bibnamefont
  {Locke}} \emph {et~al.} (\bibinfo {collaboration} {Super-Kamiokande}),\
  }\href@noop {} {\  (\bibinfo {year} {2021})},\ \Eprint
  {http://arxiv.org/abs/2112.00092} {arXiv:2112.00092 [hep-ex]} \BibitemShut
  {NoStop}%
\bibitem [{\citenamefont {Nairat}\ \emph {et~al.}()\citenamefont {Nairat},
  \citenamefont {Li},\ and\ \citenamefont {Beacom}}]{Nairat:2023xxx}%
  \BibitemOpen
  \bibfield  {author} {\bibinfo {author} {\bibfnamefont {O.}~\bibnamefont
  {Nairat}}, \bibinfo {author} {\bibfnamefont {S.~W.}\ \bibnamefont {Li}}, \
  and\ \bibinfo {author} {\bibfnamefont {J.~F.}\ \bibnamefont {Beacom}},\
  }\href@noop {} {\ }\Eprint {http://arxiv.org/abs/23XX.XXXXX}
  {arXiv:23XX.XXXXX [hep-ph]} \BibitemShut {NoStop}%
\bibitem [{\citenamefont {Gando}\ \emph {et~al.}(2012)\citenamefont {Gando}
  \emph {et~al.}}]{KamLAND:2011bnd}%
  \BibitemOpen
  \bibfield  {author} {\bibinfo {author} {\bibfnamefont {A.}~\bibnamefont
  {Gando}} \emph {et~al.} (\bibinfo {collaboration} {KamLAND}),\ }\href
  {\doibase 10.1088/0004-637X/745/2/193} {\bibfield  {journal} {\bibinfo
  {journal} {Astrophys. J.}\ }\textbf {\bibinfo {volume} {745}},\ \bibinfo
  {pages} {193} (\bibinfo {year} {2012})},\ \Eprint
  {http://arxiv.org/abs/1105.3516} {arXiv:1105.3516 [astro-ph.HE]} \BibitemShut
  {NoStop}%
\bibitem [{\citenamefont {Abe}\ \emph {et~al.}(2022{\natexlab{a}})\citenamefont
  {Abe} \emph {et~al.}}]{KamLAND:2021gvi}%
  \BibitemOpen
  \bibfield  {author} {\bibinfo {author} {\bibfnamefont {S.}~\bibnamefont
  {Abe}} \emph {et~al.} (\bibinfo {collaboration} {KamLAND}),\ }\href {\doibase
  10.3847/1538-4357/ac32c1} {\bibfield  {journal} {\bibinfo  {journal}
  {Astrophys. J.}\ }\textbf {\bibinfo {volume} {925}},\ \bibinfo {pages} {14}
  (\bibinfo {year} {2022}{\natexlab{a}})},\ \Eprint
  {http://arxiv.org/abs/2108.08527} {arXiv:2108.08527 [astro-ph.HE]}
  \BibitemShut {NoStop}%
\bibitem [{\citenamefont {Vinyoles}\ \emph {et~al.}(2017)\citenamefont
  {Vinyoles}, \citenamefont {Serenelli}, \citenamefont {Villante},
  \citenamefont {Basu}, \citenamefont {Bergstr\"om}, \citenamefont
  {Gonzalez-Garcia}, \citenamefont {Maltoni}, \citenamefont {Pe\~na Garay},\
  and\ \citenamefont {Song}}]{Vinyoles:2016djt}%
  \BibitemOpen
  \bibfield  {author} {\bibinfo {author} {\bibfnamefont {N.}~\bibnamefont
  {Vinyoles}}, \bibinfo {author} {\bibfnamefont {A.~M.}\ \bibnamefont
  {Serenelli}}, \bibinfo {author} {\bibfnamefont {F.~L.}\ \bibnamefont
  {Villante}}, \bibinfo {author} {\bibfnamefont {S.}~\bibnamefont {Basu}},
  \bibinfo {author} {\bibfnamefont {J.}~\bibnamefont {Bergstr\"om}}, \bibinfo
  {author} {\bibfnamefont {M.}~\bibnamefont {Gonzalez-Garcia}}, \bibinfo
  {author} {\bibfnamefont {M.}~\bibnamefont {Maltoni}}, \bibinfo {author}
  {\bibfnamefont {C.}~\bibnamefont {Pe\~na Garay}}, \ and\ \bibinfo {author}
  {\bibfnamefont {N.}~\bibnamefont {Song}},\ }\href {\doibase
  10.3847/1538-4357/835/2/202} {\bibfield  {journal} {\bibinfo  {journal}
  {Astrophys. J.}\ }\textbf {\bibinfo {volume} {835}},\ \bibinfo {pages} {202}
  (\bibinfo {year} {2017})},\ \Eprint {http://arxiv.org/abs/1611.09867}
  {arXiv:1611.09867 [astro-ph.SR]} \BibitemShut {NoStop}%
\bibitem [{\citenamefont {Giunti}\ and\ \citenamefont
  {Kim}(2007)}]{Giunti:2007ry}%
  \BibitemOpen
  \bibfield  {author} {\bibinfo {author} {\bibfnamefont {C.}~\bibnamefont
  {Giunti}}\ and\ \bibinfo {author} {\bibfnamefont {C.~W.}\ \bibnamefont
  {Kim}},\ }\href@noop {} {\emph {\bibinfo {title} {{Fundamentals of Neutrino
  Physics and Astrophysics}}}}\ (\bibinfo  {publisher} {Oxford University
  Press},\ \bibinfo {year} {2007})\ pp.\ \bibinfo {pages}
  {136--142}\BibitemShut {NoStop}%
\bibitem [{\citenamefont {Møller}\ \emph {et~al.}(2018)\citenamefont
  {Møller}, \citenamefont {Suliga}, \citenamefont {Tamborra},\ and\
  \citenamefont {Denton}}]{Moller:2018kpn}%
  \BibitemOpen
  \bibfield  {author} {\bibinfo {author} {\bibfnamefont {K.}~\bibnamefont
  {Møller}}, \bibinfo {author} {\bibfnamefont {A.~M.}\ \bibnamefont {Suliga}},
  \bibinfo {author} {\bibfnamefont {I.}~\bibnamefont {Tamborra}}, \ and\
  \bibinfo {author} {\bibfnamefont {P.~B.}\ \bibnamefont {Denton}},\ }\href
  {\doibase 10.1088/1475-7516/2018/05/066} {\bibfield  {journal} {\bibinfo
  {journal} {JCAP}\ }\textbf {\bibinfo {volume} {05}},\ \bibinfo {pages} {066}
  (\bibinfo {year} {2018})},\ \Eprint {http://arxiv.org/abs/1804.03157}
  {arXiv:1804.03157 [astro-ph.HE]} \BibitemShut {NoStop}%
\bibitem [{\citenamefont {Suliga}\ \emph {et~al.}(2021)\citenamefont {Suliga},
  \citenamefont {Beacom},\ and\ \citenamefont {Tamborra}}]{Suliga:2021hek}%
  \BibitemOpen
  \bibfield  {author} {\bibinfo {author} {\bibfnamefont {A.~M.}\ \bibnamefont
  {Suliga}}, \bibinfo {author} {\bibfnamefont {J.~F.}\ \bibnamefont {Beacom}},
  \ and\ \bibinfo {author} {\bibfnamefont {I.}~\bibnamefont {Tamborra}},\
  }\href@noop {} {\  (\bibinfo {year} {2021})},\ \Eprint
  {http://arxiv.org/abs/2112.09168} {arXiv:2112.09168 [astro-ph.HE]}
  \BibitemShut {NoStop}%
\bibitem [{\citenamefont {Ziegler}\ \emph {et~al.}(2022)\citenamefont
  {Ziegler}, \citenamefont {Edwards}, \citenamefont {Suliga}, \citenamefont
  {Tamborra}, \citenamefont {Horiuchi}, \citenamefont {Ando},\ and\
  \citenamefont {Freese}}]{Ziegler:2022ivq}%
  \BibitemOpen
  \bibfield  {author} {\bibinfo {author} {\bibfnamefont {J.~J.}\ \bibnamefont
  {Ziegler}}, \bibinfo {author} {\bibfnamefont {T.~D.~P.}\ \bibnamefont
  {Edwards}}, \bibinfo {author} {\bibfnamefont {A.~M.}\ \bibnamefont {Suliga}},
  \bibinfo {author} {\bibfnamefont {I.}~\bibnamefont {Tamborra}}, \bibinfo
  {author} {\bibfnamefont {S.}~\bibnamefont {Horiuchi}}, \bibinfo {author}
  {\bibfnamefont {S.}~\bibnamefont {Ando}}, \ and\ \bibinfo {author}
  {\bibfnamefont {K.}~\bibnamefont {Freese}},\ }\href {\doibase
  10.1093/mnras/stac2748} {\bibfield  {journal} {\bibinfo  {journal} {Mon. Not.
  Roy. Astron. Soc.}\ }\textbf {\bibinfo {volume} {517}},\ \bibinfo {pages}
  {2471} (\bibinfo {year} {2022})},\ \Eprint {http://arxiv.org/abs/2205.07845}
  {arXiv:2205.07845 [astro-ph.GA]} \BibitemShut {NoStop}%
\bibitem [{\citenamefont {{Schroedinger}}(1944)}]{1944Natur.153..592S}%
  \BibitemOpen
  \bibfield  {author} {\bibinfo {author} {\bibfnamefont {E.}~\bibnamefont
  {{Schroedinger}}},\ }\href {\doibase 10.1038/153592b0} {\bibfield  {journal}
  {\bibinfo  {journal} {\nat}\ }\textbf {\bibinfo {volume} {153}},\ \bibinfo
  {pages} {592} (\bibinfo {year} {1944})}\BibitemShut {NoStop}%
\bibitem [{\citenamefont {Kelly}\ \emph {et~al.}(2022)\citenamefont {Kelly},
  \citenamefont {Machado}, \citenamefont {Martinez-Soler},\ and\ \citenamefont
  {Perez-Gonzalez}}]{Kelly:2021jfs}%
  \BibitemOpen
  \bibfield  {author} {\bibinfo {author} {\bibfnamefont {K.~J.}\ \bibnamefont
  {Kelly}}, \bibinfo {author} {\bibfnamefont {P.~A.~N.}\ \bibnamefont
  {Machado}}, \bibinfo {author} {\bibfnamefont {I.}~\bibnamefont
  {Martinez-Soler}}, \ and\ \bibinfo {author} {\bibfnamefont {Y.~F.}\
  \bibnamefont {Perez-Gonzalez}},\ }\href {\doibase 10.1007/JHEP05(2022)187}
  {\bibfield  {journal} {\bibinfo  {journal} {JHEP}\ }\textbf {\bibinfo
  {volume} {05}},\ \bibinfo {pages} {187} (\bibinfo {year} {2022})},\ \Eprint
  {http://arxiv.org/abs/2110.00003} {arXiv:2110.00003 [hep-ph]} \BibitemShut
  {NoStop}%
\bibitem [{\citenamefont {Denton}\ and\ \citenamefont
  {Pestes}(2021)}]{Denton:2021rgt}%
  \BibitemOpen
  \bibfield  {author} {\bibinfo {author} {\bibfnamefont {P.~B.}\ \bibnamefont
  {Denton}}\ and\ \bibinfo {author} {\bibfnamefont {R.}~\bibnamefont
  {Pestes}},\ }\href {\doibase 10.1103/PhysRevD.104.113007} {\bibfield
  {journal} {\bibinfo  {journal} {Phys. Rev. D}\ }\textbf {\bibinfo {volume}
  {104}},\ \bibinfo {pages} {113007} (\bibinfo {year} {2021})},\ \Eprint
  {http://arxiv.org/abs/2110.01148} {arXiv:2110.01148 [hep-ph]} \BibitemShut
  {NoStop}%
\bibitem [{\citenamefont {Beacom}\ and\ \citenamefont
  {Vagins}(2004)}]{Beacom:2003nk}%
  \BibitemOpen
  \bibfield  {author} {\bibinfo {author} {\bibfnamefont {J.~F.}\ \bibnamefont
  {Beacom}}\ and\ \bibinfo {author} {\bibfnamefont {M.~R.}\ \bibnamefont
  {Vagins}},\ }\href {\doibase 10.1103/PhysRevLett.93.171101} {\bibfield
  {journal} {\bibinfo  {journal} {Phys. Rev. Lett.}\ }\textbf {\bibinfo
  {volume} {93}},\ \bibinfo {pages} {171101} (\bibinfo {year} {2004})},\
  \Eprint {http://arxiv.org/abs/hep-ph/0309300} {arXiv:hep-ph/0309300}
  \BibitemShut {NoStop}%
\bibitem [{\citenamefont {Abe}\ \emph {et~al.}(2022{\natexlab{b}})\citenamefont
  {Abe} \emph {et~al.}}]{Super-Kamiokande:2021the}%
  \BibitemOpen
  \bibfield  {author} {\bibinfo {author} {\bibfnamefont {K.}~\bibnamefont
  {Abe}} \emph {et~al.} (\bibinfo {collaboration} {Super-Kamiokande}),\ }\href
  {\doibase 10.1016/j.nima.2021.166248} {\bibfield  {journal} {\bibinfo
  {journal} {Nucl. Instrum. Meth. A}\ }\textbf {\bibinfo {volume} {1027}},\
  \bibinfo {pages} {166248} (\bibinfo {year} {2022}{\natexlab{b}})},\ \Eprint
  {http://arxiv.org/abs/2109.00360} {arXiv:2109.00360 [physics.ins-det]}
  \BibitemShut {NoStop}%
\bibitem [{\citenamefont {Harada}\ \emph {et~al.}(2023)\citenamefont {Harada}
  \emph {et~al.}}]{Super-Kamiokande:2023xup}%
  \BibitemOpen
  \bibfield  {author} {\bibinfo {author} {\bibfnamefont {M.}~\bibnamefont
  {Harada}} \emph {et~al.} (\bibinfo {collaboration} {Super-Kamiokande}),\
  }\href@noop {} {\  (\bibinfo {year} {2023})},\ \Eprint
  {http://arxiv.org/abs/2305.05135} {arXiv:2305.05135 [astro-ph.HE]}
  \BibitemShut {NoStop}%
\bibitem [{\citenamefont {Abe}\ \emph {et~al.}(2018)\citenamefont {Abe} \emph
  {et~al.}}]{Abe:2018uyc}%
  \BibitemOpen
  \bibfield  {author} {\bibinfo {author} {\bibfnamefont {K.}~\bibnamefont
  {Abe}} \emph {et~al.} (\bibinfo {collaboration} {Hyper-Kamiokande}),\
  }\href@noop {} {\  (\bibinfo {year} {2018})},\ \Eprint
  {http://arxiv.org/abs/1805.04163} {arXiv:1805.04163 [physics.ins-det]}
  \BibitemShut {NoStop}%
\bibitem [{\citenamefont {Abi}\ \emph {et~al.}(2020)\citenamefont {Abi} \emph
  {et~al.}}]{DUNE:2020ypp}%
  \BibitemOpen
  \bibfield  {author} {\bibinfo {author} {\bibfnamefont {B.}~\bibnamefont
  {Abi}} \emph {et~al.} (\bibinfo {collaboration} {DUNE}),\ }\href@noop {} {\
  (\bibinfo {year} {2020})},\ \Eprint {http://arxiv.org/abs/2002.03005}
  {arXiv:2002.03005 [hep-ex]} \BibitemShut {NoStop}%
\bibitem [{\citenamefont {Aalbers}\ \emph {et~al.}(2016)\citenamefont {Aalbers}
  \emph {et~al.}}]{Aalbers:2016jon}%
  \BibitemOpen
  \bibfield  {author} {\bibinfo {author} {\bibfnamefont {J.}~\bibnamefont
  {Aalbers}} \emph {et~al.} (\bibinfo {collaboration} {DARWIN}),\ }\href
  {\doibase 10.1088/1475-7516/2016/11/017} {\bibfield  {journal} {\bibinfo
  {journal} {JCAP}\ }\textbf {\bibinfo {volume} {11}},\ \bibinfo {pages} {017}
  (\bibinfo {year} {2016})},\ \Eprint {http://arxiv.org/abs/1606.07001}
  {arXiv:1606.07001 [astro-ph.IM]} \BibitemShut {NoStop}%
\bibitem [{\citenamefont {Cadonati}\ \emph {et~al.}(2002)\citenamefont
  {Cadonati}, \citenamefont {Calaprice},\ and\ \citenamefont
  {Chen}}]{Cadonati:2000kq}%
  \BibitemOpen
  \bibfield  {author} {\bibinfo {author} {\bibfnamefont {L.}~\bibnamefont
  {Cadonati}}, \bibinfo {author} {\bibfnamefont {F.~P.}\ \bibnamefont
  {Calaprice}}, \ and\ \bibinfo {author} {\bibfnamefont {M.~C.}\ \bibnamefont
  {Chen}},\ }\href {\doibase 10.1016/S0927-6505(01)00129-3} {\bibfield
  {journal} {\bibinfo  {journal} {Astropart. Phys.}\ }\textbf {\bibinfo
  {volume} {16}},\ \bibinfo {pages} {361} (\bibinfo {year} {2002})},\ \Eprint
  {http://arxiv.org/abs/hep-ph/0012082} {arXiv:hep-ph/0012082} \BibitemShut
  {NoStop}%
\bibitem [{\citenamefont {Laha}\ \emph {et~al.}(2014)\citenamefont {Laha},
  \citenamefont {Beacom},\ and\ \citenamefont {Agarwalla}}]{Laha:2014yua}%
  \BibitemOpen
  \bibfield  {author} {\bibinfo {author} {\bibfnamefont {R.}~\bibnamefont
  {Laha}}, \bibinfo {author} {\bibfnamefont {J.~F.}\ \bibnamefont {Beacom}}, \
  and\ \bibinfo {author} {\bibfnamefont {S.~K.}\ \bibnamefont {Agarwalla}},\
  }\href@noop {} {\  (\bibinfo {year} {2014})},\ \Eprint
  {http://arxiv.org/abs/1412.8425} {arXiv:1412.8425 [hep-ph]} \BibitemShut
  {NoStop}%
\bibitem [{\citenamefont {Lu}\ \emph {et~al.}(2016)\citenamefont {Lu},
  \citenamefont {Li},\ and\ \citenamefont {Zhou}}]{Lu:2016ipr}%
  \BibitemOpen
  \bibfield  {author} {\bibinfo {author} {\bibfnamefont {J.-S.}\ \bibnamefont
  {Lu}}, \bibinfo {author} {\bibfnamefont {Y.-F.}\ \bibnamefont {Li}}, \ and\
  \bibinfo {author} {\bibfnamefont {S.}~\bibnamefont {Zhou}},\ }\href {\doibase
  10.1103/PhysRevD.94.023006} {\bibfield  {journal} {\bibinfo  {journal} {Phys.
  Rev. D}\ }\textbf {\bibinfo {volume} {94}},\ \bibinfo {pages} {023006}
  (\bibinfo {year} {2016})},\ \Eprint {http://arxiv.org/abs/1605.07803}
  {arXiv:1605.07803 [hep-ph]} \BibitemShut {NoStop}%
\bibitem [{\citenamefont {Battistoni}\ \emph {et~al.}(2003)\citenamefont
  {Battistoni}, \citenamefont {Ferrari}, \citenamefont {Montaruli},\ and\
  \citenamefont {Sala}}]{Battistoni:2002ew}%
  \BibitemOpen
  \bibfield  {author} {\bibinfo {author} {\bibfnamefont {G.}~\bibnamefont
  {Battistoni}}, \bibinfo {author} {\bibfnamefont {A.}~\bibnamefont {Ferrari}},
  \bibinfo {author} {\bibfnamefont {T.}~\bibnamefont {Montaruli}}, \ and\
  \bibinfo {author} {\bibfnamefont {P.}~\bibnamefont {Sala}},\ }\href {\doibase
  10.1016/S0927-6505(02)00246-3} {\bibfield  {journal} {\bibinfo  {journal}
  {Astropart. Phys.}\ }\textbf {\bibinfo {volume} {19}},\ \bibinfo {pages}
  {269} (\bibinfo {year} {2003})},\ \bibinfo {note} {[Erratum: Astropart.Phys.
  19, 291--294 (2003)]},\ \Eprint {http://arxiv.org/abs/hep-ph/0207035}
  {arXiv:hep-ph/0207035} \BibitemShut {NoStop}%
\bibitem [{\citenamefont {Horiuchi}\ \emph {et~al.}(2009)\citenamefont
  {Horiuchi}, \citenamefont {Beacom},\ and\ \citenamefont
  {Dwek}}]{Horiuchi:2008jz}%
  \BibitemOpen
  \bibfield  {author} {\bibinfo {author} {\bibfnamefont {S.}~\bibnamefont
  {Horiuchi}}, \bibinfo {author} {\bibfnamefont {J.~F.}\ \bibnamefont
  {Beacom}}, \ and\ \bibinfo {author} {\bibfnamefont {E.}~\bibnamefont
  {Dwek}},\ }\href {\doibase 10.1103/PhysRevD.79.083013} {\bibfield  {journal}
  {\bibinfo  {journal} {Phys. Rev. D}\ }\textbf {\bibinfo {volume} {79}},\
  \bibinfo {pages} {083013} (\bibinfo {year} {2009})},\ \Eprint
  {http://arxiv.org/abs/0812.3157} {arXiv:0812.3157 [astro-ph]} \BibitemShut
  {NoStop}%
\bibitem [{\citenamefont {Beacom}(2010)}]{Beacom:2010kk}%
  \BibitemOpen
  \bibfield  {author} {\bibinfo {author} {\bibfnamefont {J.~F.}\ \bibnamefont
  {Beacom}},\ }\href {\doibase 10.1146/annurev.nucl.010909.083331} {\bibfield
  {journal} {\bibinfo  {journal} {Ann. Rev. Nucl. Part. Sci.}\ }\textbf
  {\bibinfo {volume} {60}},\ \bibinfo {pages} {439} (\bibinfo {year} {2010})},\
  \Eprint {http://arxiv.org/abs/1004.3311} {arXiv:1004.3311 [astro-ph.HE]}
  \BibitemShut {NoStop}%
\bibitem [{\citenamefont {Lunardini}(2016)}]{Lunardini:2010ab}%
  \BibitemOpen
  \bibfield  {author} {\bibinfo {author} {\bibfnamefont {C.}~\bibnamefont
  {Lunardini}},\ }\href {\doibase 10.1016/j.astropartphys.2016.02.005}
  {\bibfield  {journal} {\bibinfo  {journal} {Astropart. Phys.}\ }\textbf
  {\bibinfo {volume} {79}},\ \bibinfo {pages} {49} (\bibinfo {year} {2016})},\
  \Eprint {http://arxiv.org/abs/1007.3252} {arXiv:1007.3252 [astro-ph.CO]}
  \BibitemShut {NoStop}%
\bibitem [{\citenamefont {Kresse}\ \emph {et~al.}(2021)\citenamefont {Kresse},
  \citenamefont {Ertl},\ and\ \citenamefont {Janka}}]{Kresse:2020nto}%
  \BibitemOpen
  \bibfield  {author} {\bibinfo {author} {\bibfnamefont {D.}~\bibnamefont
  {Kresse}}, \bibinfo {author} {\bibfnamefont {T.}~\bibnamefont {Ertl}}, \ and\
  \bibinfo {author} {\bibfnamefont {H.-T.}\ \bibnamefont {Janka}},\ }\href
  {\doibase 10.3847/1538-4357/abd54e} {\bibfield  {journal} {\bibinfo
  {journal} {Astrophys. J.}\ }\textbf {\bibinfo {volume} {909}},\ \bibinfo
  {pages} {169} (\bibinfo {year} {2021})},\ \Eprint
  {http://arxiv.org/abs/2010.04728} {arXiv:2010.04728 [astro-ph.HE]}
  \BibitemShut {NoStop}%
\bibitem [{\citenamefont {Horiuchi}\ \emph {et~al.}(2021)\citenamefont
  {Horiuchi}, \citenamefont {Kinugawa}, \citenamefont {Takiwaki}, \citenamefont
  {Takahashi},\ and\ \citenamefont {Kotake}}]{Horiuchi:2020jnc}%
  \BibitemOpen
  \bibfield  {author} {\bibinfo {author} {\bibfnamefont {S.}~\bibnamefont
  {Horiuchi}}, \bibinfo {author} {\bibfnamefont {T.}~\bibnamefont {Kinugawa}},
  \bibinfo {author} {\bibfnamefont {T.}~\bibnamefont {Takiwaki}}, \bibinfo
  {author} {\bibfnamefont {K.}~\bibnamefont {Takahashi}}, \ and\ \bibinfo
  {author} {\bibfnamefont {K.}~\bibnamefont {Kotake}},\ }\href {\doibase
  10.1103/PhysRevD.103.043003} {\bibfield  {journal} {\bibinfo  {journal}
  {Phys. Rev. D}\ }\textbf {\bibinfo {volume} {103}},\ \bibinfo {pages}
  {043003} (\bibinfo {year} {2021})},\ \Eprint
  {http://arxiv.org/abs/2012.08524} {arXiv:2012.08524 [astro-ph.HE]}
  \BibitemShut {NoStop}%
\bibitem [{\citenamefont {Suliga}(2022)}]{Suliga:2022ica}%
  \BibitemOpen
  \bibfield  {author} {\bibinfo {author} {\bibfnamefont {A.~M.}\ \bibnamefont
  {Suliga}},\ }\enquote {\bibinfo {title} {{Diffuse supernova neutrino
  background}},}\ in\ \href {\doibase 10.1007/978-981-15-8818-1} {\emph
  {\bibinfo {booktitle} {Handbook of Nuclear Physics}}},\ \bibinfo {editor}
  {edited by\ \bibinfo {editor} {\bibnamefont {{Isao Tanihata, Hiroshi Toki,
  and Toshitaka Kajino}}}}\ (\bibinfo  {publisher} {{Springer Singapore}},\
  \bibinfo {year} {2022})\ \Eprint {http://arxiv.org/abs/2207.09632}
  {arXiv:2207.09632 [astro-ph.HE]} \BibitemShut {NoStop}%
\bibitem [{\citenamefont {Vergados}\ and\ \citenamefont
  {Ejiri}(2008)}]{Vergados:2008jp}%
  \BibitemOpen
  \bibfield  {author} {\bibinfo {author} {\bibfnamefont {J.}~\bibnamefont
  {Vergados}}\ and\ \bibinfo {author} {\bibfnamefont {H.}~\bibnamefont
  {Ejiri}},\ }\href {\doibase 10.1016/j.nuclphysb.2008.06.004} {\bibfield
  {journal} {\bibinfo  {journal} {Nucl. Phys. B}\ }\textbf {\bibinfo {volume}
  {804}},\ \bibinfo {pages} {144} (\bibinfo {year} {2008})},\ \Eprint
  {http://arxiv.org/abs/0805.2583} {arXiv:0805.2583 [hep-ph]} \BibitemShut
  {NoStop}%
\bibitem [{\citenamefont {Strigari}(2009)}]{Strigari:2009bq}%
  \BibitemOpen
  \bibfield  {author} {\bibinfo {author} {\bibfnamefont {L.~E.}\ \bibnamefont
  {Strigari}},\ }\href {\doibase 10.1088/1367-2630/11/10/105011} {\bibfield
  {journal} {\bibinfo  {journal} {New J. Phys.}\ }\textbf {\bibinfo {volume}
  {11}},\ \bibinfo {pages} {105011} (\bibinfo {year} {2009})},\ \Eprint
  {http://arxiv.org/abs/0903.3630} {arXiv:0903.3630 [astro-ph.CO]} \BibitemShut
  {NoStop}%
\bibitem [{\citenamefont {Billard}\ \emph {et~al.}(2014)\citenamefont
  {Billard}, \citenamefont {Strigari},\ and\ \citenamefont
  {Figueroa-Feliciano}}]{Billard:2013qya}%
  \BibitemOpen
  \bibfield  {author} {\bibinfo {author} {\bibfnamefont {J.}~\bibnamefont
  {Billard}}, \bibinfo {author} {\bibfnamefont {L.}~\bibnamefont {Strigari}}, \
  and\ \bibinfo {author} {\bibfnamefont {E.}~\bibnamefont
  {Figueroa-Feliciano}},\ }\href {\doibase 10.1103/PhysRevD.89.023524}
  {\bibfield  {journal} {\bibinfo  {journal} {Phys. Rev. D}\ }\textbf {\bibinfo
  {volume} {89}},\ \bibinfo {pages} {023524} (\bibinfo {year} {2014})},\
  \Eprint {http://arxiv.org/abs/1307.5458} {arXiv:1307.5458 [hep-ph]}
  \BibitemShut {NoStop}%
\bibitem [{\citenamefont {Baudis}\ \emph {et~al.}(2014)\citenamefont {Baudis},
  \citenamefont {Ferella}, \citenamefont {Kish}, \citenamefont {Manalaysay},
  \citenamefont {Marrodan~Undagoitia},\ and\ \citenamefont
  {Schumann}}]{Baudis:2013qla}%
  \BibitemOpen
  \bibfield  {author} {\bibinfo {author} {\bibfnamefont {L.}~\bibnamefont
  {Baudis}}, \bibinfo {author} {\bibfnamefont {A.}~\bibnamefont {Ferella}},
  \bibinfo {author} {\bibfnamefont {A.}~\bibnamefont {Kish}}, \bibinfo {author}
  {\bibfnamefont {A.}~\bibnamefont {Manalaysay}}, \bibinfo {author}
  {\bibfnamefont {T.}~\bibnamefont {Marrodan~Undagoitia}}, \ and\ \bibinfo
  {author} {\bibfnamefont {M.}~\bibnamefont {Schumann}},\ }\href {\doibase
  10.1088/1475-7516/2014/01/044} {\bibfield  {journal} {\bibinfo  {journal}
  {JCAP}\ }\textbf {\bibinfo {volume} {01}},\ \bibinfo {pages} {044} (\bibinfo
  {year} {2014})},\ \Eprint {http://arxiv.org/abs/1309.7024} {arXiv:1309.7024
  [physics.ins-det]} \BibitemShut {NoStop}%
\bibitem [{\citenamefont {Ruppin}\ \emph {et~al.}(2014)\citenamefont {Ruppin},
  \citenamefont {Billard}, \citenamefont {Figueroa-Feliciano},\ and\
  \citenamefont {Strigari}}]{Ruppin:2014bra}%
  \BibitemOpen
  \bibfield  {author} {\bibinfo {author} {\bibfnamefont {F.}~\bibnamefont
  {Ruppin}}, \bibinfo {author} {\bibfnamefont {J.}~\bibnamefont {Billard}},
  \bibinfo {author} {\bibfnamefont {E.}~\bibnamefont {Figueroa-Feliciano}}, \
  and\ \bibinfo {author} {\bibfnamefont {L.}~\bibnamefont {Strigari}},\ }\href
  {\doibase 10.1103/PhysRevD.90.083510} {\bibfield  {journal} {\bibinfo
  {journal} {Phys. Rev. D}\ }\textbf {\bibinfo {volume} {90}},\ \bibinfo
  {pages} {083510} (\bibinfo {year} {2014})},\ \Eprint
  {http://arxiv.org/abs/1408.3581} {arXiv:1408.3581 [hep-ph]} \BibitemShut
  {NoStop}%
\bibitem [{\citenamefont {O'Hare}(2016)}]{OHare:2016pjy}%
  \BibitemOpen
  \bibfield  {author} {\bibinfo {author} {\bibfnamefont {C.~A.~J.}\
  \bibnamefont {O'Hare}},\ }\href {\doibase 10.1103/PhysRevD.94.063527}
  {\bibfield  {journal} {\bibinfo  {journal} {Phys. Rev. D}\ }\textbf {\bibinfo
  {volume} {94}},\ \bibinfo {pages} {063527} (\bibinfo {year} {2016})},\
  \Eprint {http://arxiv.org/abs/1604.03858} {arXiv:1604.03858 [astro-ph.CO]}
  \BibitemShut {NoStop}%
\bibitem [{\citenamefont {B{\oe}hm}\ \emph {et~al.}(2019)\citenamefont
  {B{\oe}hm}, \citenamefont {Cerde{\~n}o}, \citenamefont {Machado},
  \citenamefont {Olivares-Del~Campo}, \citenamefont {Perdomo},\ and\
  \citenamefont {Reid}}]{Boehm:2018sux}%
  \BibitemOpen
  \bibfield  {author} {\bibinfo {author} {\bibfnamefont {C.}~\bibnamefont
  {B{\oe}hm}}, \bibinfo {author} {\bibfnamefont {D.}~\bibnamefont
  {Cerde{\~n}o}}, \bibinfo {author} {\bibfnamefont {P.}~\bibnamefont
  {Machado}}, \bibinfo {author} {\bibfnamefont {A.}~\bibnamefont
  {Olivares-Del~Campo}}, \bibinfo {author} {\bibfnamefont {E.}~\bibnamefont
  {Perdomo}}, \ and\ \bibinfo {author} {\bibfnamefont {E.}~\bibnamefont
  {Reid}},\ }\href {\doibase 10.1088/1475-7516/2019/01/043} {\bibfield
  {journal} {\bibinfo  {journal} {JCAP}\ }\textbf {\bibinfo {volume} {01}},\
  \bibinfo {pages} {043} (\bibinfo {year} {2019})},\ \Eprint
  {http://arxiv.org/abs/1809.06385} {arXiv:1809.06385 [hep-ph]} \BibitemShut
  {NoStop}%
\bibitem [{\citenamefont {O'Hare}(2020)}]{OHare:2020lva}%
  \BibitemOpen
  \bibfield  {author} {\bibinfo {author} {\bibfnamefont {C.~A.~J.}\
  \bibnamefont {O'Hare}},\ }\href {\doibase 10.1103/PhysRevD.102.063024}
  {\bibfield  {journal} {\bibinfo  {journal} {Phys. Rev. D}\ }\textbf {\bibinfo
  {volume} {102}},\ \bibinfo {pages} {063024} (\bibinfo {year} {2020})},\
  \Eprint {http://arxiv.org/abs/2002.07499} {arXiv:2002.07499 [astro-ph.CO]}
  \BibitemShut {NoStop}%
\bibitem [{\citenamefont {Agostini}\ \emph {et~al.}(2022)\citenamefont
  {Agostini} \emph {et~al.}}]{BOREXINO:2021efb}%
  \BibitemOpen
  \bibfield  {author} {\bibinfo {author} {\bibfnamefont {M.}~\bibnamefont
  {Agostini}} \emph {et~al.} (\bibinfo {collaboration} {BOREXINO}),\ }\href
  {\doibase 10.1103/PhysRevLett.128.091803} {\bibfield  {journal} {\bibinfo
  {journal} {Phys. Rev. Lett.}\ }\textbf {\bibinfo {volume} {128}},\ \bibinfo
  {pages} {091803} (\bibinfo {year} {2022})},\ \Eprint
  {http://arxiv.org/abs/2112.11816} {arXiv:2112.11816 [hep-ex]} \BibitemShut
  {NoStop}%
\bibitem [{\citenamefont {Vogel}\ and\ \citenamefont
  {Beacom}(1999)}]{Vogel:1999zy}%
  \BibitemOpen
  \bibfield  {author} {\bibinfo {author} {\bibfnamefont {P.}~\bibnamefont
  {Vogel}}\ and\ \bibinfo {author} {\bibfnamefont {J.~F.}\ \bibnamefont
  {Beacom}},\ }\href {\doibase 10.1103/PhysRevD.60.053003} {\bibfield
  {journal} {\bibinfo  {journal} {Phys. Rev. D}\ }\textbf {\bibinfo {volume}
  {60}},\ \bibinfo {pages} {053003} (\bibinfo {year} {1999})},\ \Eprint
  {http://arxiv.org/abs/hep-ph/9903554} {arXiv:hep-ph/9903554} \BibitemShut
  {NoStop}%
\bibitem [{\citenamefont {Apollonio}\ \emph {et~al.}(2000)\citenamefont
  {Apollonio} \emph {et~al.}}]{CHOOZ:1999hgz}%
  \BibitemOpen
  \bibfield  {author} {\bibinfo {author} {\bibfnamefont {M.}~\bibnamefont
  {Apollonio}} \emph {et~al.} (\bibinfo {collaboration} {CHOOZ}),\ }\href
  {\doibase 10.1103/PhysRevD.61.012001} {\bibfield  {journal} {\bibinfo
  {journal} {Phys. Rev. D}\ }\textbf {\bibinfo {volume} {61}},\ \bibinfo
  {pages} {012001} (\bibinfo {year} {2000})},\ \Eprint
  {http://arxiv.org/abs/hep-ex/9906011} {arXiv:hep-ex/9906011} \BibitemShut
  {NoStop}%
\end{thebibliography}%

\end{document}